\documentclass[a4paper, 11pt]{article}
\usepackage{graphicx}
\usepackage{amsmath, amssymb, amsthm}
\usepackage[hang]{subfigure}
\usepackage{overpic}
\usepackage{xcolor}
\usepackage{mathrsfs}
\usepackage{booktabs}
\usepackage{multirow}
\usepackage[normalem]{ulem}

\newcommand\bxi{\boldsymbol{\xi}}
\newcommand\bx{\boldsymbol{x}}
\newcommand\bn{\boldsymbol{n}}

\newcommand\bv{\boldsymbol{v}}

\newcommand\bg{\boldsymbol{g}}

\newcommand\bbR{\mathbb{R}}
\newcommand\bbN{\mathbb{N}}

\newcommand\dd{\,\mathrm{d}}
\newcommand\Kn{\mathit{Kn}}
\newcommand\imag{\mathrm{i}}
\newcommand \mm{\mathfrak{m}}

\newcommand\mM{\mathcal{M}}
\newcommand\mMa{\mathit{Ma}}

\newcommand\RRe{\mathrm{Re}}
\newcommand\IIm{\mathrm{Im}}

\newcommand\li{\langle i}
\newcommand\jl{j \rangle}
\newcommand\pd[2]{\dfrac{\partial {#1}}{\partial {#2}}}

\newcommand\od[2]{\dfrac{\dd {#1}}{\dd {#2}}}

\newcommand\add[1]{{ #1}}
\newcommand\delete{\bgroup\markoverwith{\textcolor{blue}{\rule[0.5ex]{2pt}{2pt}}}\ULon}
\numberwithin{equation}{section}

\graphicspath{{images/}}

{\theoremstyle{remark} \newtheorem{remark}{Remark}}

\title{Regularized 13-Moment Equations for Inverse Power Law Models}

\author{
  Zhenning Cai\thanks{Department of Mathematics, National
    University of Singapore, Level 4, Block S17, 10 Lower Kent Ridge
    Road, Singapore 119076, email: {\tt matcz@nus.edu.sg}.},~~
  Yanli Wang\thanks{Beijing Computational Science Research Center,
  Beijing, China, 100193, email: {\tt wang\_yanli@csrc.ac.cn}.}}

\begin{document}
\maketitle
\begin{abstract}
We propose a systematic methodology to derive the regularized
thirteen-moment equations in the rarefied gas dynamics for a general
class of linearized collision models. Detailed expressions of the
moment equations are written down for all inverse power law models as
well as the hard-sphere model. By linear analysis, we show that the
equations are stable near the equilibrium. The models are tested for
shock structure problems to show its capability to capture the correct
flow structure in strong nonequilibirum.

\vspace{3pt}

\noindent {\bf Keywords:} regularized 13-moment equations, inverse
power law, shock structure
\end{abstract}


\section{Introduction}
Modeling of gas dynamics has been attracting people's attention for
centuries. Even for the simplest single-species, monatomic gas, while
the classical continuum models such as Euler equations and
Navier-Stokes-Fourier equations work well in most circumstances,
people do find them inadequate when we care about some ``extreme
cases'', such as low-density regime and \add{large velocity slip or
temperature jump at a solid wall due to} gas-surface interaction.
\add{In these cases, the interaction between gas molecules are either
insufficient or severely ruined by gas-surface interactions, resulting
in the failure of the continuum models.} Although some microscopic
models, such as Boltzmann equation, Enskog equation, or even molecular
dynamics, have been validated to be accurate for most applications,
they are usually too expensive to solve due to the high
dimensionality. \add{Despite the fast computers developed nowadays, a
full three-dimensional simulation of the Boltzmann equation still
requires a huge amount of computational resources \cite{Dimarco2018},
and therefore lower-dimensional models are preferable for slip or
early-transitional flows.} Since there is a large gap between the
continuum models and the microscopic models, researchers have been
trying to find models sitting in-between, which are cheaper to
simulate than the kinetic models.

Since Euler equations and Navier-Stokes equations can be considered as
zeroth-order and first-order approximations of the Boltzmann equation
in the continuum limit \cite{Struchtrup2005}, various attempts have
been made to derive higher-order approximations. For example, by
Chapman-Enskog expansion \cite{Enskog, Chapman}, one obtains Burnett
equations and super Burnett equations as third- and fourth-order
approximations \cite{Burnett1936, Reinecke, Shavaliyev}; by Grad's
expansion, equations for stress tensor and heat fluxes can be derived
to provide better closure than the Navier-Stokes and Fourier laws
\add{so that the models are suitable for a wider range of Knudsen
numbers} \cite{Grad1949, Grad1958}; by the assumption of maximum
entropy, Euler equations can be extended to include 14 (or more)
moments \cite{Muller, Dreyer, Levermore}. However, these attempts show
that going beyond Navier-Stokes is quite nontrivial: Burnett and super
Burnett equations are linearly unstable \cite{Bobylev1982}; Grad's
method has the hyperbolicity problem and the convergence problem
\cite{Grad13toR13, Muller}; equations by maximum entropy are still
difficult to solve numerically due to an ill-posed optimization
problem hidden in the equations \cite{Tallec}. These deficiencies have
been severely restricting the applications of these models.

In spite of this, a number of new thoughts have been introduced for
this classical modeling problem. In the current century, all these
classical models are re-studied. Burnett equations have been fixed to
regain linear stability \cite{Bobylev2006, Bobylev2008}; the
hyperbolicity problem of Grad's equations is fixed in \cite{Cai2015a};
approximations of maximum-entropy equations have been proposed which
have explicit analytical expressions \cite{McDonald2013}. At the same
time, Grad's old idea hidden in his notes \cite{Grad1958} has been
picked up to build new models called regularized moment equations
\cite{Struchtrup2003}. Although all these models are quite new, based
on current studies, we find the regularized moment equations to be
interesting due to its relatively complete theory (boundary conditions
\cite{Torrilhon2008}, H-theorem \cite{Struchtrup2007, Torrilhon2012})
and a number of numerical studies \cite{Torrilhon2006}. However, the
complete regularized 13-moment equations have been derived only for
Maxwell molecules. In \cite{Torrilhon2004}, it has been demonstrated
by the example of plane shock structure that the equations derived for
Maxwell molecules are not directly applicable to the hard-sphere
model, \add{which indicates that collision models need to be taken
into account during the model derivation. This inspires us to go
beyond Maxwell molecules, and study more realistic interaction models
directly. In this work, we will focus on inverse power potentials,
which cover both Maxwell molecules and hard-sphere molecules (as the
limit), and have been verified by experiments to be realistic for a
number of gases \cite[Table 8.1]{Torrens}.}

As far as we know, the only regularized moment model derived for
non-Maxwell monatomic molecules is \cite{Struchtrup2013}, which is
the fully linearized equations for hard-sphere molecules. Such a model
cannot be applied in nonlinear regimes such as shock waves. In this
work, we are going to extend the work \cite{Struchtrup2013} and write
down equations for all inverse power law models linearized about the
\emph{local Maxwellian}. \add{The long derivation of the R13 equations
is done by our automated Mathematica code. In \cite{Gupta2012}, the
authors already used computer algebra systems to derive complicated
moment equations, which turns out to be much more efficient than using
pen and paper.} Plane shock wave structures will be computed based on
these 13-moment models to show better results than a simple alteration
of the Maxwell model.

The rest of this paper is organized as follows. In Section
\ref{sec:full_R13}, we first introduce the explicit expressions of the
R13 moment equations, and then show the linear stability analysis.  In
Section \ref{sec:Moment}, the derivation of the R13 moment equations
is presented. Some numerical experiments verifying the capability of
the R13 system are carried out in Section \ref{sec:numerical} and some
concluding remarks are made in Section \ref{sec:conclusion}. A brief
introduction to the Boltzmann equation, the expressions of the
infinite moment equations and the concrete form of the right-hand side
of the R13 equations are given in the appendices.

\section{R13 moment equations for linearized IPL model}
\label{sec:full_R13}
In this section, we are going to present the regularized 13-moment
equations for the IPL model, followed by their linear stability and
dispersion relations. Before that, we start from a quick review of
some properties of the IPL model.

\subsection{A brief review of the IPL model}
The IPL model \add{contains a class of potentials that are frequently
studied in the gas kinetic theory (c.f. \cite{Bird, Cowling, Harris,
Struchtrup}). It} assumes that the potential between two molecules is
proportional to an inverse power of the distance between them:
\begin{displaymath}
\varphi(r) = \frac{\kappa}{1-\eta} r^{1-\eta},
\end{displaymath}
where $\kappa$ specifies the intensity of the force between particles.
Based on this assumption, the viscosity coefficient of the gas in
equilibrium is proportional to a certain power of the temperature of
the gas, which is usually written by $\mu_{\mathrm{ref}} (\theta /
\theta_{\mathrm{ref}})^{\omega}$, where $\theta$ is the temperature
represented in the unit of specific energy:
\begin{displaymath}
\theta = \frac{k_B}{\mm} T
\end{displaymath}
with $T$ being the temperature in Kelvin and $k_B$ being the Boltzmann
constant, and $\mu_{\mathrm{ref}}$ is the reference viscosity
coefficient at temperature $\theta_{\mathrm{ref}}$. The notation $\mm$
is the mass of a single molecule, and the viscosity index $\omega$ is
related to $\eta$ by $\omega = (\eta + 3) / (2\eta - 2)$. When $\eta =
5$, the model is Maxwell molecules, whose viscosity index is $1$; when
$\eta \rightarrow \infty$, the IPL model reduces to the hard-sphere
model, whose viscosity index is $1/2$. Detailed introduction to the
IPL model based on kinetic models is presented in Appendix
\ref{sec:linearized_Bol}.

Below we use the symbol $\mu$ to denote a more familiar ``first
approximation'' of the viscosity coefficient, which is obtained by the
truncated series expansion using Sonine polynomials
\cite{Vincenti1966Introduction}. For IPL models, the value of $\mu$
can be obtained by the following formula \cite{Bird}:
\begin{equation}
  \label{eq:mu}
  \mu = \frac{5 \mm (k_B T / (\mm \pi))^{1/2} (2 k_B T / \kappa)^{2/(\eta - 1)}}{8A_2(\eta)\Gamma[4 - 2/(\eta -1)]},
\end{equation}
with $A_2(\eta)$ being a constant depending only on $\eta$. Some of
the values of this constant are given in Table \ref{tab:coe_A2}.

\begin{table}[ht!]
  \centering
  \def\arraystrech{1.5}
  {\footnotesize
  \begin{tabular}[ht]{c||c|c|c|c|c}
    $\eta$ & $5$ & $7$ & $10$ & $17$ & $\infty$\\
    \hline \hline 
   $A_2(\eta)$ & $0.4362$ & $0.3568$ & $0.3235$ & $0.3079$ & $0.3333$
  \end{tabular} }
  \caption{Coefficients $A_2(\eta)$ for different $\eta$.}
  \label{tab:coe_A2}
\end{table}

\subsection{R13 moment equations}
\label{sec:R13equations}
As the main result of this paper, the R13 moment equations for general
IPL models will be presented in this section. For convenience, the
equations are to be written down using ``primitive variables'', which
are density $\rho$, velocity $v_i$, temperature $\theta$, tracefree
stress tensor $\sigma_{ij}$, and heat flux $q_i$. All the indicies run
from $1$ to $3$. Due to the constraint
$\sigma_{11} + \sigma_{22} + \sigma_{33} = 0$, these variables amount
to 13 quantities as in Grad \cite{Grad1949}. Below we are going to use
the Einstein summation convection without using superscripts. For
instance, the above constraint will be written as $\sigma_{ii} = 0$.

With these 13 variables, the equations for $\rho$, $v_i$ and $\theta$
can be written by
\begin{equation}
  \label{eq:Euler}
  \begin{aligned}
  \od{\rho}{t} + \rho \pd{v_k}{x_k} = 0, \\
  \rho \od{v_i}{t} + \theta \pd{\rho}{x_i} + \rho \pd{\theta}{x_i} + \pd{\sigma_{ik}}{x_k} = 0,  \\
  \frac{3}{2}\rho\od{\theta}{t} + \rho \theta \pd{v_k}{x_k}  +
  \pd{q_k}{x_k} + \sigma_{kl} \pd{v_k}{x_l} = 0,
\end{aligned}
\end{equation}
which are in fact the conservation laws of mass, momentum and
\add{total} energy represented by primitive variables. \add{More
precisely, from the above equations, one can derive the equations for
the momentum density $\rho v_i$ with momentum flux $\rho (v_i v_k +
\theta \delta_{ik}) + \sigma_{ik}$, and for the energy density
$\frac{1}{2} \rho (v_i v_i + 3 \theta)$ with energy flux $\frac{1}{2}
\rho v_k (v_i v_i + 5 \theta) + \sigma_{ik} v_i + q_k$.} To close the
above system, the evolution of the stress tensor $\sigma_{kl}$ and the
heat flux $q_k$ needs to be specified. The system \eqref{eq:Euler}
turns out to be Euler equations if $\sigma_{ik}$ and $q_k$ are set to
be zero.  Finer models based on Chapman-Enskog expansion, such as
Navier-Stokes-Fourier equations, Burnett equations and super-Burnett
equations, represent $\sigma_{ik}$ and $q_k$ using derivatives of
$\rho$, $v_i$ and $\theta$. Following Grad \cite{Grad1949}, the
13-moment equations describe the evolution of $\sigma_{ik}$ and $q_k$
by supplementing \eqref{eq:Euler} with additional equations. \add{Here
we adopt the form used in \cite[eqs. (6.5)(6.6)]{Struchtrup} and write
down these equations as:\footnote{In \cite[eqs.
(6.5)(6.6)]{Struchtrup}, the author uses the notations $u_{ijk}^0$,
$u_{ik}^1$ and $w^2$, while we use the notations $m_{ijk}^{(\eta)}$,
$R_{ij}^{(\eta)}$ and $\Delta^{(\eta)}$. The relations are
\[ m_{ijk}^{(\eta)} = u_{ijk}^0, \quad R_{ij}^{(\eta)} = u_{ij}^1 - (7
+ 2C^{(\eta)}) \theta \sigma_{ij}, \quad \Delta^{(\eta)} = w^2. \]}}
\begin{equation}
  \label{eq:R13_sigma_q}
\begin{aligned}
  \od{\sigma_{ij}}{t} + \sigma_{ij} \pd{v_k}{x_k} +
  \frac{4}{5}\pd{q_{\langle i}}{x_{j\rangle}} + 2 \rho \theta
  \pd{v_{\langle i}}{x_{j\rangle}} + 2\sigma_{k\langle
    i}\pd{v_{j\rangle}}{x_k} + \pd{m_{ijk}^{(\eta)}}{x_k}
  =  \Sigma_{ij}^{(\eta,1)} + \Sigma_{ij}^{(\eta,2)}, &\\
  \od{q_i}{t} + \frac{5}{2}\rho\theta \pd{\theta}{x_i}
  +\frac{5}{2}\sigma_{ik}\pd{\theta}{x_k}+ \theta
  \pd{\sigma_{ik}}{x_k} - \theta\sigma_{ik} \pd{\ln \rho}{x_k} +
  \frac{7}{5}q_k \pd{v_i}{x_k} + \frac{2}{5}q_k\pd{v_k}{x_i} +
  \frac{7}{5} q_i\pd{v_k}{x_k} -
  \frac{\sigma_{ij}}{\rho} \pd{\sigma_{jk}}{x_k} &  \\
  {} + C^{(\eta)}\left(\sigma_{ik}\pd{\theta}{x_k} +
    \theta\pd{\sigma_{ik}}{x_k}\right) +
  \frac{1}{2}\pd{R_{ik}^{(\eta)}}{x_k} +
  \frac{1}{6}\pd{\Delta^{(\eta)}}{x_i} + m_{ijk}^{(\eta)}\pd{v_j}{x_k}
  = Q_{i}^{(\eta,1)}  + Q_{i}^{(\eta, 2)},
\end{aligned} 
\end{equation}
where the angular brackets represent the symmetric and tracefree part
of a tensor \add{($T_{\langle i j \rangle} = \frac{1}{2} (T_{ij} + T_{ji})
- \frac{1}{3} \delta_{ij} T_{kk}$ for any two-tensor $T$)}, and some
values of the constant $C^{(\eta)}$ are listed in Table
\ref{tab:coe_C}. In \eqref{eq:R13_sigma_q}, the newly introduced
variables $m_{ijk}^{(\eta)}$, $R_{ik}^{(\eta)}$ and $\Delta^{(\eta)}$
are the moments contributing to second- and higher-order terms in the
Chapman-Enskog expansion,\footnote{This holds for any molecule
potentials. We refer the readers to \cite[eq. (8,14)]{Struchtrup},
where our $R_{ij}^{(\eta)}$ is denoted by $w_{ij}^1$.} and the
right-hand sides $\Sigma_{ij}^{(\eta,1)}, \Sigma_{ij}^{(\eta,2)}$ and
$Q_i^{(\eta,1)}, Q_i^{(\eta,2)}$ come from the collision between gas
molecules. Here we assume that the collision is linearized about the
local Maxwellian. \add{Up to now, the system is exact but still not
closed. The main contribution of this work is to close the system by
providing expressions of $m_{ijk}^{(\eta)}$, $R_{ik}^{(\eta)}$,
$\Delta^{(\eta)}$ and the right-hand sides using the 13 moments. Note
that the moments $m_{ijk}^{(\eta)}$, $R_{ik}^{(\eta)}$,
$\Delta^{(\eta)}$ are quantities appearing in Grad's 26-moment theory.
More specifically, $m_{ijk}^{(\eta)}$ is the three-tensor formed by
all tracefree third-order moments, and $R_{ik}^{(\eta)}$ and
$\Delta^{(\eta)}$ are fourth-order moments. Below we first provide the
expressions for $\Sigma_{ij}^{(\eta,1)}$ and $Q_i^{(\eta,1)}$:}
\begin{equation}
  \label{eq:S_Q_G13}
  \begin{aligned}
    \Sigma_{ij}^{(\eta, 1)}  &= D_{0}^{(\eta)} \frac{\theta \rho}{\mu}
    \sigma_{ij} + D_{1}^{(\eta)} \left( \pd{v_{\li}}{x_k} \sigma_{\jl
        k} + \pd{v_k}{x_{\li}} \sigma_{\jl k} \right) + D_{2}^{(\eta)}
    \pd{v_k}{x_k} \sigma_{ij} + D_{3}^{(\eta)} \rho
    \theta \pd{v_{\li}}{x_{\jl}}, \\
   & \qquad + D_{4}^{(\eta)} q_{\li} \pd{\ln\theta}{x_{\jl}}  + D_{5}^{(\eta)}
    q_{\li}\pd{\ln \rho}{x_{\jl}} + D_{6}^{(\eta)} \frac{\partial q_{\li}}{\partial
      x_{\jl} } , 
    \\
    Q_i^{(\eta, 1)}   &=E_{0}^{(\eta)} \frac{\theta \rho}{\mu}q_i +
    E_1^{(\eta)}\sigma_{ik}\pd{\theta}{x_k} + E_2^{(\eta)} \theta\sigma_{ik}\pd{\ln \rho}{x_k} 
    + E_3^{(\eta)} q_k\left(\pd{v_k}{x_i} + \pd{v_i}{x_k}\right) \\
    & \qquad + E_4^{(\eta)} q_i\pd{v_k}{x_k}  + E_5^{(\eta)} \theta
    \pd{\sigma_{ki}}{x_k}  + E_6^{(\eta)} \theta \rho \pd{\theta}{x_i},
  \end{aligned}
\end{equation}
where the coefficients $D_i^{(\eta)}$ and $E_i^{(\eta)}$ are partially
tabulated in Table \ref{tab:coe_R13}. Note that we have intentionally
split the right-hand sides in \eqref{eq:R13_sigma_q} into two parts,
so that the ``generalized Grad 13-moment (GG13) equations'' can be
extracted from
\eqref{eq:R13_sigma_q} by setting
\begin{equation} \label{eq:GG13}
m_{ijk}^{(\eta)} = R_{ik}^{(\eta)} = \Delta^{(\eta)} =
  \Sigma_{ij}^{(\eta,2)} = Q_i^{(\eta,2)} = 0.
\end{equation}
The GG13 equations are introduced \cite{Struchtrup2005} by the order
of magnitude method, and its fully linearized version for hard spheres
has been derived in \cite{Struchtrup2013}.

\begin{table}[ht!]
  \centering
  \def\arraystrech{1.5}
  {\footnotesize
  \begin{tabular}[ht]{c||c|c|c|c|c}
    $\eta$ & $5$ & $7$ & $10$ & $17$  & $\infty$\\
    \hline \hline 
   $C^{(\eta)}$ & $0$ & $-0.0715$  & $-0.1193$ & $ -0.1615$ & $-0.2161$
  \end{tabular} }
  \caption{Coefficients $C^{(\eta)}$ for different $\eta$. }
  \label{tab:coe_C}
\end{table}

To give the R13 equations, we need to close \eqref{eq:R13_sigma_q} by
specifying $m_{ijk}^{(\eta)}$, $R_{ik}^{(\eta)}$, $\Delta^{(\eta)}$,
$\Sigma_{ij}^{(\eta, 2)}$, and $Q_i^{(\eta, 2)}$. The closure depends
on the specific form of the collision model. Here we again assume that
the collision is linearized about the local Maxwellian. Thus the R13
theory gives the following closure:
\begin{equation}
  \label{eq:R13_mdelR}
  \begin{aligned}
    m_{ijk}^{(\eta)} & = \frac{\mu}{ \theta \rho} \left(A_1^{(\eta)}
      q_{\langle i} \pd{v_{j}}{x_{k\rangle}} + A_2^{(\eta)}\theta
      \sigma_{\langle ij}
      \pd{\ln \rho}{x_{k\rangle}} + A_3^{(\eta)} \sigma_{\langle ij}
      \pd{\theta}{x_{k\rangle}} + A_4^{(\eta)} \theta
      \pd{\sigma_{\langle ij}}{x_k\rangle}\right), \\
    \Delta^{(\eta)} & = \frac{\mu}{\theta \rho}\left(B_1^{(\eta)}
      \theta \pd{q_k}{x_k} + B_2^{(\eta)} \theta \sigma_{ij}
      \pd{v_i}{x_j} + B_3^{(\eta)} q_k \pd{\theta}{x_k} + B_4^{(\eta)}
      \theta q_k
      \pd{\ln \rho}{x_k}  \right),  \\
    R_{ij}^{(\eta)} & = C_0^{(\eta)} \theta \sigma_{ij} +
    \frac{\mu}{\theta \rho} \left( C_1^{(\eta)} \theta
      \pd{q_{\langle i}}{x_{j\rangle}} + C_2^{(\eta)} \theta
      \sigma_{k\langle i} \pd{v_{j\rangle}}{x_k} +
      C_3^{(\eta)} \theta^2 \rho \pd{v_{\langle i}}{x_{j\rangle}} + C_4^{(\eta)}
      q_{\langle i} \pd{\theta}{x_{j\rangle}} + C_5^{(\eta)} \theta
      q_{\langle i} \pd{\ln
        \rho }{x_{j\rangle}} \right).
  \end{aligned}
\end{equation} 
Some values of the coefficients $A_i^{(\eta)}$, $B_i^{(\eta)}$, and
$C_i^{(\eta)}$ are given in Table \ref{tab:coe_R13}. The full
expressions of $\Sigma_{ij}^{(\eta, 2)}$ and $Q_i^{(\eta, 2)}$ are
quite lengthy and we provide them in Appendix \ref{sec:SQ_2}. A simple
case is $\eta = 5$, for which we have
\begin{equation}
  \label{eq:S_Q_5}
  \begin{aligned}
    \Sigma_{ij}^{(5, 2)} & = 0, \qquad Q_i^{(5, 2)} = 0,
  \end{aligned}
\end{equation}
and the corresponding model matches the one derived for Maxwell
molecules in \cite{Struchtrup2003} (with terms nonlinear in
$\sigma_{ij}$ and $q_i$ removed since we use the linearized collision
model).

\begin{table}[ht]
  \centering
  \def\arraystrech{1.5}
  {\footnotesize
  \begin{tabular}[ht]{r||r||r|r|r|r|r|r|r}
    $\eta$ & & 0&1 & 2 &3 & 4 & 5 & 6\\
    \hline     \hline 
    \multirow{3}*{$5$} & $A_i^{(\eta)}$& $\times$ &$-1.6$ & $2$ & $0$& $-2$&  $\times$ & $\times$\\
           &      $B_i^{(\eta)}$ & $\times$ & $-12$ & $-12$ & $-30$& $12$&  $\times$ & $\times$\\ 
           &      $C_i^{(\eta)}$ & $0$ & $-4.8$ & $-6.8571$ &  $0$ & $-4.8$&  $4.8$ & $\times$ \\
           &      $D_i^{(\eta)}$ & $-1.0$ & $0$ & $0$ &  $0$ & $0$&  $0$ & $0$ \\
           &      $E_i^{(\eta)}$ & $-0.6667$ & $0$ & $0$ &  $0$ & $0$&  $0$ & $0$ \\
    \hline \hline 
    \multirow{3}*{$7$} & $A_i^{(\eta)}$&  $\times$ & $-1.4430$ & $2.0094$ & $0.2417$ & $-1.9679$ & $\times$  & $\times$ \\
           &      $B_i^{(\eta)}$ &  $\times$ & $-11.4061$ & $-11.2200$ & $-27.5608$ & $12.0836$ & $\times$  & $\times$\\
           &      $C_i^{(\eta)}$ &   $-0.1226$ & $-4.7155$ & $-6.6789$ & $-0.2456$ & $-3.7894$ & $4.7705$ &  $\times$ \\
           &      $D_i^{(\eta)}$ &$-0.9967$ & $0.0404$ & $-0.0269$ & $0.0030$ & $0.0254$ & $-0.0576$ & $0.0569$ \\
           &      $E_i^{(\eta)}$ & $-0.6636$ & $0.1075$ & $0.0005$ & $0.0441$ & $-0.0294$ & $0.0$ & $0.0046$ \\
    \hline \hline 
    \multirow{3}*{$10$} & $A_i^{(\eta)}$& $\times$ &$-1.3445$ & $2.0266$ & $0.40152$ & $-1.9562$&  $\times$ & $\times$\\
           &      $B_i^{(\eta)}$ & $\times$  &$-11.0838$ & $-10.7755$ & $-26.1567$ & $12.2346$&  $\times$& $\times$\\ 
           &      $C_i^{(\eta)}$ &  $-0.2051$ & $-4.6797$ & $-6.5941$ & $-0.4122$ & $-3.1530$ & $4.7727$ & $\times$\\
           &      $D_i^{(\eta)}$ &  $-0.9909$ &$ 0.0669$ & $-0.0446$ & $0.0085$ & $0.0347$ & $-0.0967$ & $0.0948$  \\
           &      $E_i^{(\eta)}$ & $-0.6582$ & $0.1757$ & $0.0015$ & $0.0733$ & $-0.0489$ & $-0.0001$ & $0.0129$ \\
    \hline \hline 
    \multirow{3}*{$17$} & $A_i^{(\eta)}$& $\times$ & $-1.2609$ & $2.0493$ & $0.5429$ & $-1.9523$ &  $\times$& $\times$ \\
           &      $B_i^{(\eta)}$ & $\times$ & $-10.8444$ & $-10.4280$ & $-25.0494$ & $12.4328$ &  $\times$ & $\times$\\ 
           &      $C_i^{(\eta)}$ &   $-0.2784$ & $-4.6617$ & $-6.5415$ & $-0.5618$ & $-2.6123$ & $4.7895$& $\times$ \\
           &      $D_i^{(\eta)}$ & $-0.9834$ & $0.0900$ & $-0.0600$& $0.0156$ & $0.0379$ & $-0.1316$ & $0.1279$ \\
           &      $E_i^{(\eta)}$ & $-0.6512$ &$ 0.2332$ & $0.0028$ & $0.0987$ & $-0.0658$ & $-0.0002$ & $0.0237$ \\
    \hline \hline 
    \multirow{3}*{$\infty$} & $A_i^{(\eta)}$& $\times$ &$-1.1542$ & $2.0902$ & $0.7315$ & $-1.9562$ &  $\times$ & $\times$\\
           &      $B_i^{(\eta)}$ & $\times$ & $-10.5892$ & $-10.0287$ & $-23.7646$ & $12.7856$ &  $\times$ & $\times$\\ 
           &      $C_i^{(\eta)}$ &  $-0.3749$ & $-4.6572$ & $-6.5039$ & $-0.7619$ &  $-1.9262$ & $4.8328$& $\times$ \\
           &      $D_i^{(\eta)}$ & $-0.9703$ & $0.1195$ & $-0.0797$ & $0.0282$ & $0.0347$ & $-0.1773$ & $0.1707$ \\
           &      $E_i^{(\eta)}$ &  $-0.6392$ & $0.3049$ & $0.0052$ & $0.1313$ & $-0.0875$ & $-0.0003$ & $0.0427$ \\
  \end{tabular} 
}
\caption{Coefficient of $A_i^{(\eta)}$, $B_i^{(\eta)}$, $C_i^{(\eta)}$ , $D_i^{(\eta)}$ and $E_i^{(\eta)}$for different $\eta$.}   
\label{tab:coe_R13}
\end{table}

\add{%
\subsection{Discussion on the order of accuracy} \label{sec:order}
One possible way to describe the accuracy of the moment models in the
near-continuum regime is to use the notion of ``order of accuracy''
\cite{Struchtrup2005,Struchtrup}. In such a regime, the Knudsen number
$\Kn$, i.e. the ratio of the mean free path to the characteristic
length of the problem, is regarded as a small number.  Thus
Chapman-Enskog expansion can be applied, and all non-equilibrium
moments are expanded into power series of $\Kn$, e.g.,
\begin{equation} \label{eq:CE_sigmaq}
\begin{aligned}
\sigma_{ij} &= \Kn\,\sigma_{ij}^{(1)} + \Kn^2 \sigma_{ij}^{(2)} + \Kn^3 \sigma_{ij}^{(3)} + \cdots, \\
q_i &= \Kn\,q_i^{(1)} + \Kn^2 q_i^{(2)} + \Kn^3 q_i^{(3)} + \cdots.
\end{aligned}
\end{equation}
By asymptotic analysis, all these terms can be represented by the
conservative variables and their derivatives. Truncating the above
series up to the term $\Kn^k$ and inserting the result into
\eqref{eq:Euler}, one obtains moment equations with $k$th order of
accuracy. By this approach, the models derived from Chapman-Enskog
expansion up to zeroth to third order are, respectively, Euler
equations, Navier-Stokes-Fourier equations, Burnett equations, and
super-Burnett equations. These equations contain only equilibrium
variables: density, velocity and temperature.

In the 13-moment model, one can also assume $\Kn$ is small and apply
the expansion \eqref{eq:CE_sigmaq} to obtain models including only
equilibrium variables. Suppose the second-order Chapman-Enskog
expansion of a moment model agrees with the Burnett equations, while
its third-order Chapman-Enskog expansion differs from super-Burnett
equations, then we say that the moment model has the second-order
accuracy (or Burnett order). For example, Grad's 13-moment equations
have the first-order accuracy for general IPL potentials, but have
second-order accuracy for Maxwell molecules; GG13 equations are
extensions to Grad's 13-moment theory to achieve second-order accuracy
for all molecule potentials. In general, the expansion
\eqref{eq:CE_sigmaq} is usually obtained by multiplying the equations
of $\sigma_{ij}$ and $q_i$ in the moment system by $\Kn$. Therefore in
most cases, a 13-moment model has $k$th-order accuracy if the
equations for $\sigma_{ij}$ and $q_i$ \eqref{eq:R13_sigma_q} are
accurate up to the $(k-1)$th order. R13 equations have the third-order
accuracy, as the closure \eqref{eq:R13_mdelR} provides the equations
\eqref{eq:R13_sigma_q} exact second-order contributions.

In the literature, there exist some similar 13-moment models obtained
by other approaches to include second-order derivatives in the
equations for $\sigma_{ij}$ and $q_i$. For instance, the relaxed
Burnett equations \cite{Jin2001} are also derived for arbitrary
interaction potentials, and as mentioned in \cite{Jin2001,
Struchtrup}, these equations have second-order accuracy. Another
similar model is the NCCR (Nonlinear Coupled Constitutive Relations)
equations \cite{Myong1999}. These equations do not include information
from the Burnett order, and therefore they have the first-order
accuracy and distinguish different interaction models by viscosity and
heat conductivity coefficients. The full R13 models for Maxwell
molecules and the BGK model, which have the third-order accuracy, have
been derived in \cite{Struchtrup2003} and \cite{Struchtrup}, and the
linear R13 equations for the hard-sphere model have been derived in
\cite{Struchtrup2013}.
}

\subsection{Linear stability and dispersion}
For the newly proposed R13 equations for IPL models, we are going to
check some of its basic properties in this work. In this section, we
focus on the linear properties including its stability in time and
space, and the dispersion and damping of sound waves.

Following \cite{Struchtrup2003,Struchtrup,Struchtrup2013}, we apply
the analysis to one-dimensional linear dimensionless equations. The
linearization is performed about a global equilibrium state with
density $\rho_0$, zero velocity, and temperature $\theta_0$. The
derivation of the one-dimensional linear dimensionless equations
consists of the following steps:
\begin{enumerate}
\item Introduce the \emph{small} dimensionless variables $\hat{\rho}$,
  $\hat{\theta}$, $\hat{v}_i$, $\hat{\sigma}_{ij}$, and $\hat{q}_i$ by
  \begin{equation}
    \label{eq:dimension_var}
    \rho = \rho_0(1 + \hat{\rho}), \quad \theta = \theta_0(1 +
    \hat{\theta}), \quad v_i = \sqrt{\theta_0} \hat{v}_i,  \quad 
    \sigma_{ij} = \rho_0 \theta_0 \hat{\sigma}_{ij}, \quad  q_i =
    \rho_0 \sqrt{\theta_0}^3 \hat{q}_i.
  \end{equation}
\item Let $L$ be the characteristic length, and define the
  dimensionless space and time variables by
  \begin{equation} \label{eq:space_time}
  x_i = L \hat{x}_i, \quad t = \frac{L}{\sqrt{\theta_0}}\hat{t}.
  \end{equation}
\item Substitute \eqref{eq:dimension_var}\eqref{eq:space_time} into
  the R13 equations \eqref{eq:Euler}\eqref{eq:R13_sigma_q} and
  \eqref{eq:R13_mdelR}, and drop all the terms nonlinear in the
  variables with hats introduced in \eqref{eq:dimension_var}.
\item Reduce the resulting equations to the one-dimensional system by
  dropping all the terms with derivatives with respect to $x_2$ and
  $x_3$, and setting
  \begin{equation}
  \begin{gathered}
  \hat{v}_1 = \hat{v}, \quad \hat{q}_1 = \hat{q}, \quad
  \hat{\sigma}_{11} = \hat{\sigma}, \quad
  \hat{\sigma}_{22} = \hat{\sigma}_{33} = -\frac{1}{2} \hat{\sigma}, \\
  \hat{v}_2 = \hat{v}_3 = \hat{q}_2 = \hat{q}_3 =
    \hat{\sigma}_{12} = \hat{\sigma}_{23} = \hat{\sigma}_{13} = 0.
  \end{gathered}
  \end{equation}
\end{enumerate}
The resulting equations can be written down more neatly if we
introduce the Knudsen number
\begin{equation} \label{eq:Kn}
  \Kn = \frac{\mu_0 \sqrt{\theta_0}}{\rho_0 \theta_0 L},
\end{equation}
where $\mu_0$ is the viscosity coefficient at temperature $\theta_0$.
For all IPL models, the one-dimensional linear dimensionless equations
have the form:
\begin{equation}
  \label{eq:1d}
  \begin{aligned}
    & \pd{\hat{\rho}}{\hat{t}} + \pd{\hat{v}}{\hat{x}} = 0, \\
    & \pd{\hat{v}}{\hat{t}} + \pd{\hat{\theta}}{\hat{x}} +
    \pd{\hat{\rho}}{\hat{x}} +
    \pd{\hat{\sigma}}{\hat{x}} = 0, \\
    & \pd{\hat{\theta}}{\hat{t}} + \frac{2}{3} \pd{\hat{q}}{\hat{x}} +
    \frac{2}{3} \pd{\hat{v}}{\hat{x}} = 0, \\
    & \uline{ \pd{\hat{\sigma}}{\hat{t}} +
      \frac{8}{15}\pd{\hat{q}}{\hat{x}}} +
    \frac{4}{3}\pd{\hat{v}}{\hat{x}} + \uuline{\Kn
      \alpha_1^{(\eta)}\pd{^2\hat{\sigma}}{\hat{x}^2}} =
    \frac{\alpha_2^{(\eta)}}{\Kn} \hat{\sigma} + \uline{
      \frac{\alpha_3^{(\eta)}}{\Kn} \hat{\sigma} +
      \alpha_4^{(\eta)}\pd{\hat{q}}{\hat{x}}
      + \alpha_5^{(\eta)}\pd{\hat{v}}{\hat{x}}} + \\
    & \qquad \uuline{ \frac{\alpha_6^{(\eta)}}{\Kn} \hat{\sigma} +
      \alpha_7^{(\eta)}\pd{\hat{q}}{\hat{x}} +
      \alpha_8^{(\eta)}\pd{\hat{v}}{\hat{x}} + \Kn
      \left(\alpha_9^{(\eta)} \pd{^2\hat{\theta}}{\hat{x}^2} +
        \alpha_{10}^{(\eta)}\pd{^2\hat{\rho}}{\hat{x}^2} +
        \alpha_{11}^{(\eta)}\pd{^2\hat{\sigma}}{\hat{x}^2} \right)},
    \\
    &\uline{ \pd{\hat{q}}{\hat{t}} + \beta_1^{(\eta)}
      \pd{\hat{\sigma}}{\hat{x}}} +
    \frac{5}{2}\pd{\hat{\theta}}{\hat{x}} + \uuline{\beta_2^{(\eta)}
      \pd{\hat{\sigma}}{\hat{x}} + \beta_3^{(\eta)}
      \Kn\pd{^2\hat{q}}{\hat{x}^2} + \beta_4^{(\eta)} \Kn
      \pd{^2\hat{v}}{\hat{x}^2}} = \frac{\beta_5^{(\eta)}}{\Kn}
    \hat{q} + \\
    & \qquad \uline{\frac{\beta_6^{(\eta)}}{\Kn} \hat{q} +
      \beta_7^{(\eta)}\pd{\hat{\theta}}{\hat{x}} + \beta_8^{(\eta)}
      \pd{\hat{\sigma}}{\hat{x}}} +
    \uuline{\frac{\beta_9^{(\eta)}}{\Kn} \hat{q} +
      \beta_{10}^{(\eta)}\pd{\hat{\theta}}{\hat{x}} +
      \beta_{11}^{(\eta)} \pd{\hat{\sigma}}{\hat{x}} + \Kn \left(
        \beta_{12}^{(\eta)} \pd{^2 \hat{q}}{\hat{x}^2} +
        \beta_{13}^{(\eta)} \pd{^2\hat{v}}{\hat{x}} \right)},
  \end{aligned}
\end{equation}
where $\alpha_i^{(\eta)}$ and $\beta_i^{(\eta)}$ depend only on
$\eta$, and their values for some choices of $\eta$ are listed in
Table \ref{tab:coe_linear}. In \eqref{eq:1d}, if we replace all the
terms with double underlines by zero, we obtain the linearized GG13
equations. Furthermore, if we set all the terms with both single and
double underlines to be zero, then the result is the linearized
Navier-Stokes-Fourier equations. \add{The left-hand side of
\eqref{eq:1d} comes from the advection, and the right-hand side comes
from the collision. By Table \ref{tab:coe_linear}, it can be clearly
seen that when $\eta = 5$ (Maxwell molecules), due to the simplicity
of the collision operator, all the underlined terms on the right-hand
side disappear.}

\begin{table}[ht!]
  \centering
   \def\arraystrech{1.5}
  {\footnotesize
    \begin{tabular}[ht]{r||r||r|r|r|r|r|r|r}
      $\eta$ & & 1 & 2 &3 & 4 & 5 & 6 & 7\\ 
      \hline     \hline 
      \multirow{2}*{$5$} & $\alpha_i^{(\eta)}$& $-1.2$ & $-1$ & $0$ 
                          & $0$ &  $0$ & $0$ & $0$ \\
             &      $\beta_i^{(\eta)}$ & $1$ & $0$ & $-3.6$ & $0$
                              &$-2/3$  & $0$ & $0$\\ 
      \hline 
      \multirow{2}*{$7$} & $\alpha_i^{(\eta)}$& $-1.1808$ & $-0.9983$ & $0.0015$ & $0.0380$ & $0.0020$ & $0.0013$ & $0.0217$   \\ 
             &      $\beta_i^{(\eta)}$ &  $0.9285$ & $-0.0613$ & $-3.4729$ & $-0.0819$ & $-0.6648$ & $0.0012$ &$ 0.0046$ \\
      \hline 
      \multirow{2}*{$10$} & $\alpha_i^{(\eta)}$&  $-1.1737$ & $-0.9951$ & $0.0042$ & $0.0632$ & $0.0056$ & $0.0037$ & $0.0362$  \\ 
             & $\beta_i^{(\eta)}$ &  $0.8806$ & $-0.1026$ & $-3.4072$ & $-0.1374$ & $-0.6616$ & $0.0034$ & $0.0129$ \\  
      \hline 
      \multirow{2}*{$17$} & $\alpha_i^{(\eta)}$& $-1.1714$ & $-0.9911$ & $0.0077$ & $0.0853$ & $0.0104$ & $0.0068$ & $0.0490$   \\ 
             &      $\beta_i^{(\eta)}$ &$0.8385$ & $-0.1392$ & $-3.3613$ & $-0.1873$ & $-0.6574$ & $0.0062$ & $0.0237$ \\ 
      \hline
      \multirow{2}*{$\infty$} & $\alpha_i^{(\eta)}$& $-1.1737$ & $-0.9842$ & $0.0139$ & $0.1138$ & $0.0188$ & $0.0124 $& $0.0659$ \\
             &      $\beta_i^{(\eta)}$ & $0.7839$ & $-0.1875$ & $-3.3173$ & $-0.2540 $& $-0.6503$ & $0.0111$ & $0.0427$   \\
      \hline \hline
      $\eta$ & & 8 & 9 &10 & 11 & 12 & 13 &  \\ 
      \hline     
      \multirow{2}*{$5$} & $\alpha_i^{(\eta)}$& $0$ & $0$ & $0$ & $0$& $\times$ &$\times$ \\
             &      $\beta_i^{(\eta)}$ & $0$ & $0$ & $0$ & $0$& $0$  & $0$\\ 
      \hline 
      \multirow{2}*{$7$} & $\alpha_i^{(\eta)}$ & $0.0017$ & $0.0809$ & $0.0017$ & $0.0105$& $\times$ &$\times$ \\
             &      $\beta_i^{(\eta)}$  & $0.0$ & $0.0008$ & $0.0031$ & $-0.0003$ & $-0.1440$ & $0.0011$\\ 
      \hline 
      \multirow{2}*{$10$} & $\alpha_i^{(\eta)}$ & $0.0049$ & $0.1352$ & $0.0050$ & $0.0202$ & $\times$  &$\times$ \\
             &      $\beta_i^{(\eta)}$ & $-0.0001$ & $0.0023$ & $0.0087$ & $-0.0008$ & $-0.2225$ & $0.0036$\\ 
      \hline 
      \multirow{2}*{$17$} & $\alpha_i^{(\eta)}$ & $0.0091$ & $0.1840$ & $0.0093$ & $0.0308$ &$\times$ &$\times$ \\
             &      $\beta_i^{(\eta)}$ & $-0.0002$ & $0.0042$ & $0.0161$ & $-0.0012$ & $-0.2814$ & $0.0076$\\
      \hline 
      \multirow{2}*{$\infty$} & $\alpha_i^{(\eta)}$   &  $0.0168$ & $0.2499$ & $0.0173$ & $0.0476$ & $\times$ &$\times$\\
             &      $\beta_i^{(\eta)}$  &   $-0.0003$ & $0.0077$ & $0.0295$ & $-0.0015$ & $-0.3451$ & $0.0160$
    \end{tabular}
  }
  \caption{Coefficients of the linearized R13 system for different $\eta$.}
  \label{tab:coe_linear}
\end{table}

For simplicity, we will omit the hats on the variables hereafter. In
general, the linear GG13 or R13 system has the form
\begin{equation}
  \label{eq:linear_sys}
  \pd{u_A}{t} +  \mathcal{A}_1^{(\eta)} \pd{u_A}{x} +
  \mathcal{A}_2^{(\eta)} \pd{^2 u_A}{x^2} + \mathcal{A}_3^{(\eta)} u_A
  = 0,
\end{equation}
where $u_A = (\rho , v, \theta, \sigma, q)^T$ and the matrices
$\mathcal{A}_i^{(\eta)}$ are constant matrices which can be observed
from \eqref{eq:1d}. To study the linear waves, we
consider the plane wave solution:
\begin{equation}
  \label{eq:linear_wave}
  u_A(x, t) = \tilde{u}_A\exp[\imag (\Omega t - kx)], 
\end{equation}
where $\tilde{u}_A$ is the initial amplitude of the wave, $\Omega$ is
frequency and $k$ is the wave number. Inserting the above solution
into \eqref{eq:linear_sys} yields
\begin{equation}
  \label{eq:linear_sys_sol}
    \mathcal{G}^{(\eta)}\tilde{u}_A = 0, \quad \text{where} \quad
  \mathcal{G}^{(\eta)} =
  \left(\imag \Omega - \imag k \mathcal{A}_1^{(\eta)}  - k^2 
    \mathcal{A}_2^{(\eta)}  + \mathcal{A}_3^{(\eta)}\right). 
\end{equation}
and the existence of nontrivial solutions requires
\begin{equation}
  \label{eq:dispersion}
  \det[\mathcal{G}^{(\eta)}] = 0.
\end{equation}
From \eqref{eq:dispersion}, we can get the relation between $\Omega$
and $k$, and thus all the desired properties such as the amplification
and dispersion of the linear waves can naturally be obtained. \add{In
our analysis below, we choose $\Kn = 1$ to get quantitative results.}

\add{%
\begin{remark}
The equations \eqref{eq:1d} in the case $\eta = \infty$ can be used to
compare with the results in \cite{Struchtrup2013} for cross-checking.
Small deviation between our coefficients and the coefficients in
\cite{Struchtrup2013} can be observed. For example, in
\cite{Struchtrup2013}, the value of $\alpha_2^{(\infty)}$ is $-0.98632$,
while our analysis gives $\alpha_2^{(\infty)} = -0.9842$. We believe
that such discrepancies are due to different truncation when inverting
the collision operator during the derivation. According to the
method reported in \cite[eq. (23)]{Struchtrup2013}, our result is
probably more accurate since we preserve more terms in the truncation.
Details are to be given in Section \ref{sec:Moment}.
\end{remark}
}

\subsubsection{Linear stability in time and space} 
\begin{figure}[!ht]
\centering
\subfigure[G13]{%
  \begin{overpic}
  [width=.45\textwidth, clip]{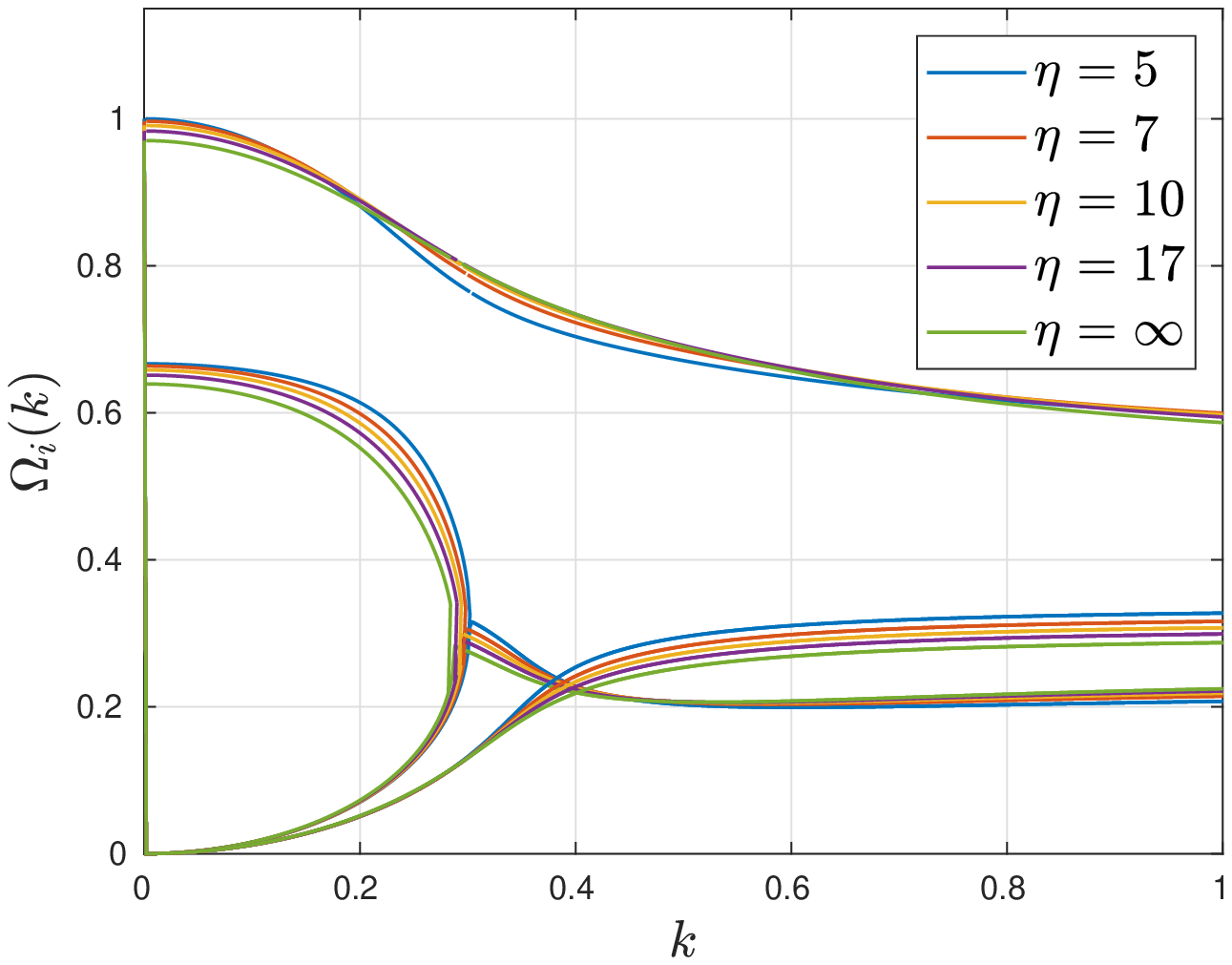}  
  \end{overpic}
  } \quad
\subfigure[R13]{%
  \begin{overpic}
    [width=.45\textwidth, clip]{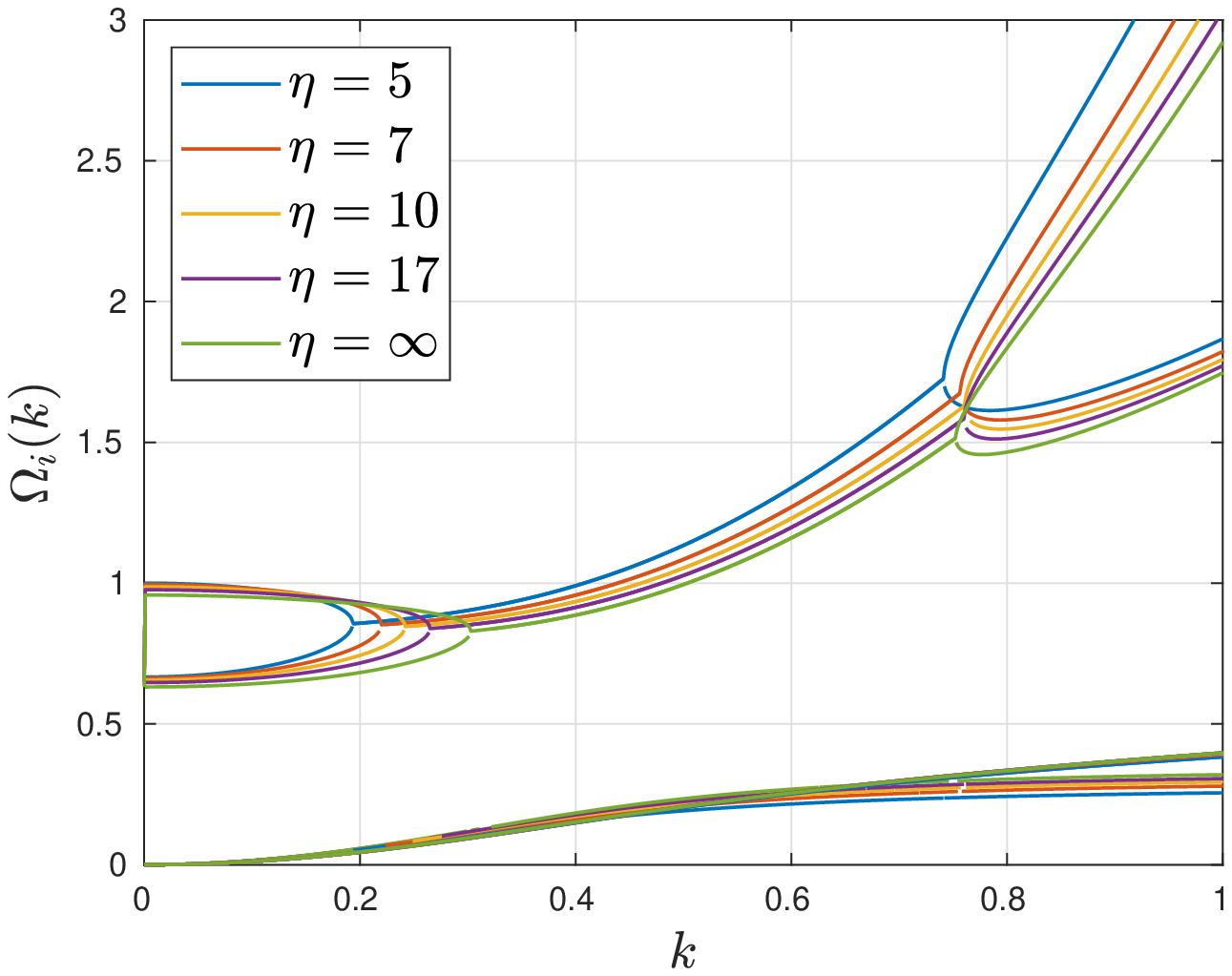}    
  \end{overpic}
}  
\caption{Damping coefficients $\Omega_i(k)$ of the G13 and R13 systems
  for different $\eta$.}
\label{fig:stability_time}
\end{figure}

We first discuss the linear stability of the generalized G13 and R13
systems in time and space. For the time stability, we require that the
norm of the amplitude decreases with time for any given wave number $k
\in \mathbb{R}$. Precisely, if we assume $\Omega = \Omega_r(k) +
\imag\Omega_i(k)$, the time stability requires $\Omega_i(k) \geqslant
0$. Figure \ref{fig:stability_time} shows possible values of
$\Omega_i(k)$ on the complex plane. Note that $\Omega_i(k)$ is a
multi-valued function since \eqref{eq:dispersion} may have multiple
solutions for a given $k$. It is observed that for all choices $\eta$,
the values of $\Omega(k)$ always locate on the upper half of the
complex plane, indicating the linear stability for both R13 and
generalized G13 equations.

\begin{figure}[!ht]
\centering
\subfigure[G13]{%
  \begin{overpic}
  [width=.45\textwidth,height = 0.35\textwidth, clip]{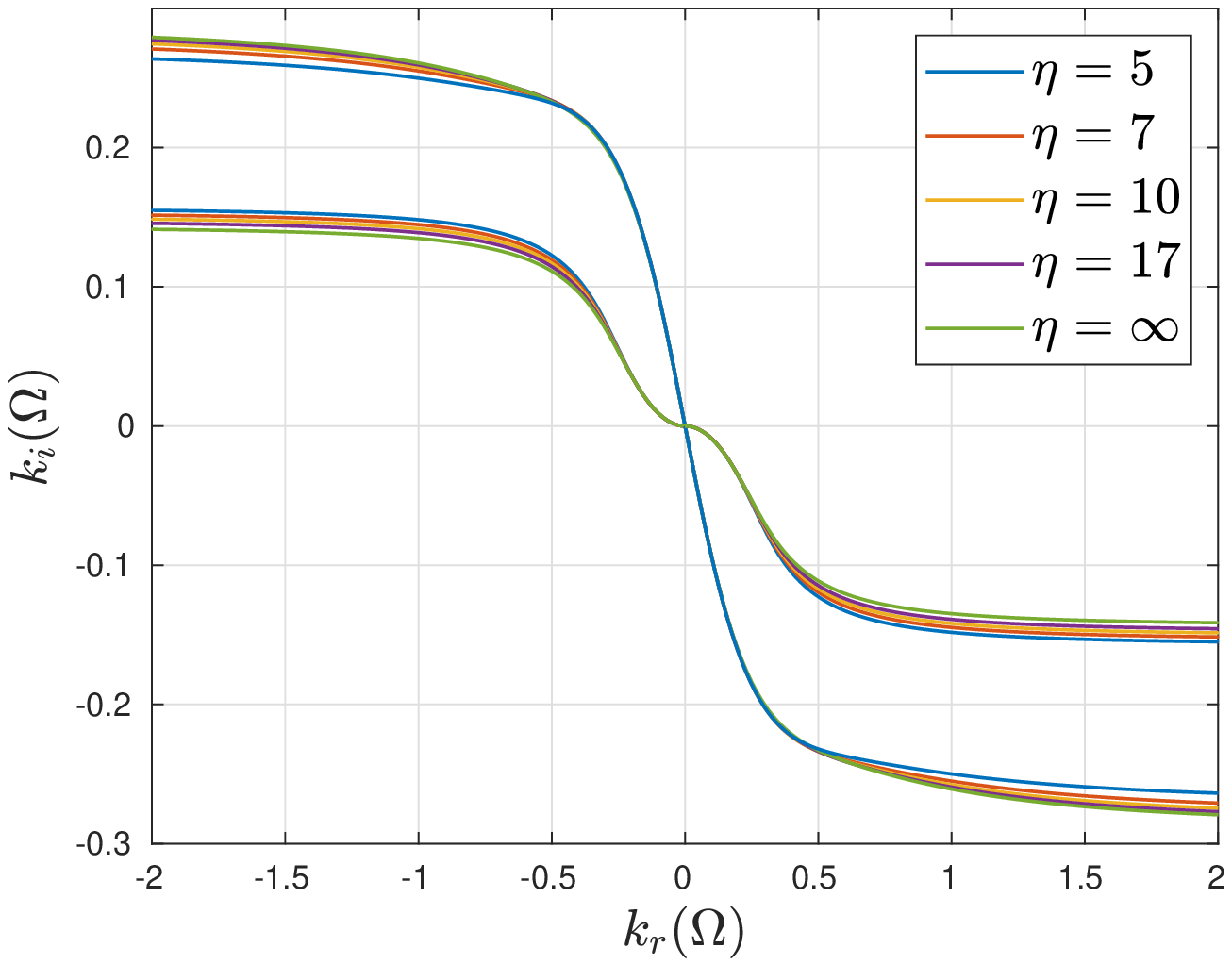}  
  \end{overpic}
  } \quad
\subfigure[R13]{%
  \begin{overpic}
    [width=.45\textwidth, height = 0.35\textwidth,clip]{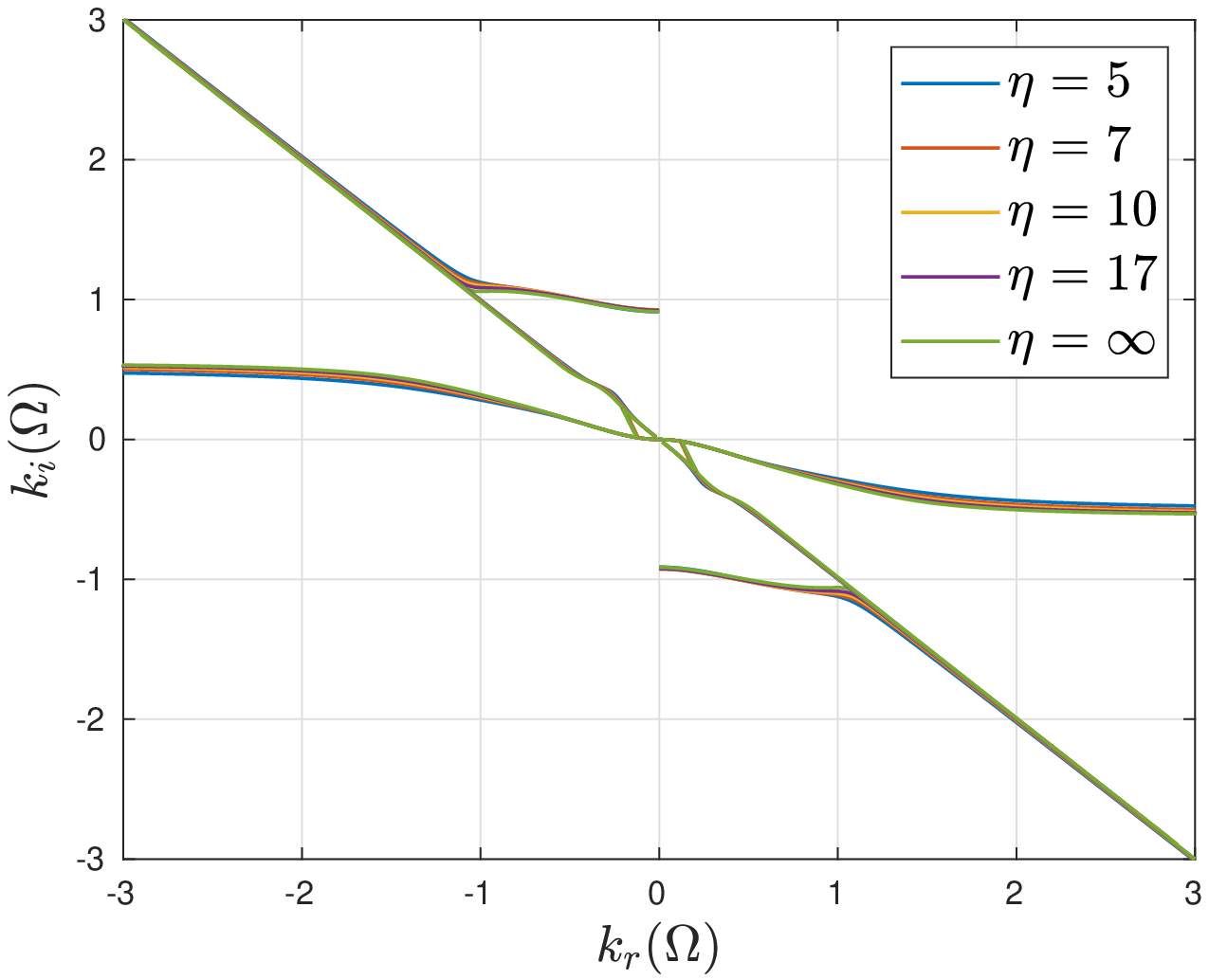}    
  \end{overpic}
} 
\caption{The solutions $k(\Omega)$ of the dispersion relation in the
  complex plane with $\Omega$ as parameter of the G13 and R13 systems
  for different $\eta$.}
\label{fig:stability_space}
\end{figure}

For the stability in space, we require that for a given wave
frequency, the amplitude should not increase along the direction of
wave propagation. Now we assume that $\Omega \in \mathbb{R}$ is given,
and let $k = k_r(\Omega) + \imag k_i(\Omega)$. Then the wave is stable
in space if $k_r(\Omega) k_i(\Omega) \leqslant 0$. Figure
\ref{fig:stability_space} shows the values of $k$ on the complex plane
with $\Omega$ as the parameter. The results show that all the curves
do not enter the upper right or the lower left quadrant for both
R13 and generalized G13 equations, showing the spatial stability for
both models. Again, such a stability result holds for all $\eta$
considered in our experiments.

\subsubsection{Dispersion and damping} 
\begin{figure}[!ht]
\centering
\subfigure[GG13]{%
  \begin{overpic}
    [width=.45\textwidth, clip]{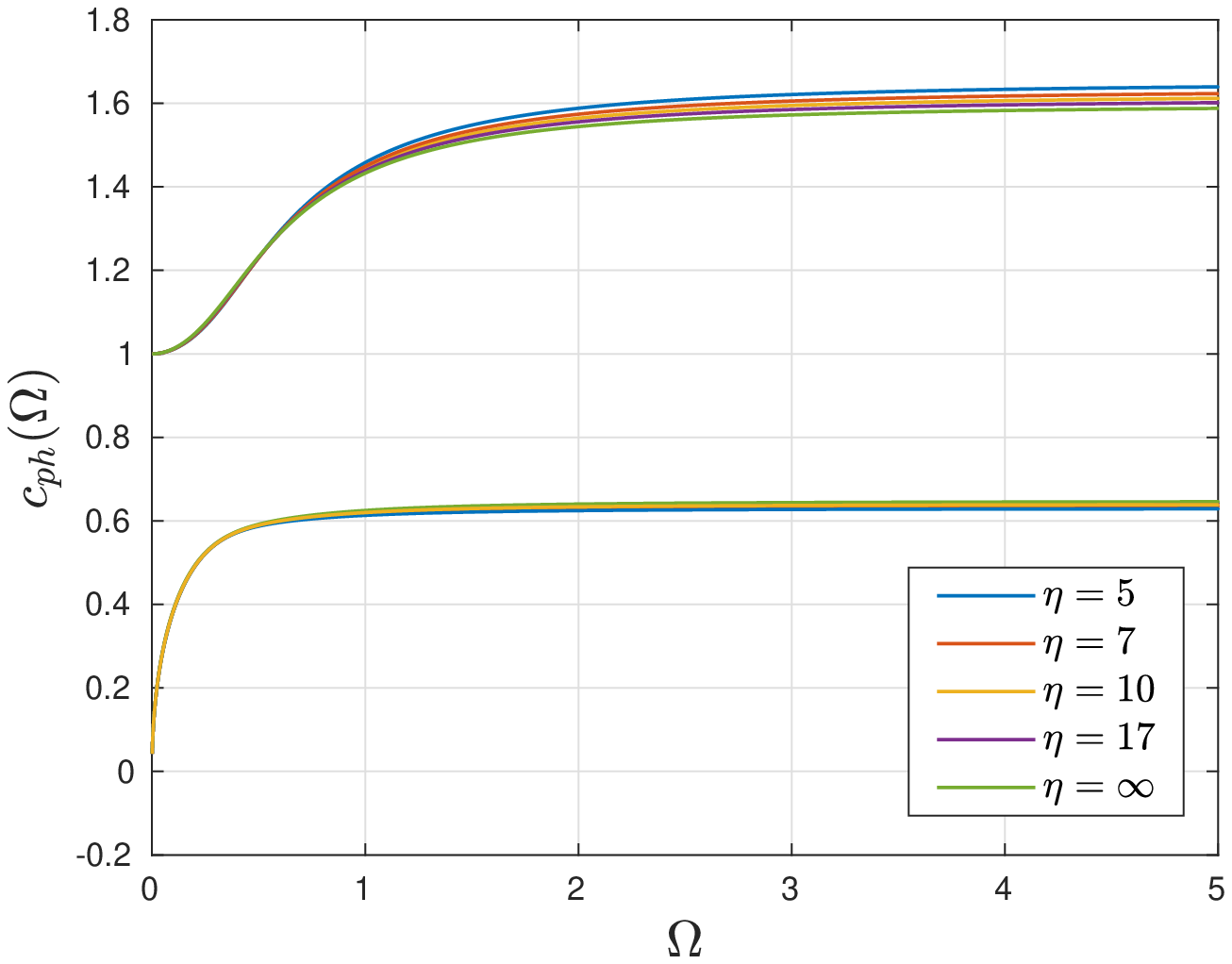}  
  \end{overpic}
  } \quad
\subfigure[R13]{%
  \begin{overpic}
    [width=.44\textwidth, clip]{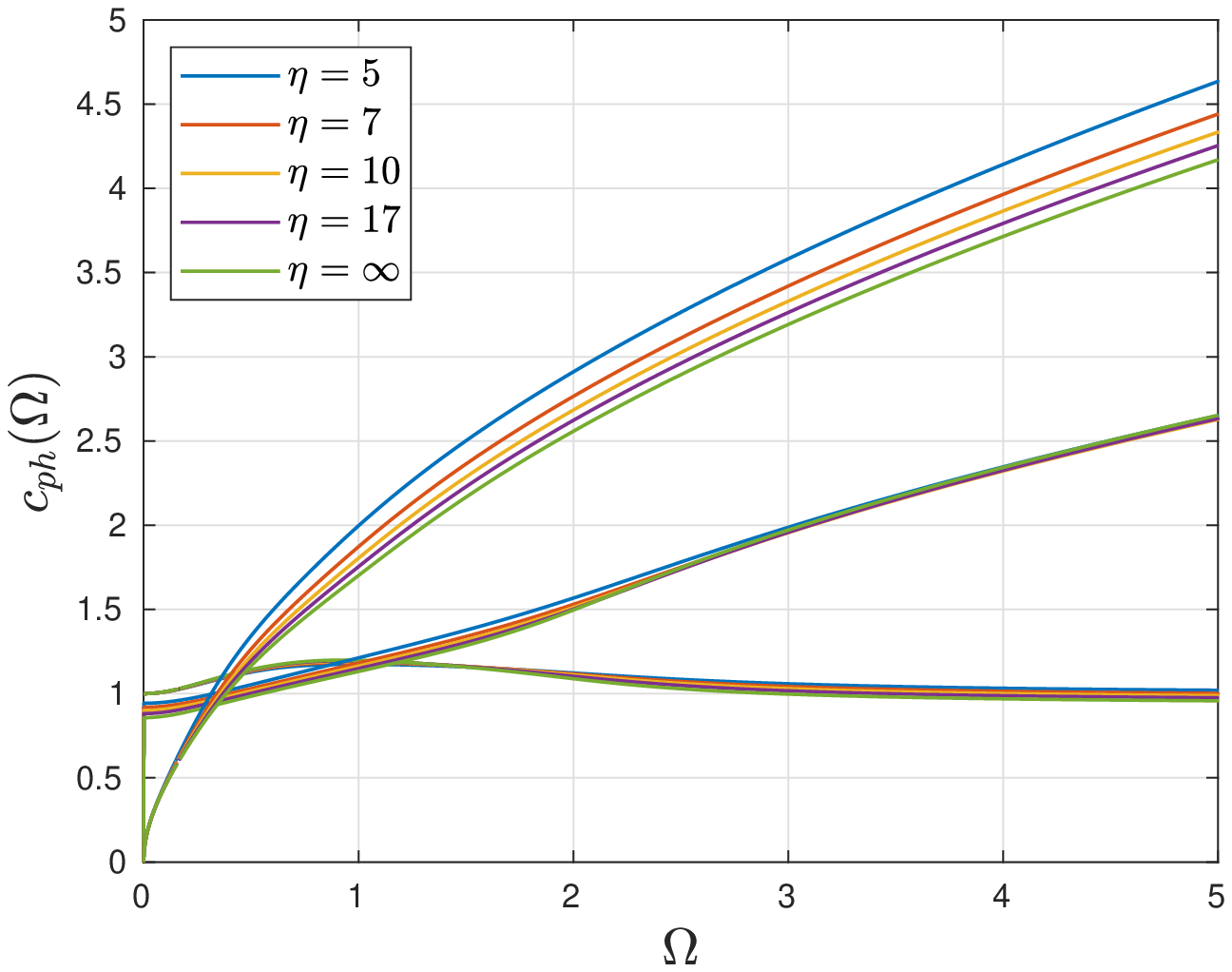}    
  \end{overpic}
} 
\caption{Phase speed $c_{ph}$ over frequency $\Omega$ of the G13 and
  R13 systems for different $\eta$.}
\label{fig:velocity}
\end{figure}

We proceed by discussing the phase speeds as functions of frequency
for the R13 and GG13 systems. For a given wave frequency $\Omega$, we
define the damping rate $\alpha$ and the wave speed $v_{ph}$ by
\begin{equation}
  \label{eq:wave_solution}
  \alpha = -k_i(\Omega), \quad v_{ph} = \frac{\Omega}{k_r(\Omega)}.
\end{equation}
For the Euler equations, where $\sigma = q = 0$ in \eqref{eq:1d}, the
absolute phase velocity is $|v_{ph}| = \sqrt{5/3}$ for all wave
frequency $\Omega$. For R13 and GG13 equations, $v_{ph}$ depends on
$\Omega$, causing the dispersion of sound waves. Here we define the
dimensionless phase speed $c_{ph} = v_{ph} / \sqrt{5/3}$ and plot
$c_{ph}$ as a function of the frequency $\Omega$ in Figure
\ref{fig:velocity} for R13 and GG13 equations. Note that
$c_{ph}(\Omega)$ is also a multi-valued function, and in Figure
\ref{fig:velocity}, we only plot the positive phase velocities. For
GG13 equations, the phase velocity has an upper limit, indicating the
hyperbolic nature of the system, while R13 equations can achieve
infinitely large phase velocities. In general, the phase speeds do not
change much as $\eta$ varies, which predicts similar behavior of sound
waves in different monatomic gases.

\begin{figure}[!ht]
\centering
\subfigure[inverse phase velocity]{%
  \label{fig:ipv}
  \begin{overpic}
  [width=.45\textwidth,  clip]{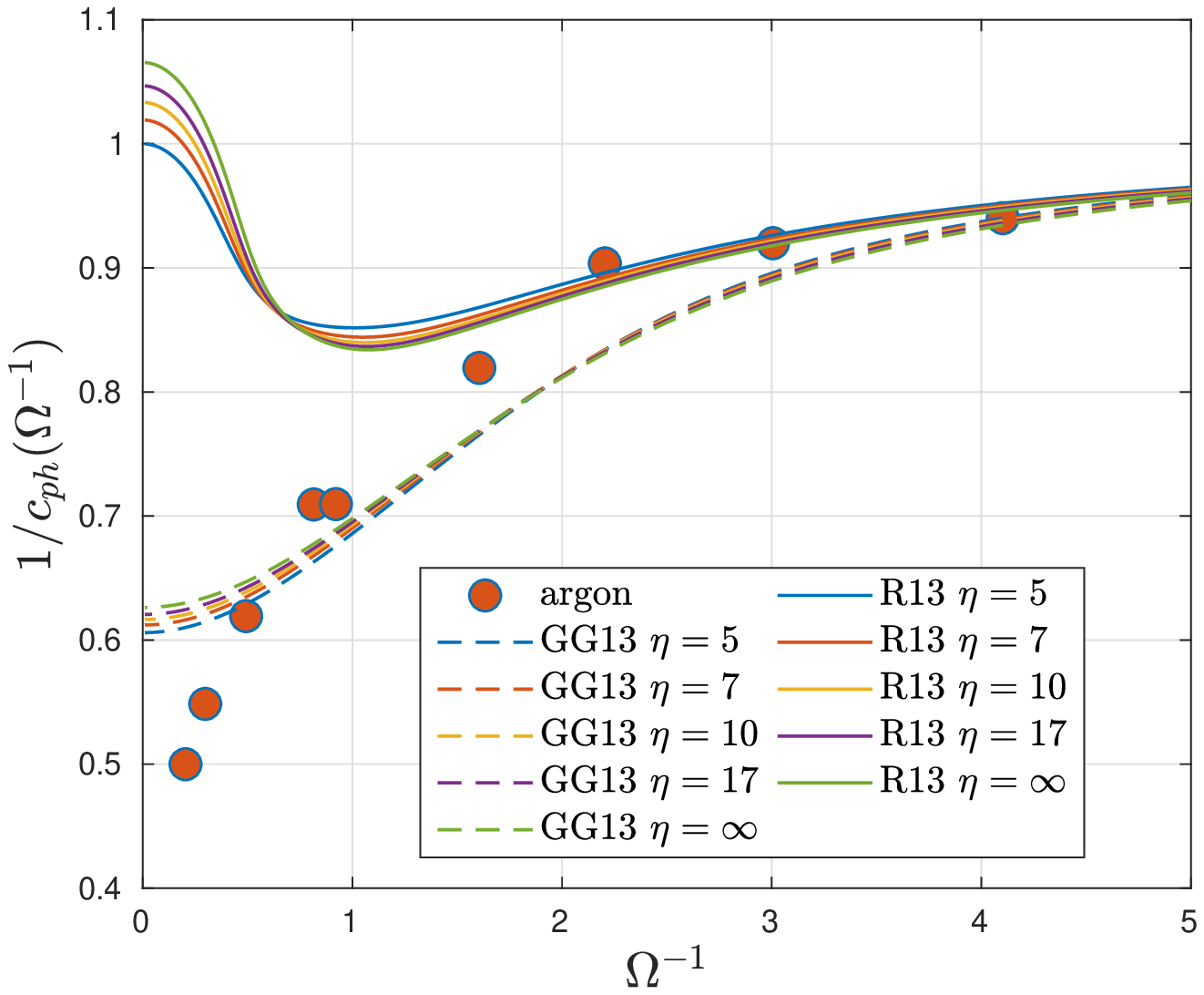}  
  \end{overpic}
  } \quad
\subfigure[damping]{%
  \label{fig:damping}
  \begin{overpic}
    [width=.45\textwidth,   clip]{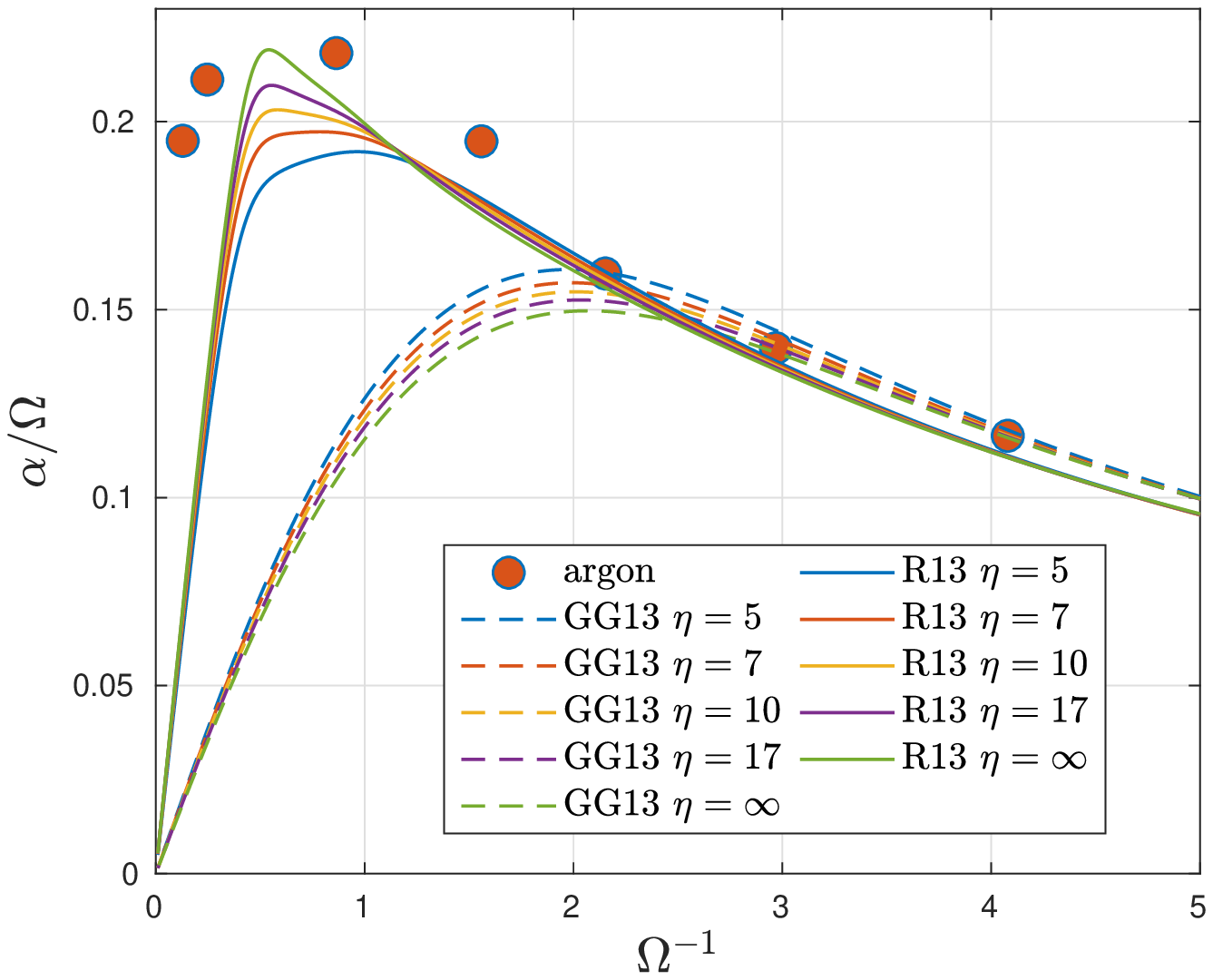}    
  \end{overpic}
} 
\caption{Inverse phase speed and damping over frequency $\Omega$ of
  the G13 and R13 systems for different $\eta$. \add{The bullets are
  the experimental results for argon \cite{Meyer1957}}.}
\label{fig:weakest}
\end{figure}

To study the phase speeds for large frequency waves, we plot the
inverse wave speed $1/c_{ph}$ as a function of the inverse frequency
$1/\Omega$ in Figure \ref{fig:ipv}, and the reduced damping rate
$\alpha / \Omega$ is plotted in Figure \ref{fig:damping} also as a
function of $1/\Omega$. In these figures, only the mode with the
weakest damping is given. Figure \ref{fig:ipv} shows that for GG13
equations, the phase velocity increases monotonically as the frequency
increases, while for R13 equations, the wave slows down as $\Omega$
reaches a value close to $1$. Such an observation agrees with the
results in \cite{Struchtrup2003,Struchtrup2013}, while the
experimental results for argon \cite{Meyer1957} suggest the monotonicity of
the phase velocity, which is closer to the prediction of GG13
equations. \add{Note that here we set the Knudsen number $\Kn$ to be
$1$. For another Knudsen number, the actual frequency of the wave
should be $\Omega / \Kn$. This means that if we consider a wave with a
fixed actual frequency travelling in the gas with a low Knudsen
number, we need to focus on large values of $\Omega^{-1}$. Indeed,
when $\Omega^{-1} > 1$, R13 models give better approximation of the
phase velocity, while for small $\Omega^{-1}$, which corresponds to
large Knudsen numbers, there is no guarantee whether R13 or GG13 is
superior, and the reason why GG13 provides better prediction requires
further investigation.  Nevertheless, for the damping rate shown in
Figure \ref{fig:damping}, R13 equations give significantly better
agreement with the experimental data for the whole range of the
frequency.}

\section{Derivation of moment systems}
\label{sec:Moment}
In this section, we provide the detailed procedure to derive GG13 and
R13 equations. In general, both models can be derived from infinite
moment equations by the method of order of magnitude. Following
\cite{Grad1949}, the moments can be considered as the coefficients in
the series expansion of the distribution function in the gas kinetic
theory. The distribution function is a function of position $\bx$,
particle velocity $\bxi$, and time $t$, which is a mesoscopic
description of fluid states in statistical physics. The moment method
proposed by Grad \cite{Grad1949} is one of the methods to derive
macroscopic models from the kinetic theory. Our starting point is the
same as \cite{Grad1949}, but we adopt the form used in
\cite{Kumar1966}, which expands the distribution function
$f(\bx,\bxi,t)$ as
\begin{equation}
  \label{eq:basis_fun}
  f(\bx,\bxi,t) = \sum_{l=0}^{+\infty} \sum_{m=-l}^l \sum_{n=0}^{+\infty}
    f_{lmn}(\bx, t) \psi_{lmn}(\bx,\bxi,t),
\end{equation}
where $\psi_{lmn}(\cdot)$ is the basis function based on Sonine
polynomials and spherical harmonics, the detailed form of which is
listed in Appendix \ref{sec:burnett}. Here $\bv = (v_1, v_2, v_3)^T$
is the velocity vector. The coefficients $f_{lmn}$ satisfy
$\overline{f_{lmn}} = (-1)^m f_{l,-m,n}$, and they are related to
Grad's 13 moments by
\begin{equation}
  \label{eq:macro_coe}
  \begin{gathered}
    f_{000} = \rho, \qquad f_{1m0} = 0, \quad m = -1, 0, 1, \qquad
    f_{001} = 0, \\
    \sigma_{11} = \sqrt{2}\RRe(f_{220}) - f_{200}/\sqrt{3}, \quad
    \sigma_{12} = -\sqrt{2}\IIm(f_{220}), \quad \sigma_{13} =
    -\sqrt{2}\RRe(f_{210}), \\
    \sigma_{22} = -\sqrt{2}\RRe(f_{220}) - f_{200}/\sqrt{3}, \quad
    \sigma_{23} = \sqrt{2}\IIm(f_{210}),\quad \sigma_{33} =
    2f_{200}/\sqrt{3}, \\
    q_1 = \sqrt{5}\RRe(f_{111}), \qquad q_2 = -\sqrt{5}\IIm(f_{111}),
    \qquad q_3 = -\sqrt{5/2}f_{101}.
  \end{gathered}
\end{equation}
These relations indicate the equivalence between Grad's 13-moments and
the following 13 variables:
\begin{equation} \label{eq:13m}
  f_{000}, \: v_1, \: v_2, \: v_3, \: \theta, \: f_{220}, \: f_{210},
  \: f_{200}, \: f_{2,-1,0}, \: f_{2,-2,0}, \: f_{111}, \: f_{101}, \:
  f_{1,-1,1}.
\end{equation}
Below we focus only on the derivation of equations for these
quantities.

The exact evolution equations for $f_{lmn}$ have been derived from the
Boltzmann equation with linearized collision operator in
\cite{Cai2018N}. In general, the equations for other $f_{lmn}$ have the
form
\begin{equation}
  \label{eq:moment_eq}
  \pd{f_{lmn}}{t} + S_{lmn} + T_{lmn} = \frac{\rho\theta}{\mu}
  \sum_{n'=0}^{+\infty} a_{lnn'}\theta^{n-n'}f_{lmn'}, 
\end{equation}
where $S_{lmn}$ contains time derivatives and $T_{lmn}$ contains
spatial derivatives. More precisely, $S_{lmn}$ is the linear
combination of the terms
\begin{equation} \label{eq:S_lmn}
\pd{\theta}{t} f_{l,m',n-1}, \quad \pd{v_i}{t}f_{l-1,m',n}, \quad
  \text{and} \quad \pd{v_i}{t} f_{l+1,m',n-1}
\end{equation}
with $i = 1,2,3$ and $m' = m-1,m,m+1$, and $T_{lmn}$ has the form
\begin{displaymath}
T_{lmn} = \sum_{l',m',n'} \left( \alpha_{lmn}^{l'm'n'}
  (\nabla_{\bx} \bv, \nabla_{\bx} \theta, \bv, \theta) f_{l'm'n'} +
  \sum_{i=1}^3 \beta_{lmn,i}^{l'm'n'}(\bv, \theta) \pd{f_{l'm'n'}}{x_i}
\right),
\end{displaymath}
which shows that $T_{lmn}$ is linear in all the coefficients
$f_{l'm'n'}$, while the linear coefficients are nonlinear functions of
$\bv$, $\theta$ and their spatial derivatives. The convection term
$T_{lmn}$ also has the following properties:
\begin{description}
\item[(P1)] The differential operator appears only once in each coefficient
  $\alpha_{lmn}^{l'm'n'} (\nabla_{\bx} \bv, \nabla_{\bx} \theta, \bv,
  \theta)$.
\item[(P2)] The coefficients $\alpha_{lmn}^{l'm'n'} (\nabla_{\bx} \bv,
  \nabla_{\bx} \theta, \bv, \theta)$ and $\beta_{lmn,i}^{l'm'n'}(\bv,
  \theta)$ are nonzero only if
  \begin{displaymath}
  l + 2n -3 \leqslant  l' + 2n' \leqslant l + 2n +1, \quad l-2\leqslant l' \quad
    \text{and} \quad n - 2 \leqslant n' \leqslant n+1.
  \end{displaymath}
\end{description}
The second property (P2) will play an important role in the derivation
of moment equations.

The precise expressions of $S_{lmn}$ and $T_{lmn}$ will be given in
Appendix \ref{sec:burnett}, and on the right-hand side of
\eqref{eq:moment_eq}, $a_{lnn'}$ are pure numbers for all IPL models.
Note that \eqref{eq:moment_eq} has already included the conservation
laws \eqref{eq:Euler}, which can be obtained by setting $(l,m,n)$ to
be
\begin{displaymath}
(0,0,0), \quad (1,-1,0), \quad (1,0,0), \quad (1,1,0), \quad (0,0,1).
\end{displaymath}
The subsequent derivation may include tedious formulas, and in our
implementation, all the calculations are done by the computer algebra system
Wolfram Mathematica. Below we only describe the algorithm we use in the
Mathematica code, and will not write out the lengthy intermediate results
in the calculational process.

\add{%
\begin{remark}
Note that complex basis functions $\psi_{lmn}$ are introduced in the
expansion \eqref{eq:basis_fun}, resulting in complex coefficients
$f_{lmn}$, which seems to complicate the derivation. In Grad's
original formulation \cite{Grad1949}, basis functions based on
Cartesian coordinates are considered, so that all the coefficients
are real. However, in Grad's original expansion using Hermite
polynomials, one cannot find 13 coefficients which matches exactly all
the moments in his 13-moment equations. Our basis functions, which are
similar to the ones in \cite{Kumar}, correspond to the 13 moments
very well (see \eqref{eq:13m}). Meanwhile, our basis functions are
based on spherical coordinates, so that the rotational invariance of
the collision operator can be easily utilized to simplify the
calculation.  Based on spherical coordinates, complex basis functions
can also be avoided by using real spherical harmonics instead of complex
spherical harmonics \cite[Section 1.5.2]{Bayin}. However, real
spherical harmonics do not have simple recurrence formulas such as
\cite[Eqs. (15.150)(15.151)]{Arfken}, which will result in a more
complicated form of $S_{lmn}$ and $T_{lmn}$.
\end{remark}

\subsection{General idea for the moment closure}
Before providing the details of the derivation, we would first like to 
explain the general methodology of the moment closure. To close the
advection term, one can observe from \eqref{eq:R13_sigma_q} that we
need to provide expressions for the moments $m_{ijk}$, $R_{ik}$ and
$\Delta$, which correspond to the coefficients $f_{3m0}$, $f_{1m1}$
and $f_{002}$, respectively. To close the collision term, it can be
seen from \eqref{eq:moment_eq} that we have to provide expressions for
infinite terms $f_{1mn}$ and $f_{2mn}$ for any positive integer $n$.
In our implementation, this is done by a truncation of the infinite
series in \eqref{eq:moment_eq}, whose fast convergence has already
been demonstrated in \cite{Cai2015}, so that only a finite number of
coefficients need to be considered.

The derivation of R13 equations is mostly similar to the
Chapman-Enskog expansion. However, there are two key differences:
\begin{description}
\item[(K1)] All the moments are to be represented by the 13 moments and
  their derivatives, while only 5 equilibrium moments are involved in
  the Chapman-Enskog expansion;
\item[(K2)] We expect that the equations have the Burnett order and
  involve only second-order derivatives, while third-order derivatives
  are involved in Burnett equations.
\end{description}
Due to the first difference, there are in principle infinite versions
of R13 equations in the Burnett order, since the leading order terms
in $\sigma_{ij}$ and $q_i$ can be represented by equilibrium
variables:
\begin{equation} \label{eq:NSF}
\sigma_{ij} \approx -\frac{2\mu}{\alpha_2^{(\eta)}} \pd{v_{\li}}{x_{\jl}},
  \qquad q_i \approx -\frac{5\mu}{2\beta_5^{(\eta)}} \pd{\theta}{x_i},
\end{equation}
as already resulted in many different versions of R13 equations for
Maxwell molecules \cite{Timokhin2017}. In our derivation, since
$\sigma_{ij}$ and $q_i$ are already included in the system, we will
avoid using \eqref{eq:NSF} to make any replacement except in the
derivation of first-order expressions. More precisely, according to
Chapman-Enskog expansion, the first-order term of the distribution
function can be completely represented by the following quantities:
\begin{displaymath}
\rho, \, v_i, \, \theta, \, \pd{v_{\li}}{x_{\jl}}, \, \pd{\theta}{x_i}.
\end{displaymath}
At this step, we will apply \eqref{eq:NSF} to replace the derivatives
of $v_i$ and $\theta$ by $\sigma_{ij}$ and $q_i$. By such replacement,
one order of derivative can be eliminated, so that (K2) can be
automatically achieved.
}

\subsection{Chapman-Enskog expansion of the coefficients}
\label{sec:derivation}
\add{This section is devoted to the details of the asymptotic
analysis. As mentioned in the previous section,} the idea of
Chapman-Enskog expansion is utilized here to derive models with
different orders of accuracy. To begin with, we introduce the scaling
$t = t'/\epsilon$ and $x_i = x_i'/\epsilon$, and rewrite the equations
\eqref{eq:moment_eq} with time and spatial variables $t'$ and $x_i'$.
In the resulting equations, a factor $\epsilon^{-1}$ is introduced to
the right-hand side of \eqref{eq:moment_eq}. In this section, we will
work on the scaled equations, and the prime symbol on $t'$ and $x_i'$
will be omitted. Based on such a transform, we write down the
asymptotic expansion for all the moments as
\begin{equation}
  \label{eq:CE}
  f_{lmn} = f_{lmn}^{(0)} + \epsilon f_{lmn}^{(1)} + \epsilon^2
  f_{lmn}^{(2)} + \epsilon^3 f_{lmn}^{(3)} + \cdots.
\end{equation}
In the original Chapman-Enskog expansion of the distribution function
$f = f^{(0)} + \epsilon f^{(1)}$, we require that $f^{(0)}$ is the
local Maxwellian $\mathcal{M}$. The corresponding assumption for the
coefficients is 
\begin{equation}
  \label{eq:order0}
  f_{lmn}^{(0)} = \left\{
    \begin{array}{cc}
      \rho, & {\rm if}~ (l, m, n) = (0,0,0), \\
      0, & {\rm otherwise},
    \end{array}
\right.
\end{equation}
By \eqref{eq:CE}, the term  $S_{lmn}$ and $T_{lmn}$ can be expanded
correspondingly:
\begin{equation}
  S_{lmn} = S_{lmn}^{(0)} + \epsilon S_{lmn}^{(1)} + \epsilon^2
  S_{lmn}^{(2)} + \epsilon^3 S_{lmn}^{(3)} + \cdots, \quad 
   T_{lmn} = T_{lmn}^{(0)} + \epsilon T_{lmn}^{(1)} + \epsilon^2
  T_{lmn}^{(2)} + \epsilon^3 T_{lmn}^{(3)} + \cdots.
\end{equation}
Note that the above expansion is straightforward since both $S_{lmn}$
and $T_{lmn}$ are linear in all the coefficients $f_{lmn}$. Thus the
moment equations turn out to be
\begin{equation}
  \label{eq:max_ite}
  \begin{aligned}
  & \pd{(f_{lmn}^{(0)} + \epsilon f_{lmn}^{(1)} + \epsilon^2
      f_{lmn}^{(2)} + \epsilon^3 f_{lmn}^{(3)} + \cdots)}{t} & \\
  & \qquad  + (S_{lmn}^{(0)} + \epsilon S_{lmn}^{(1)} + \epsilon^2
    S_{lmn}^{(2)} + \epsilon^3 S_{lmn}^{(3)} + \cdots) + (T_{lmn}^{(0)}
    + \epsilon T_{lmn}^{(1)} + \epsilon^2
    T_{lmn}^{(2)} + \epsilon^3 T_{lmn}^{(3)} + \cdots) &  \\
 &  \qquad \qquad  = \frac{1}{\epsilon} \left(\frac{\rho\theta}{\mu}
      \sum_{n'=0}^{+\infty}
      a_{lnn'}\theta^{n-n'} \left(f_{lmn'}^{(0)} + \epsilon
        f_{lmn'}^{(1)} + \epsilon^2 f_{lmn'}^{(2)} + \epsilon^3
        f_{lmn'}^{(3)} + \cdots \right) \right).
  \end{aligned}
\end{equation} 
Matching the terms with the same orders with respect to $\epsilon$,
one obtains
\begin{equation}
  \label{eq:max_ite_1}
  \pd{f_{lmn}^{(k)}}{t} + S_{lmn}^{(k)} + T_{lmn}^{(k)} =
    \frac{\rho\theta}{\mu}\sum_{n'=0}^{+\infty}
    a_{lnn'}\theta^{n-n'} f_{lmn'}^{(k+1)}, \quad k \geqslant 0.
\end{equation}
In Chapman-Enskog expansion, due to the assumption \eqref{eq:order0},
the equation \eqref{eq:max_ite_1} is only applied to the case $(l,n)
\neq (0,0), (1,0), (0,1)$, in which the right-hand side of
\eqref{eq:max_ite_1} is nonzero and \eqref{eq:max_ite_1} can provide
us expressions for $f_{lmn}^{(k+1)}$.  For more details about the
Chapman-Enskog expansion, we refer the readers to textbooks such as
\cite{Struchtrup2005}. In what follows, we are going to introduce a
generalized version of the Chapman-Enskog expansion involving 13
moments in the assumption, which will be carried out below by studying
each $k$ incrementally. Note that the idea of the following method is
also applicable for more general collision models.

\subsubsection{First order ($k=0$)} 
When $(l,n) \neq (0,0), (1,0), (0,1)$, using \eqref{eq:order0} and the
fact that $S_{lmn}$ is a linear combination of \eqref{eq:S_lmn}, one
can see that $f_{lmn}^{(0)} = S_{lmn}^{(0)} = 0$. Thus when $k=0$, the
equation \eqref{eq:max_ite_1} becomes
\begin{equation}
  \label{eq:order1}
  T_{lmn}^{(0)} =\frac{\rho\theta}{\mu} \sum_{n'=0}^{+\infty}
    a_{lnn'}\theta^{n-n'} f_{lmn'}^{(1)}.
\end{equation}
If $l = 0$, $n > 1$ or $l \geqslant 3$, the property (P2) and
\eqref{eq:order0} show that $T_{lmn}^{(0)} = 0$, meaning that
\begin{equation}
  \label{eq:order1_1}
  f_{lmn}^{(1)} = 0, \quad {\rm if}~ l \geqslant 3~{\rm or}~ l = 0. 
\end{equation}
Below we focus on the cases $l=1$ and $l=2$. Note that $f_{1m0} = 0$,
by which we can rewrite \eqref{eq:order1} as
\begin{equation}
  \label{eq:order1_2}
  T_{1mn}^{(0)} = \frac{\rho\theta}{\mu}\sum_{n'=1}^{+\infty}
  a_{1nn'}\theta^{n-n'} f_{1mn'}^{(1)}, \qquad
  T_{2mn}^{(0)} =  \frac{\rho\theta}{\mu}\sum_{n'=0}^{+\infty}
  a_{2nn'}\theta^{n-n'} f_{2mn'}^{(1)}. 
\end{equation}
Again by the property (P2) and \eqref{eq:order0}, we see that
$T_{1mn}^{(0)} = 0$ for all $n > 1$ and $T_{2mn}^{(0)} = 0$ for all $n
> 0$. For any given $m$, the values of $f_{1mn'}^{(1)}$ and
$f_{2mn'}^{(1)}$ can be solved from \eqref{eq:order1_2}, and the
result has the form:
\begin{equation}
  \label{eq:NavierStokesFourier}
  f_{1mn}^{(1)} = \frac{\mu}{\rho\theta}A_{1n} \theta^{n-1} T_{1m1}^{(0)},
  \qquad 
  f_{2mn}^{(1)} = \frac{\mu}{\rho\theta} A_{2n} \theta^n T_{2m0}^{(0)},
\end{equation}
where $A_{1n}$ and $A_{2n}$ are pure numbers. In principle, obtaining
\eqref{eq:NavierStokesFourier} requires solving an infinite matrix. In our
implementation, this is approximated by a cutoff of the right-hand
sides of \eqref{eq:order1_2} up to $n' \leqslant 9$.

Until now, our calculation is completely the same as the classical
Chapman-Enskog expansion. In \eqref{eq:NavierStokesFourier}, all first
order quantities can be represented by the conservative quantities and
their derivatives (hidden in the expression of $T_{1m1}^{(0)}$ and
$T_{2m0}^{(0)}$). However, to derive 13-moment equations, we are
required to represent the distribution functions using more moments
and less derivatives. For example, in Grad's 13-moment theory, no
derivatives are included in the ansatz of the distribution function.
In our derivation, this can be achieved by writing
\eqref{eq:NavierStokesFourier} as
\begin{equation}
  \label{eq:order1_3}
  f_{1mn}^{(1)} = \frac{A_{1n}}{A_{11}}\theta^{n-1}f_{1m1}^{(1)}, \qquad 
  f_{2mn}^{(1)} = \frac{A_{2n}}{A_{20}}\theta^{n}f_{2m0}^{(1)}.
\end{equation}
Since $f_{2m0}$ and $f_{1m1}$ are included in the 13-moment theory, we
mimic the assumption of Chapman-Enskog expansion \eqref{eq:order0} and
set
\begin{equation}
  \label{eq:sq1}
  f_{1m1} = \epsilon
  f_{1m1}^{(1)}, \quad f_{2m0} = \epsilon f_{2m0}^{(1)}, \qquad 
  f_{1m1}^{(k)} =f_{2m0}^{(k)} = 0, \quad k
  \geqslant 2.
\end{equation}
By \eqref{eq:order1_3} and \eqref{eq:sq1}, we can write down the
approximation of the distribution function up to first order in
$\epsilon$ using the 13 moments: 
\begin{equation} \label{eq:f0_f1}
  \begin{split}
  f(\bx,\bxi,t) \approx \mathcal{M}(\bx,\bxi,t) &+
    \sum_{m=-1}^1 f_{1m1}(\bx,t) \sum_{n=1}^{+\infty}
      \frac{A_{1n}}{A_{11}} [\theta(\bx,t)]^{n-1} \psi_{1mn}(\bx,\bxi,t) \\
  & + \sum_{m=-2}^2 f_{2m0}(\bx,t) \sum_{n=0}^{+\infty}
      \frac{A_{2n}}{A_{20}} [\theta(\bx,t)]^n \psi_{2mn}(\bx,\bxi,t),
  \end{split}
\end{equation}
where we have written all the parameters $t,\bx$ and $\bxi$ for
clarification. Note that in contrast to the Chapman-Enskog expansion
of the distribution function, no derivatives are involved in the above
expression. \add{The equation \eqref{eq:f0_f1} is also different from
Grad's ansatz for the 13-moment theory, since \eqref{eq:f0_f1} can be
written equivalently as
\begin{equation} \label{eq:f0_f1_sigmaq}
\begin{split}
f(\bxi) &\approx \Bigg[ 1 -
  \sum_{i=1}^3 \frac{(\xi_i - v_i) q_i}{\theta^2}
  \sum_{n=1}^{+\infty} \frac{A_{1n}}{A_{11}}
    \sqrt{\frac{3\pi^{1/2} n!}{10\Gamma(n+5/2)}}
    L_n^{(3/2)} \left( \frac{\bxi - \bv}{2\theta} \right) \\
& \qquad + \sum_{i=1}^3 \sum_{j=1}^3
  \frac{\sigma_{ij}(\xi_i - v_i)(\xi_j - v_j)}{\theta^2}
  \sum_{n=0}^{+\infty} \frac{A_{2n}}{A_{20}}
    \sqrt{\frac{15\pi^{1/2} n!}{32\Gamma(n+7/2)}}
    L_n^{(5/2)} \left( \frac{\bxi - \bv}{2\theta} \right) \Bigg]
    \mathcal{M}(\bxi).
\end{split}
\end{equation}
Here $L_n^{(\alpha)}$ is the Laguerre polynomial defined in
\eqref{eq:Laguerre}, and we have omitted the variables $t$ and $\bx$
for conciseness. Grad's ansatz for the 13-moment theory can be
obtained by truncating both infinite sums in \eqref{eq:f0_f1_sigmaq}
by preserving only their first terms, and therefore Grad's 13-moment
theory does not fully represent the first-order term of the
distribution function for general collision models.} Due to the
assumption \eqref{eq:sq1}, when we apply \eqref{eq:max_ite_1},
equations for all the 13 moments should be excluded, i.e., $(l,n) \neq
(0,0), (0,1), (1,0), (1,1), (2,0)$.

\begin{remark}
The solvability of \eqref{eq:order1_2} relies on the existence of the
spectral gap for the linearized Boltzmann collision operator. For IPL
models, this has been proven in \cite{Mouhot2007}. In particular, when
$\eta = 5$, we have $A_{1n} = \delta_{1n} / a_{111}$ and $A_{2n} =
\delta_{0n} / a_{200}$. In this case, the first two orders
\eqref{eq:f0_f1} are exactly the ansatz in Grad's 13-moment theory.
\end{remark}

\subsubsection{Second order ($k=1$)} 
Now we set $k = 1$ in \eqref{eq:max_ite_1}. Since $f_{000} = f_{001} =
f_{1m0} = 0$ and $f_{2m0}^{(k)} = f_{1m1}^{(k)} = 0$ for $k\geqslant
2$, the result can be written as
\begin{equation}
  \label{eq:order2}
  \pd{f_{lmn}^{(1)}}{t} + S_{lmn}^{(1)} + T_{lmn}^{(1)} =
    \frac{\rho\theta}{\mu}\sum_{n'=n_0(l)}^{+\infty}
    a_{lnn'}\theta^{n-n'} f_{lmn'}^{(2)},
\end{equation} 
where
\begin{equation}
  n_0(l) = \left\{ \begin{array}{ll}
    2, & \text{if } l = 0,1, \\
    1, & \text{if } l = 2, \\
    0, & \text{if } l \geqslant 3.
  \end{array} \right.
\end{equation}
Since $f_{lmn}^{(1)}$ has been fully obtained in the previous section
(see \eqref{eq:order1_1} and \eqref{eq:sq1}), the expressions of
$S_{lmn}^{(1)}$ and $T_{lmn}^{(1)}$ can be naturally obtained. Thus
the left-hand side of \eqref{eq:order2} can already be represented by
the 13 moments. To obtain the second-order contributions
$f_{lmn}^{(2)}$, for any $l$ and $m$, we just need to solve the
infinite linear system \eqref{eq:order2}, and the general result is
\begin{equation}
  \label{eq:order2_2}
  f_{lmn}^{(2)} =\frac{\mu}{\rho\theta}
  \sum_{n' = n_0(l)}^{+\infty} B_{lnn'}\theta^{n-n'}
    \left(\pd{f_{lmn'}^{(1)}}{t} +S_{lmn'}^{(1)}+ T_{lmn'}^{(1)} \right), 
\end{equation}
where $B_{lnn'}$ are all pure numbers. In our implementation, we again
truncate the system \eqref{eq:order2} at $n' = \lfloor 10-l/2
\rfloor$. For the purpose of deriving R13 equations, we only need
$f_{lmn}^{(2)}$ up to $l = 4$.

The right-hand side of \eqref{eq:order2_2} still contains time
derivatives, which are not desired. In general, they can all be
replaced by spatial derivatives. Note that $\partial_t f_{lmn}^{(1)}$
is nonzero only if $l = 1$ or $l = 2$. In these cases, we have
\begin{equation} \label{eq:td_f}
\begin{aligned}
\frac{\partial f_{1mn'}^{(1)}}{\partial t} &=
  \frac{A_{1n'}}{A_{11}} \left(
    \theta^{n'-1} \frac{\partial f_{1m1}^{(1)}}{\partial t} +
    (n'-1) \theta^{n'-2} \pd{\theta}{t} f_{1m1}^{(1)}
  \right), \\
\frac{\partial f_{2mn'}^{(1)}}{\partial t} &=
  \frac{A_{2n'}}{A_{20}} \left(
    \theta^{n'} \frac{\partial f_{2m0}^{(1)}}{\partial t} +
    n' \theta^{n'-1} \pd{\theta}{t} f_{2m0}^{(1)}
  \right).
\end{aligned}
\end{equation}
We first focus on $\partial_t f_{1m1}^{(1)}$ and $\partial_t
f_{2m0}^{(1)}$. Taking $\partial_t f_{1m1}^{(1)}$ as an example, we do
the following calculation:
\begin{equation} \label{eq:cal1}
\begin{split}
& \frac{\partial f_{1m1}^{(1)}}{\partial t} +
  \frac{1}{\epsilon} \left( S_{1m1}^{(0)} + T_{1m1}^{(0)} \right) +
  \left( S_{1m1}^{(1)} + T_{1m1}^{(1)} \right) \\
={} & \frac{1}{\epsilon} \frac{\rho \theta}{\mu}
    \sum_{n=1}^{+\infty} a_{11n} \theta^{1-n} f_{1mn}^{(1)} +
    \frac{\rho \theta}{\mu}
      \sum_{n=1}^{+\infty} a_{11n} \theta^{1-n} f_{1mn}^{(2)}
  + \mathcal{O}(\epsilon) \\
={} & \frac{1}{\epsilon} \frac{\rho \theta}{\mu}
  \frac{f_{1m1}^{(1)}}{A_{11}} +
  \sum_{n=2}^{+\infty} a_{11n} \theta^{1-n}
    \sum_{n'=2}^{+\infty} B_{1nn'} \theta^{n-n'} \left(
      \pd{f_{lmn'}^{(1)}}{t} +S_{lmn'}^{(1)}+ T_{lmn'}^{(1)}
    \right) + \mathcal{O}(\epsilon) \\
={} & \frac{1}{\epsilon} \frac{\rho \theta}{\mu} \frac{f_{1m1}^{(1)}}{A_{11}} +
  \frac{1}{A_{11}} \sum_{n=2}^{+\infty} \sum_{n'=2}^{+\infty}
    a_{11n} B_{1nn'} A_{1n'} \left(
      \pd{f_{1m1}^{(1)}}{t} + \frac{n'-1}{\theta} \pd{\theta}{t} f_{1m1}^{(1)}
    \right) \\
& \qquad \qquad + \sum_{n'=2}^{+\infty}
    \left( \sum_{n=2}^{+\infty} a_{11n} B_{1nn'} \right) \theta^{1-n'}
    \left( S_{lmn'}^{(1)} + T_{lmn'}^{(1)} \right) +
    \mathcal{O}(\epsilon),
\end{split}
\end{equation}
from which it can be solved that
\begin{equation} \label{eq:td_f1}
\begin{split}
\frac{\partial f_{1m1}^{(1)}}{\partial t} &=
  \left( 1 - \sum_{n'=2}^{+\infty} \lambda_{1n'} \right)^{-1}
  \Bigg[ \frac{1}{\epsilon} \left(
    \frac{\rho\theta}{\mu} \frac{f_{1m1}^{(1)}}{A_{11}} - T_{1m1}^{(0)}
  \right) \\
& \qquad + \sum_{n'=2}^{+\infty}
    \frac{(n'-1)\lambda_{1n'}}{\theta} \pd{\theta}{t} f_{1m1}^{(1)} +
  \sum_{n'=1}^{+\infty} C_{1n'} \theta^{1-n'}
    \left( S_{1mn'}^{(1)} + T_{1mn'}^{(1)} \right) \Bigg] +
    \mathcal{O}(\epsilon),
\end{split}
\end{equation}
where we have used $S_{1m1}^{(0)} = 0$, and the newly introduced
constants are
\begin{equation}
C_{ln'} = \left\{ \begin{array}{ll}
  -1, & \text{if } n'=n_0(l)-1, \\
  \displaystyle \sum_{n=n_0(l)}^{+\infty} a_{l,n_0(l)-1,n} B_{lnn'},
    & \text{if } n' \geqslant n_0(l),
\end{array} \right. \qquad
\lambda_{ln'} = \frac{C_{ln'} A_{ln'}}{A_{l,n_0(l)-1}}.
\end{equation}
Similarly, we can obtain
\begin{equation} \label{eq:td_f2}
\begin{split}
\frac{\partial f_{2m0}^{(1)}}{\partial t} &=
  \left( 1 - \sum_{n'=1}^{+\infty} \lambda_{2n'} \right)^{-1}
  \Bigg[ \frac{1}{\epsilon} \left(
    \frac{\rho\theta}{\mu} \frac{f_{2m0}^{(1)}}{A_{20}} -
    T_{2m0}^{(0)}
  \right) \\
& \qquad + \sum_{n'=1}^{+\infty}
    \frac{n'\lambda_{2n'}}{\theta} \pd{\theta}{t} f_{2m0}^{(1)} +
  \sum_{n'=0}^{+\infty} C_{2n'} \theta^{-n'}
    \left( S_{2mn'}^{(1)} + T_{2mn'}^{(1)} \right) \Bigg] +
    \mathcal{O}(\epsilon),
\end{split}
\end{equation}

As a summary, the time derivatives in \eqref{eq:order2_2} can be
replaced with spatial derivatives by the following operations:
\begin{itemize}
\item If $l\neq 1$ and $l\neq 2$, set the derivative $\partial_t
  f_{lmn'}^{(1)}$ to be zero. If $l=1$ or $l=2$, replace $\partial_t
  f_{1mn'}^{(1)}$ by \eqref{eq:td_f}\eqref{eq:td_f1}\eqref{eq:td_f2}.
\item After replacement, the results still include the time
  derivatives of $\bv$ and $\theta$. These terms can be replaced by
  conservation laws \eqref{eq:Euler}. In fact, we can use the fact
  that $\sigma_{kl}$ and $q_k$ are $\mathcal{O}(\epsilon)$ terms to
  rewrite the conservation laws of momentum and energy as
  \begin{equation} \label{eq:v_theta}
  \frac{\partial v_i}{\partial t} = -v_k \pd{v_i}{x_k} -
    \frac{\theta}{\rho} \pd{\rho}{x_i} - \pd{\theta}{x_i}
    + \mathcal{O}(\epsilon), \qquad
  \frac{\partial \theta}{\partial t} = -\frac{2}{3} \theta \pd{v_k}{x_k}
    -v_k \pd{\theta}{x_k} + \mathcal{O}(\epsilon),
  \end{equation}
  and use these equations for substitution.
\item After the above replacements, the $\mathcal{O}(\epsilon)$ terms
  can be dropped.
\end{itemize}
By now, we have presented all the coefficients $f_{lmn}^{(2)}$ by the
13 moments and their spatial derivatives. When $l=1,2$, the results
include a coefficient $1/\epsilon$ coming from \eqref{eq:td_f1} and
\eqref{eq:td_f2}. More precisely, $f_{1mn}^{(2)}$ and $f_{2mn}^{(2)}$
have the form
\begin{align}
\label{eq:f1}
f_{1mn}^{(2)} &= \frac{1}{\epsilon}
  \theta^{n-1} \left(
    \sum_{n'=2}^{+\infty} B_{1nn'} \frac{A_{1n'}}{A_{11}}
  \right) \!
  \left( 1 - \sum_{n'=2}^{+\infty} \lambda_{1n'} \right)^{-1} \!\!
  \left( \frac{f_{1m1}^{(1)}}{A_{11}} -
    \frac{\mu}{\rho\theta} T_{1m1}^{(0)}
  \right) + W_{1mn}^{(2)}, ~~ n \geqslant 2, \\
\label{eq:f2}
f_{2mn}^{(2)} &= \frac{1}{\epsilon}
  \theta^n \left(
    \sum_{n'=1}^{+\infty} B_{2nn'} \frac{A_{2n'}}{A_{20}}
  \right)
  \left( 1 - \sum_{n'=1}^{+\infty} \lambda_{2n'} \right)^{-1}
  \left( \frac{f_{2m0}^{(1)}}{A_{20}} -
    \frac{\mu}{\rho\theta} T_{2m0}^{(0)}
  \right) + W_{2mn}^{(2)}, \quad n \geqslant 1,
\end{align}
where $W_{1mn}^{(2)}$ and $W_{2mn}^{(2)}$ are terms independent of
$\epsilon$. Note that only first-order derivatives have been
introduced into $f_{lmn}^{(2)}$, while in the original Chapman-Enskog
expansion, the second-order (Burnett-order) term $f^{(2)}$ includes
second-order derivatives.

\subsubsection{Third order ($k = 2$)}
Similar to the case $k=1$, when $k=2$, we can solve the linear system
\eqref{eq:max_ite_1} to get
\begin{equation}
  \label{eq:order3_1}
  f_{lmn}^{(3)} =\frac{\mu}{\rho\theta}\sum_{n'=n_0(l)}^{+\infty}
  B_{lnn'} \theta^{n - n'}
    \left(\pd{f_{lmn'}^{(2)}}{t} + S_{lmn'}^{(2)} + T_{lmn'}^{(2)} \right),
  \qquad n \geqslant n_0(l).
\end{equation}
After inserting the expression of $f_{lmn}^{(2)}$ into the above
equation, we again need to deal with the time derivatives. The time
derivatives appearing in $S_{lmn'}^{(2)}$ can again be replaced by
\eqref{eq:v_theta} without $\mathcal{O}(\epsilon)$ terms. Below we
focus only the time derivative of $f_{lmn'}^{(2)}$.

When $l \neq 1$ and $l \neq 2$, the second-order term $f_{lmn'}^{(2)}$
does not include the coefficient $1/\epsilon$, and therefore
$\partial_t f_{lmn'}^{(2)}$ can be computed by inserting the
expression of $f_{lmn'}^{(2)}$, expanding the time derivative, and
then replacing the time derivative of each moment in \eqref{eq:13m} by
\eqref{eq:v_theta}\eqref{eq:td_f1}\eqref{eq:td_f2} and the continuity
equation $\partial_t \rho = -\mathrm{div} (\rho \bv)$. After
replacement, we can also safely drop the $\mathcal{O}(\epsilon)$
terms. The treatment for $l=1,2$ is more complicated, and the process
will be detailed below.

When $l=1$ and $n \geqslant 2$ (the case $l=2, n \geqslant 1$ is
similar), by \eqref{eq:f1},
\begin{equation} \label{eq:df2_dt}
\begin{split}
\pd{f_{1mn'}^{(2)}}{t} &= \frac{1}{\epsilon} \theta^{n'-1}
  \frac{D_{1n'}}{A_{11}} \pd{f_{1m1}^{(1)}}{t} -
  \frac{1}{\epsilon} \theta^{n'-1} D_{1n'} \frac{\partial}{\partial t}
      \left( \frac{\mu}{\rho \theta} T_{1m1}^{(0)} \right) \\
&\quad + \frac{1}{\epsilon} (n'-1) \theta^{n'-2} D_{1n'} \left(
    \frac{f_{1m1}^{(1)}}{A_{11}} - \frac{\mu}{\rho \theta} T_{1m1}^{(0)}
  \right) \pd{\theta}{t} +
  \pd{W_{1mn'}^{(2)}}{t},
\end{split}
\end{equation}
where
\begin{equation}
D_{1n} = \left(
    \sum_{n'=2}^{+\infty} B_{1nn'} \frac{A_{1n'}}{A_{11}}
  \right) \left( 1 - \sum_{n'=2}^{+\infty} \lambda_{1n'} \right)^{-1}.
\end{equation}
The last term $\partial_t W_{1mn'}^{(2)}$ is independent of
$\epsilon$, and therefore the time derivative can also be replaced by
conservation laws and \eqref{eq:td_f1}\eqref{eq:td_f2} without
$\mathcal{O}(\epsilon)$ terms. However, for the first three terms on
the right-hand side of \eqref{eq:df2_dt}, due to the existence of
$1/\epsilon$, when replaced by spatial derivatives, the
$\mathcal{O}(\epsilon)$ terms have to be taken into account to capture
the $\mathcal{O}(1)$ contribution. Note that $T_{1m1}^{(0)}$ also
appears in \eqref{eq:NavierStokesFourier}, where the first equation
for $n = 1$ represents Fourier's law. We know that $(\rho \theta)^{-1}
T_{1m1}^{(0)}$ is essentially the spatial derivative of $\theta$
multiplied by a constant. Therefore the second and third terms involve
only the time derivative of density $\rho$ and temperature $\theta$,
and they can be replaced exactly by the conservation laws
\eqref{eq:Euler}. Thus the only troublesome term is again $\partial_t
f_{1m1}^{(1)}$. Before discussing this term, we first implement all
the aforementioned replacements and write the result as
\begin{displaymath} \label{eq:f3}
f_{1mn}^{(3)} = \frac{1}{\epsilon} \frac{\mu}{\rho\theta}
  \frac{\theta^{n-1}}{A_{11}} \left(
    \sum_{n'=2}^{+\infty} B_{1nn'} D_{1n'}
  \right) \pd{f_{1m1}^{(1)}}{t} + R_{1mn}^{(3)},
\end{displaymath}
where $R_{1mn}^{(3)}$ is the collection of terms which does not
include any time derivatives.

By now, we can carry out the calculation similar to \eqref{eq:cal1}:
\begin{equation} \label{eq:cal2}
\begin{split}
& \frac{\partial f_{1m1}^{(1)}}{\partial t} +
  \frac{1}{\epsilon} \left( S_{1m1}^{(0)} + T_{1m1}^{(0)} \right) +
  \left( S_{1m1}^{(1)} + T_{1m1}^{(1)} \right) +
  \epsilon \left( S_{1m1}^{(2)} + T_{1m1}^{(2)} \right) \\
={} & \frac{1}{\epsilon} \frac{\rho \theta}{\mu}
    \sum_{n=1}^{+\infty} a_{11n} \theta^{1-n} f_{1mn}^{(1)} +
    \frac{\rho \theta}{\mu}
      \sum_{n=1}^{+\infty} a_{11n} \theta^{1-n} f_{1mn}^{(2)} +
    \epsilon \frac{\rho \theta}{\mu}
      \sum_{n=1}^{+\infty} a_{11n} \theta^{1-n} f_{1mn}^{(3)}
  + \mathcal{O}(\epsilon^2) \\
={} & \frac{1}{\epsilon} \frac{\rho \theta}{\mu}
  \frac{f_{1m1}^{(1)}}{A_{11}} +
  \frac{\rho \theta}{\mu}
    \sum_{n=2}^{+\infty} a_{11n} \theta^{1-n} f_{1mn}^{(2)} \\
& \qquad {} + \sum_{n=2}^{+\infty} \frac{a_{11n}}{A_{11}}
    \sum_{n'=2}^{+\infty} B_{1nn'} D_{1n'} \pd{f_{1m1}^{(1)}}{t} +
  \epsilon \frac{\rho\theta}{\mu} \sum_{n=2}^{+\infty} a_{11n}
    \theta^{1-n} R_{1mn}^{(3)} + \mathcal{O}(\epsilon^2),
\end{split}
\end{equation}
Solving $\partial_t f_{1m1}^{(1)}$ from the above equation, we get
\begin{equation} \label{eq:df1_dt}
\begin{split}
\frac{\partial f_{1m1}^{(1)}}{\partial t} &= \left(
  1 - \frac{1}{A_{11}} \sum_{n'=2}^{+\infty} C_{1n'} D_{1n'}
\right)^{-1} \Bigg[
  \frac{1}{\epsilon} \left(
    \frac{\rho \theta}{\mu} \frac{f_{1m1}^{(1)}}{A_{11}} - T_{1m1}^{(0)}
  \right) \\
& \qquad + \left( \frac{\rho \theta}{\mu} \sum_{n=2}^{+\infty}
  a_{11n} \theta^{1-n} f_{1mn}^{(2)}
  - \left( S_{1m1}^{(1)} + T_{1m1}^{(1)} \right) \right) \\
& \qquad + \epsilon \left(
  \frac{\rho \theta}{\mu} \sum_{n=2}^{+\infty}
    a_{11n} \theta^{1-n} R_{1mn}^{(3)}
  - \left( S_{1m1}^{(2)} + T_{1m1}^{(2)} \right) \right)
\Bigg] + \mathcal{O}(\epsilon^2).
\end{split}
\end{equation}
In \eqref{eq:df1_dt}, the time derivatives in $S_{1m1}^{(2)}$ can be
replaced by \eqref{eq:v_theta} with $\mathcal{O}(\epsilon)$ terms
dropped, while the time derivatives in $S_{1m1}^{(1)}$ had to be
replaced by complete conservation laws \eqref{eq:Euler}. The last step
is to substitute the above equation into \eqref{eq:f3}, and the
$\mathcal{O}(\epsilon^2)$ term in \eqref{eq:df1_dt} can now be
discarded. This completes the calculation of $f_{1mn}^{(3)}$.

The calculation of $f_{2mn}^{(3)}$ follows exactly the same procedure,
and the details are omitted. The above procedure shows that the
expression of $f_{lmn}^{(3)}$ includes second-order derivatives of the
13 moments.

\subsection{Thirteen moment equations}
With all the moments up to third order calculated, we are ready to write down
the 13-moment equations. Note that all the 13-moment equations include the
conservation laws \eqref{eq:Euler}. Therefore we focus only on the
equations for $\sigma_{ij}$ and $q_j$, or equivalently, $f_{2m0}$ and
$f_{1m1}$. Since $f_{2m0} = \epsilon f_{2m0}^{(1)}$ and $f_{1m1} =
\epsilon f_{1m1}^{(1)}$, these equations are to be obtained by
truncation of \eqref{eq:max_ite}. Two different truncations are
considered below, which correspond to GG13 equations and R13
equations, respectively.

\subsubsection{Generalized Grad's 13-moment equations}
The derivation of GG13 equations is basically a truncation of
\eqref{eq:max_ite} up to the first order. The result reads
\begin{equation}
  \label{eq:gg13}
  \begin{aligned}
    & \epsilon \pd{f_{2m0}^{(1)}}{t} + \epsilon S_{2m0}^{(1)}+
    T_{2m0}^{(0)} + \epsilon T_{2m0}^{(1)} = \frac{\rho\theta}{\mu}
    \left(\frac{1}{A_{20}} f_{2m0}^{(1)} + \epsilon \sum_{n'=1}^{+\infty}
      a_{20n'}\theta^{-n'} f_{2mn'}^{(2)} \right), \\
    & \epsilon \pd{f_{1m1}^{(1)}}{t} + \epsilon S_{1m1}^{(1)}+
    T_{1m1}^{(0)} + \epsilon T_{1m1}^{(1)} = \frac{\rho\theta}{\mu}
    \left(\frac{1}{A_{11}} f_{1m1}^{(1)} + \epsilon \sum_{n'=2}^{+\infty}
      a_{11n'}\theta^{1-n'} f_{1mn'}^{(2)} \right),
  \end{aligned}
\end{equation}
where we have used
\begin{align}
  \sum_{n'=0}^{+\infty} a_{20n'} \theta^{-n'} f_{2mn'}^{(1)} &=
    \sum_{n'=0}^{+\infty} a_{20n'} \frac{A_{2n'}}{A_{20}} f_{2m0}^{(1)} =
    \frac{1}{A_{20}} f_{2m0}^{(1)}, \\
  \sum_{n'=1}^{+\infty} a_{11n'} \theta^{1-n'} f_{1mn'}^{(1)} &=
    \sum_{n'=1}^{+\infty} a_{11n'} \frac{A_{1n'}}{A_{11}} f_{1m1}^{(1)} =
    \frac{1}{A_{11}} f_{1m1}^{(1)}.
\end{align}
All the terms in \eqref{eq:gg13} have been represented by the 13 moments
\eqref{eq:13m} and their derivatives in Section \ref{sec:derivation}, and the
time derivatives in $S_{2m0}^{(1)}$ can be replaced by \eqref{eq:v_theta} with
$\mathcal{O}(\epsilon)$ terms discarded. The final step is to revert the
scaling of space and time introduced in the beginning of Section
\ref{sec:derivation}, which can be simply achieved by setting $\epsilon$ to be
$1$.

This set of equations are called generalized Grad's 13-moment (GG13)
equations as proposed in \cite{Struchtrup2013}. Similar to Grad's
13-moment equations, the GG13 equations are also first-order
quasi-linear equations. However, they are different from Grad's
13-moment equations in two ways: (1) in Grad's 13-moment equations, the
terms $T_{1m1}^{(1)}$ and $T_{2m0}^{(1)}$ come directly from the
truncation of the distribution function, while in generalized Grad's
13-moment equations, they include information from inversion of the
linearized collision operator; (2) in Grad's 13-moment equations, the
collision term does not include information from $f_{1mn'}$ with $n' >
1$ or $f_{2mn'}$ with $n' > 0$. Due to such differences, as mentioned
in Section \ref{sec:order}, Grad's 13-moment equations are accurate
only up to first order for general IPL models.

\add{%
Now we compare the equations \eqref{eq:gg13} with the equations
\eqref{eq:R13_sigma_q}. On the left-hand side, we can see from
\eqref{eq:gg13} that no second-order terms are involved. Since the
moments $m_{ijk}$ ($f_{3m0}$), $R_{ik}$ ($f_{2m1}$) and $\Delta$
($f_{002}$) are all $O(\epsilon^2)$ terms, they are simply set to be
zero in GG13 equations. The right-hand side is much more complicated
due to the involved formulas for second-order terms. We would just
like to point out that the coefficient $D_0^{(\eta)}$ in
\eqref{eq:S_Q_G13} does not equal the coefficient $1/A_{20}$ in
\eqref{eq:gg13}, since the same term also appears in $f_{2mn}^{(2)}$,
as is shown in \eqref{eq:f2}. Due to the similar reason, the
coefficient $E_0^{(\eta)}$ in \eqref{eq:S_Q_G13} does not equal
$1/A_{11}$ in \eqref{eq:gg13}.
}

\subsubsection{Regularized 13-moment equations}
To gain one more order of accuracy, we need to keep the second-order terms in
\eqref{eq:max_ite}, and the result is
\begin{equation}
  \label{eq:f1f2_R13}
  \begin{aligned}
    & \epsilon \pd{ f_{2m0}^{(1)}}{t}+ \epsilon S_{2m0}^{(1)} +
    \epsilon^2 S_{2m0}^{(2)} + T_{2m0}^{(0)} + \epsilon T_{2m0}^{(1)}
    + \epsilon^2
    T_{2m0}^{(2)} = \\
    & \qquad \qquad \frac{\rho\theta}{\mu} \left(\frac{1}{A_{20}} f_{2m0}^{(1)}
      + \sum_{n'=1} a_{2nn'}\theta^{n-n'} \left(
        \epsilon f_{2mn'}^{(2)} + \epsilon^2
        f_{2mn'}^{(3)}  \right) \right), \\
    & \epsilon \pd{ {f}_{1m1}^{(1)}}{t} + \epsilon S_{1m1}^{(1)} +
    \epsilon^2 S_{1m1}^{(2)} + T_{1m1}^{(0)} + \epsilon T_{1m1}^{(1)}
    + \epsilon^2
    T_{1m1}^{(2)} = \\
    & \qquad \qquad \frac{\rho\theta}{\mu} \left(\frac{1}{A_{11}} {f}_{1m1}^{(1)}
      + \sum_{n'=2} a_{1nn'}\theta^{n-n'} \left(
        \epsilon f_{1mn'}^{(2)} + \epsilon^2
        f_{1mn'}^{(3)}\right) \right).
  \end{aligned}
\end{equation}
Again, the time derivatives for velocity and temperature in
$S_{2m0}^{(i)}$ and $S_{1m1}^{(i)}, i = 1, 2$ need to be replaced by spatial
derivatives. In order to preserve the second-order terms, the time
derivatives in $S_{2m0}^{(1)}$ and $S_{1m1}^{(1)}$ need to be
substituted by the complete conservation laws \eqref{eq:Euler}, where
as in $S_{2m0}^{(2)}$ and $S_{1m1}^{(2)}$, the replacement of time
derivatives are done by using \eqref{eq:v_theta} and discarding
$\mathcal{O}(\epsilon)$ terms.  Afterwards, we set $\epsilon$ to be
$1$, and the result is regularized 13-moment equations.

\add{%
By replacing all the coefficients with the primitive variables
$\sigma_{ij}$ and $q_i$, the equations \eqref{eq:R13_sigma_q} with 
\eqref{eq:S_Q_G13}\eqref{eq:R13_mdelR} and
\eqref{eq:S_2}\eqref{eq:Q_2} can be obtained. Compared with the linear
R13 equations obtained in \cite{Struchtrup2013}, much more information
is included in this nonlinear version. For instance, in
\eqref{eq:R13_mdelR}, one can see that all the first three terms in
$m_{ijk}^{(\eta)}$ are nonlinear and all the last three terms in
$\Delta^{(\eta)}$ are nonlinear. Clearly these terms cannot be ignored
in problem such as the structure of plane shock waves, which will be
studied in the following section.
}


\section{Numerical examples}
\label{sec:numerical}
In this section, we are going to test the behavior of the nonlinear R13
equations by computing the structure of one-dimensional plane shock
waves, which is a benchmark problem in the gas kinetic theory. It
involves strong nonequilibrium, but does not have any boundary
condition, which makes it suitable for testing the ability of
describing nonequilibrium processes for our models.

For one-dimensional flow, the moments satisfy
\begin{equation} \label{eq:1dflow}
v_2 = v_3 = 0, \quad \sigma_{12} = \sigma_{13} = \sigma_{23} = 0, \quad
  \sigma_{22} = \sigma_{33} = -\frac{1}{2} \sigma_{11}, \quad q_2 = q_3 = 0.
\end{equation}
For simplicity, we use the notation $v = v_1$, $\sigma = \sigma_{11}$
and $q = q_1$. Then thirteen moments are reduced to five moments
$\rho$, $v$, $\theta$, $\sigma$ and $q$. Below we write down the R13
model for these quantities in the form of balance laws:
\begin{equation}
  \label{eq:1d_conservative}
  \begin{aligned}
   &  \rho_t + (\rho v)_x = 0, \\
   &  (\rho v)_t + (\rho v^2 + \rho\theta + \sigma)_x = 0, \\
   & \left(\frac{1}{2}\rho v^2+\frac{3}{2}\rho\theta\right)_t+ \left(q
      +
      \frac{1}{2}\rho v^3  + \frac{5}{2}\rho v \theta + v\sigma\right)_x = 0, \\
  &  (\rho v^2 + \rho\theta + \sigma)_t + \left(\frac{6}{5}q + \rho v^3
    + 3\rho v \theta + 3v \sigma + m^{(\eta)}\right)_x = \Sigma^{(\eta)}_{\mathrm{1D}}, \\
   &  \left(q + \frac{1}{2}\rho v^3 + \frac{5}{2}\rho v \theta +
      v\sigma\right)_t + (\mathrm{FQ})_x = Q^{(\eta)}_{\mathrm{1D}} + v \Sigma^{(\eta)}_{\mathrm{1D}}, 
  \end{aligned}
  \end{equation}
where 
\begin{equation}
  \label{eq:FQ}
  \mathrm{FQ} =  \frac{16}{5} v q + \frac{1}{2}\rho v^4 + 4v^2\theta +
     \frac{5}{2}\rho \theta^2  + \frac{5}{2}v^2 \sigma + \left(\frac{7}{2} -
     \sqrt{\frac{14}{3}} C_{\mathrm{1D}}^{(\eta)} \right) \theta \sigma + m^{(\eta)}  v + \frac{1}{2}R^{(\eta)} +
     \frac{1}{6}\Delta^{(\eta)},
\end{equation}
with $m^{(\eta)}, R^{(\eta)}, \Delta^{(\eta)}$ being $m^{(\eta)}_{111}, R^{(\eta)}_{11}$ and $\Delta^{(\eta)}$ in
\eqref{eq:R13_mdelR} substituted by \eqref{eq:1dflow}. On the
right-hand side, $\Sigma^{(\eta)}_{\mathrm{1D}}$ and
$Q^{(\eta)}_{\mathrm{1D}}$ are, respectively, $\Sigma_{11}^{(\eta,1)}
+ \Sigma_{11}^{(\eta,2)}$ and $Q_{1}^{(\eta,1)} +Q_{1}^{(\eta,2)}$ in
\eqref{eq:R13_sigma_q} subject to \eqref{eq:1dflow}. The constant
$C_{\mathrm{1D}}^{(\eta)}$ depends only on $\eta$, and some of its
values are listed in Table \ref{tab:coe_C_1d}.
\begin{table}[ht!]
  \centering
  \def\arraystrech{1.5}
  {\footnotesize
  \begin{tabular}[ht]{c||c|c|c|c|c}
    $\eta$ & $5$ & $7$ & $10$ & $17$ & $\infty$\\
    \hline \hline 
    $C_{\mathrm{1D}}^{(\eta)}$ & $0$ & $0.0331$  & $0.0553$ & $0.0748$ & $0.1$
  \end{tabular} }
  \caption{Coefficient $C_{\mathrm{1D}}^{(\eta)}$ for different $\eta$. }
  \label{tab:coe_C_1d}
\end{table}

The structure of plane shock waves with Mach number $\mMa$ can be obtained by
setting the initial data to be
\begin{equation}
  \label{eq:initial_shock}
  (\rho, v, \theta, \sigma, q) = \left\{
    \begin{array}[c]{cc}
      (\rho_l, v_l, \theta_l, 0, 0), & x < 0, \\
      (\rho_r, v_r, \theta_r, 0, 0), & x > 0, 
    \end{array}
\right.
\end{equation}
where 
\begin{gather*}
  \rho_l = 1, \quad v_l = \sqrt{5/3}\mMa, \quad  \theta_l = 1, \\
  \rho_r = \frac{4\mMa^2}{\mMa^2 + 3}, \quad v_r =
  \sqrt{\frac{5}{3}}\frac{\mMa^2 + 3}{4\mMa}, \quad \theta_r = \frac{5\mMa^2
  - 1}{4\rho_r}.
\end{gather*}

To solve \eqref{eq:1d_conservative} and \eqref{eq:initial_shock}
numerically, the finite volume method is adopted. Since the left-hand
side of \eqref{eq:1d_conservative} has the form of a conservation law,
we apply the HLL scheme in the discretization. The right-hand side
provides the non-conservative part, for which central difference
method is used to approximate both the first and second derivatives.
For the time discretization, we use the classical forward Euler method
in all the examples. The DSMC results for variable hard sphere models
with the same viscosity index are used as reference solutions
\cite{Bird}.

\begin{figure}[!ht]
\centering
\subfigure[$\mMa=1.55$, $\bar{\rho}, \bar{v}, \bar{\theta}$]{%
  \begin{overpic}
    [width = .45\textwidth, clip]{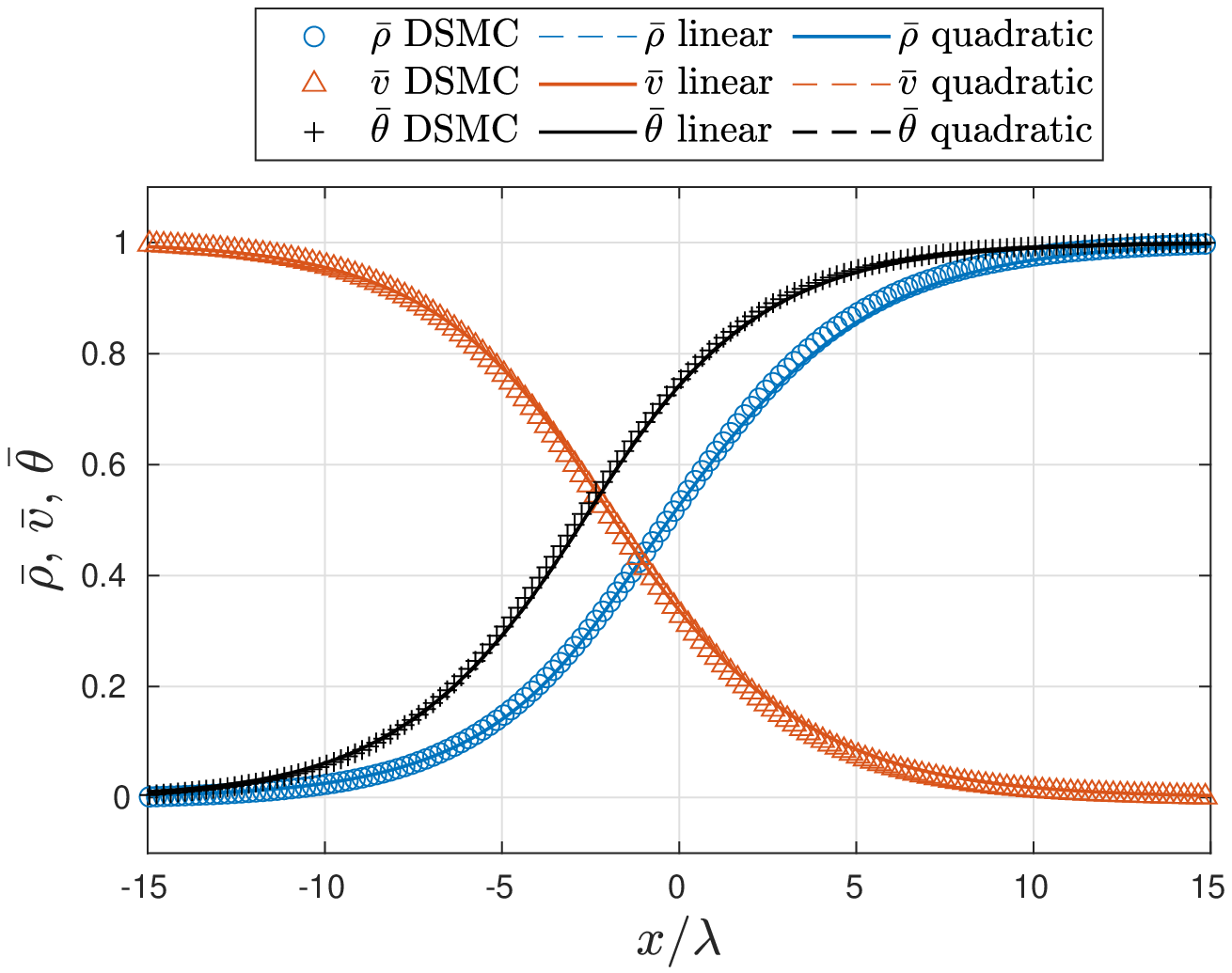}  
  \end{overpic}
  } 
\subfigure[$\mMa=9.0$,  $\bar{\rho}, \bar{v}, \bar{\theta}$]{%
  \begin{overpic}
    [width = .45\textwidth, clip]{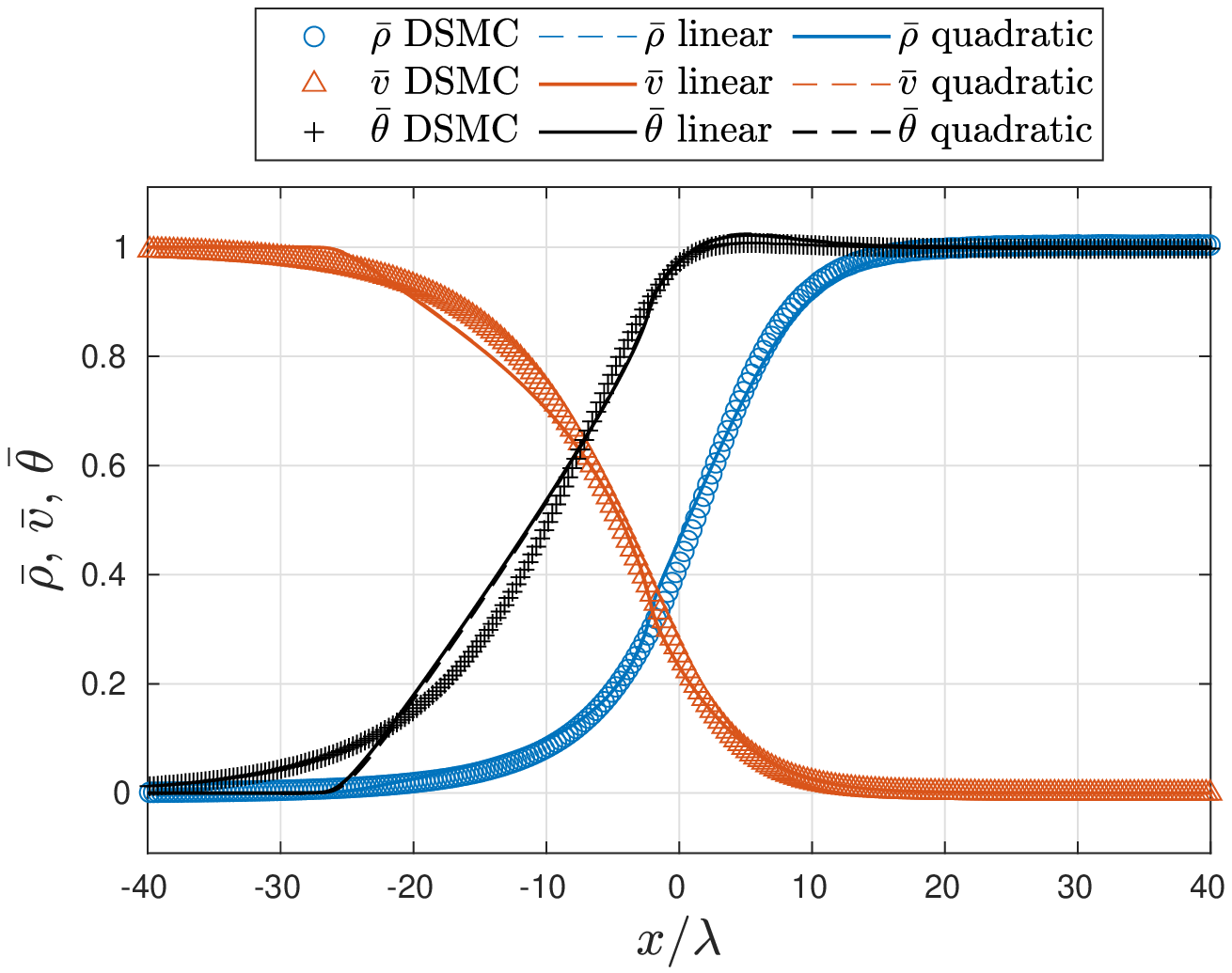}  
  \end{overpic}
  }
  \caption{Normalized density, velocity, and temperature of shock
    structures for the Maxwell molecules model and Mach numbers
    $\mMa = 1.55, 9$. DSMC solutions for the variable hard sphere
    model are provided as reference results. The horizontal axis is
    $x/\lambda$ with $\lambda$ being the mean free path.The solid
    lines are the numerical results for the linearized collision model
    and the dashed lines are those for the quadratic collision
    model.}
\label{fig:quad_shock_rhoutheta}
\end{figure}

\begin{figure}[!ht]
\centering
\subfigure[$\mMa=1.55$, $\sigma, q$]{%
  \begin{overpic}[width = .5\textwidth, clip]{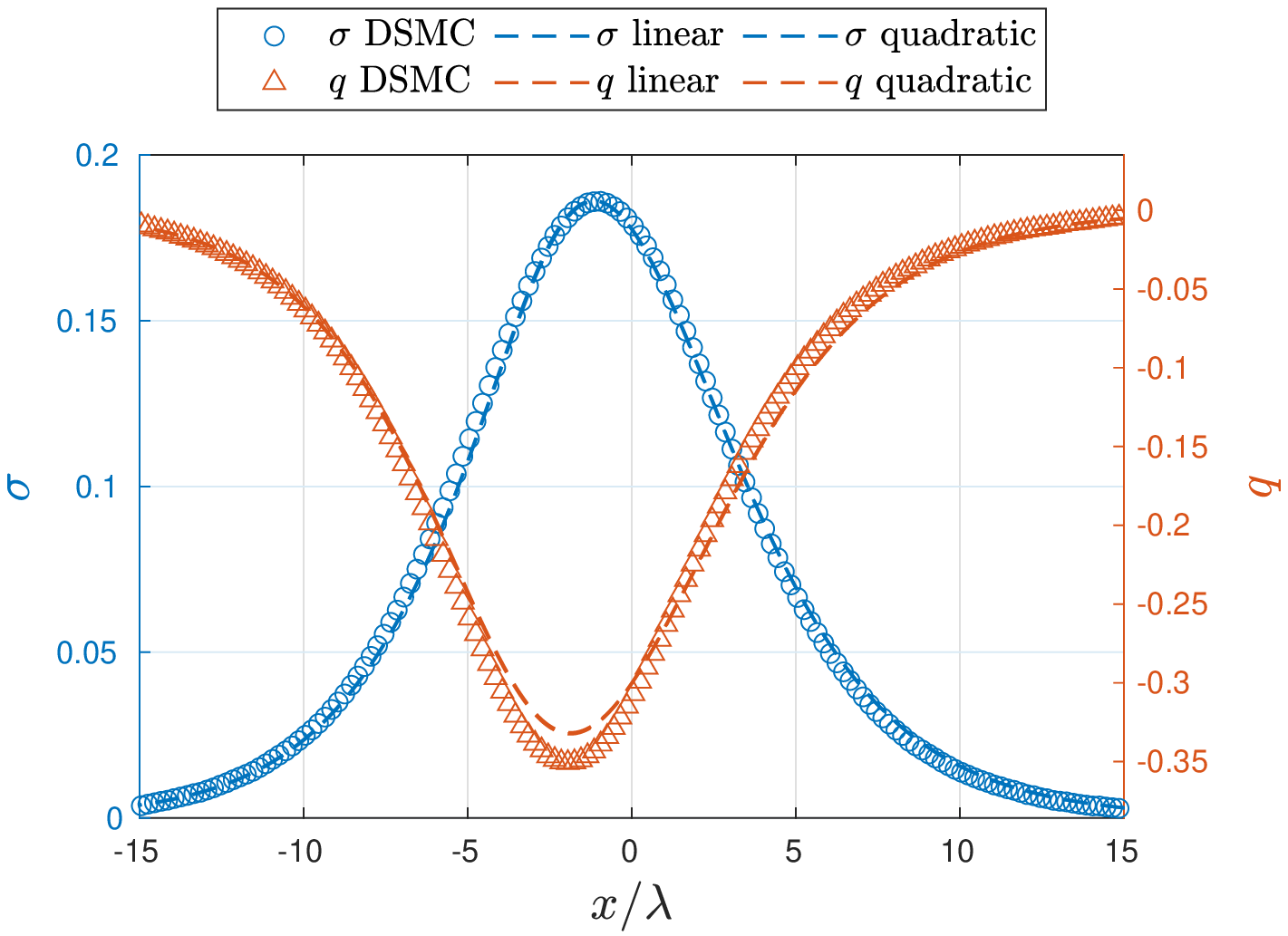}  
  \end{overpic}
  }%
  \subfigure[$\mMa=9.0$, $\sigma, q$]{%
  \begin{overpic}
  [ width = .5\textwidth,clip]{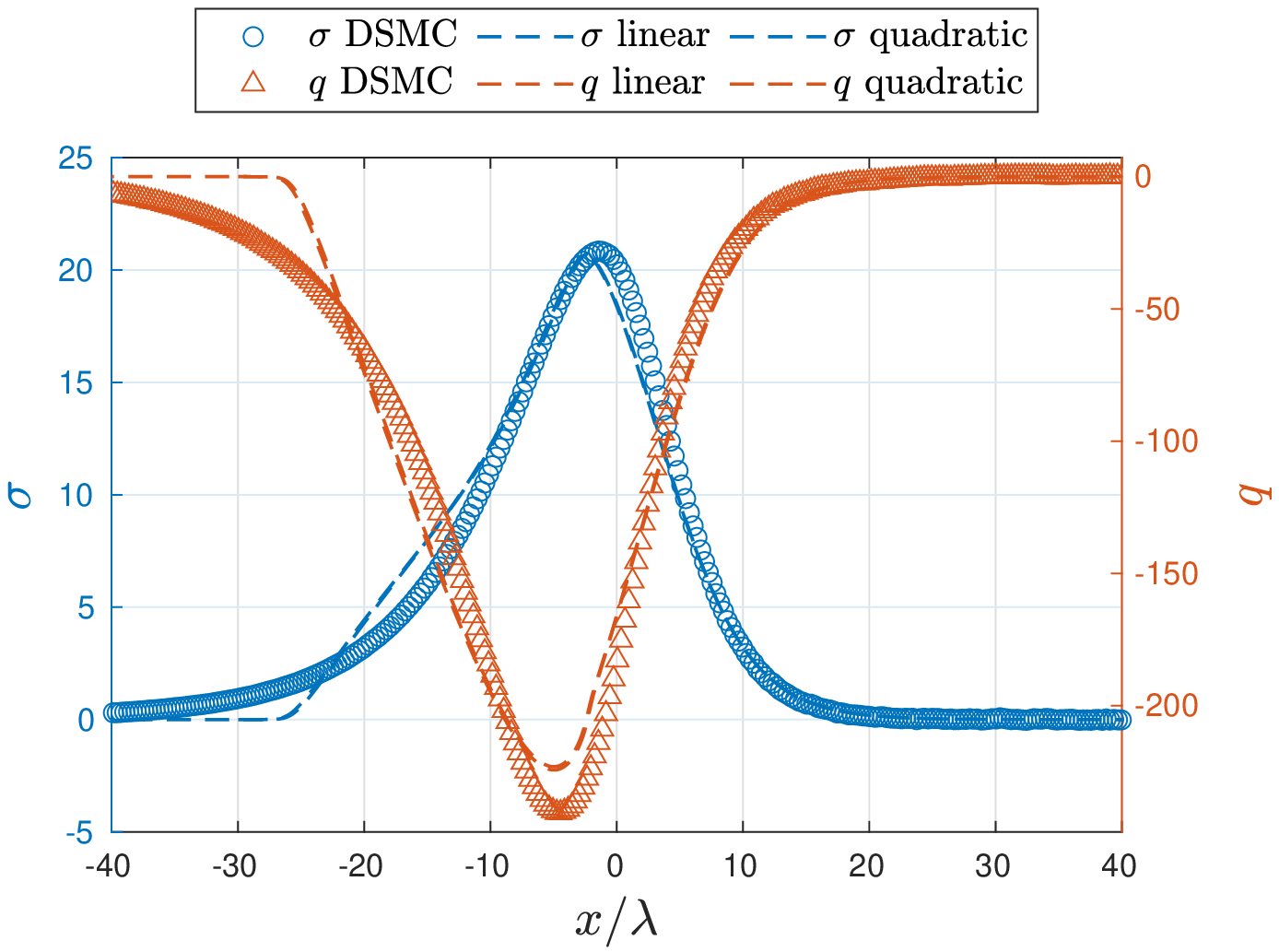}  
  \end{overpic}
  }
  \caption{The stress $\sigma$ and heat flux $q$ of shock structures
    for the Maxwell molecules model and Mach numbers
    $\mMa = 1.55, 9$. DSMC results for the variable hard
    sphere model are provided as references. The horizontal axis is
    $x/\lambda$ with $\lambda$ being the mean free path. The left
    $y$-axis corresponds to the stress and the right $y$-axis
    corresponds to the heat flux. The solid lines are the numerical
    results for the linearized collision model and the dashed lines
    are those for the quadratic collision model. }
\label{fig:quad_shock_sigmaq}
\end{figure}

\add{%
\subsection{Shock structure for Maxwell molecules}
This section is devoted to shock structure computation of Maxwell
molecules. The same computation has been carried out in
\cite{Timokhin2017}, and the main purpose of this section is the
verification of our numerical method and the comparison between linearized
and quadratic collision terms. Note that the R13 equations for Maxwell
molecules with quadratic collision terms are already available to us
\cite{Struchtrup}. Two Mach numbers $1.55$ and $9.0$ are tested, and
we show the numerical results for all the moments in Figures
\ref{fig:quad_shock_rhoutheta} and \ref{fig:quad_shock_sigmaq}, where
the density, velocity and temperature are given in their normalized
form:
\begin{displaymath}
  \bar{\rho} = \frac{\rho - \rho_l}{|\rho_r - \rho_l|}, \qquad
  \bar{v} = \frac{v - v_r}{|v_l - v_r|}, \qquad
  \bar{\theta} = \frac{\theta - \theta_l}{|\theta_r - \theta_l|}.
\end{displaymath}
For the small Mach number $1.55$, both linearized and quadratic
collision terms provide good agreement with DSMC results, except for a
slight underestimation of the peak heat flux. Surprisingly, when the
Mach number reaches $9.0$, R13 equations with linearized and quadratic
collision terms still provide almost identical shock structure.
Similar to the results in \cite{Timokhin2017}, our profiles also show
some typical structures for R13 solutions with high Mach numbers, such
as kinks in the profiles and a too fast decay in the low-density
region, which indicates the correctness of our simulation. As is
stated in \cite{Timokhin2017}, the underpredicted heat flux is due to
the loss of some fourth-order terms in the regularized moment
equations. Nevertheless, these results indicate that quadratic
collision terms do not contribute too much to the shock structure for
Mach number lower than $9.0$.
}

\subsection{Shock structures for different Mach numbers}

In this experiment, we test the approximability of the R13 model by
varying the Mach number for a non-Maxwell collision model. Four Mach
numbers $\mMa = 1.55, 3.0, 6.5, 9.0$ are taken into account, and we
consider the inverse power law model with $\eta = 10$ and the
hard-sphere model ($\eta = \infty$) in our tests.

\begin{figure}[!ht]
\centering
\subfigure[$\mMa=1.55$, $\bar{\rho}, \bar{v}, \bar{\theta}$]{%
  \begin{overpic}
  [width=.45\textwidth,clip]{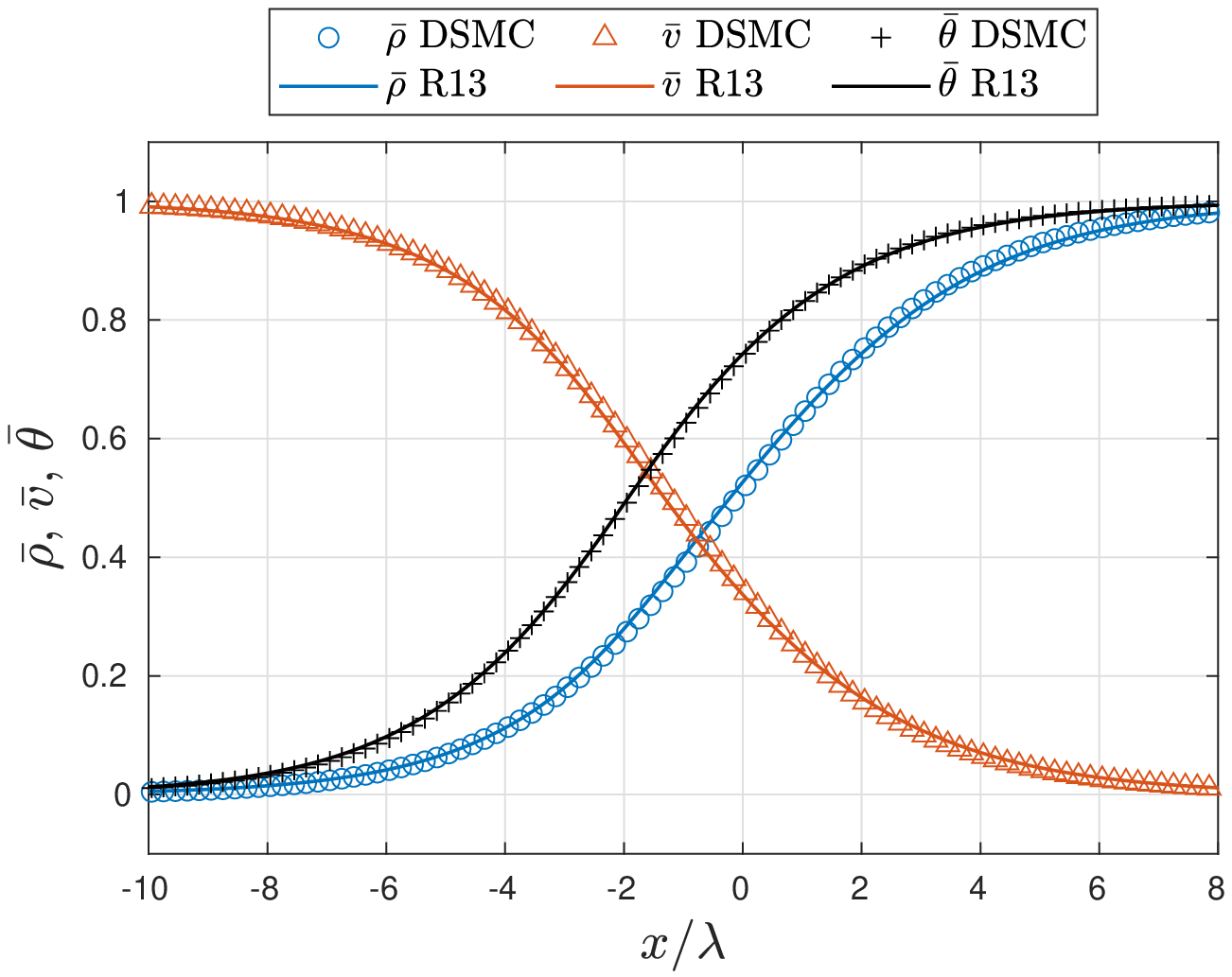}  
  \end{overpic}
  } \quad
  \subfigure[$\mMa=3.0$, $\bar{\rho}, \bar{v}, \bar{\theta}$]{%
  \begin{overpic}
  [width=.45\textwidth, clip]{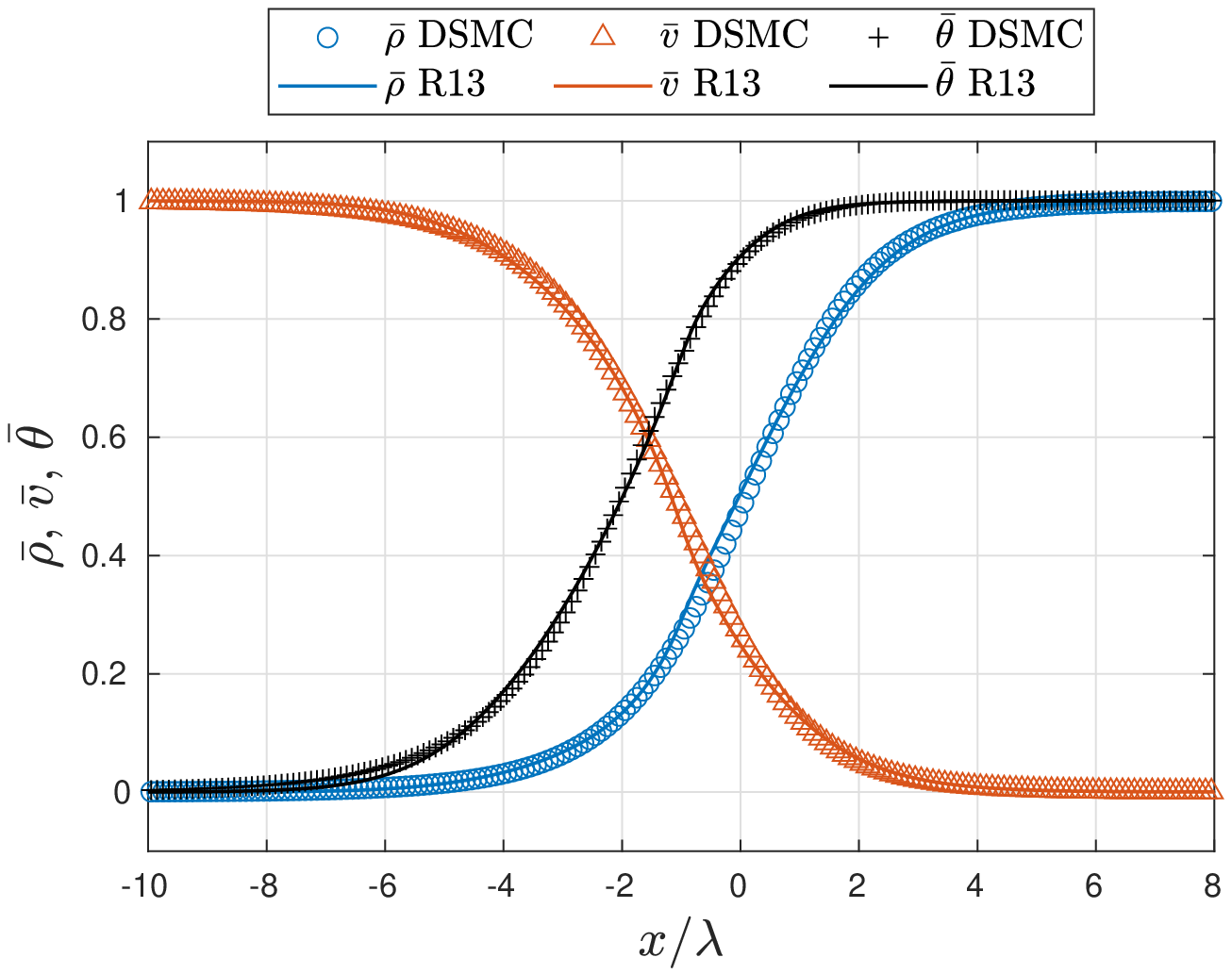}  
  \end{overpic}
  }\\
  \subfigure[$\mMa=6.5$, $\bar{\rho}, \bar{v}, \bar{\theta}$]{%
  \begin{overpic}
  [width=.45\textwidth,clip]{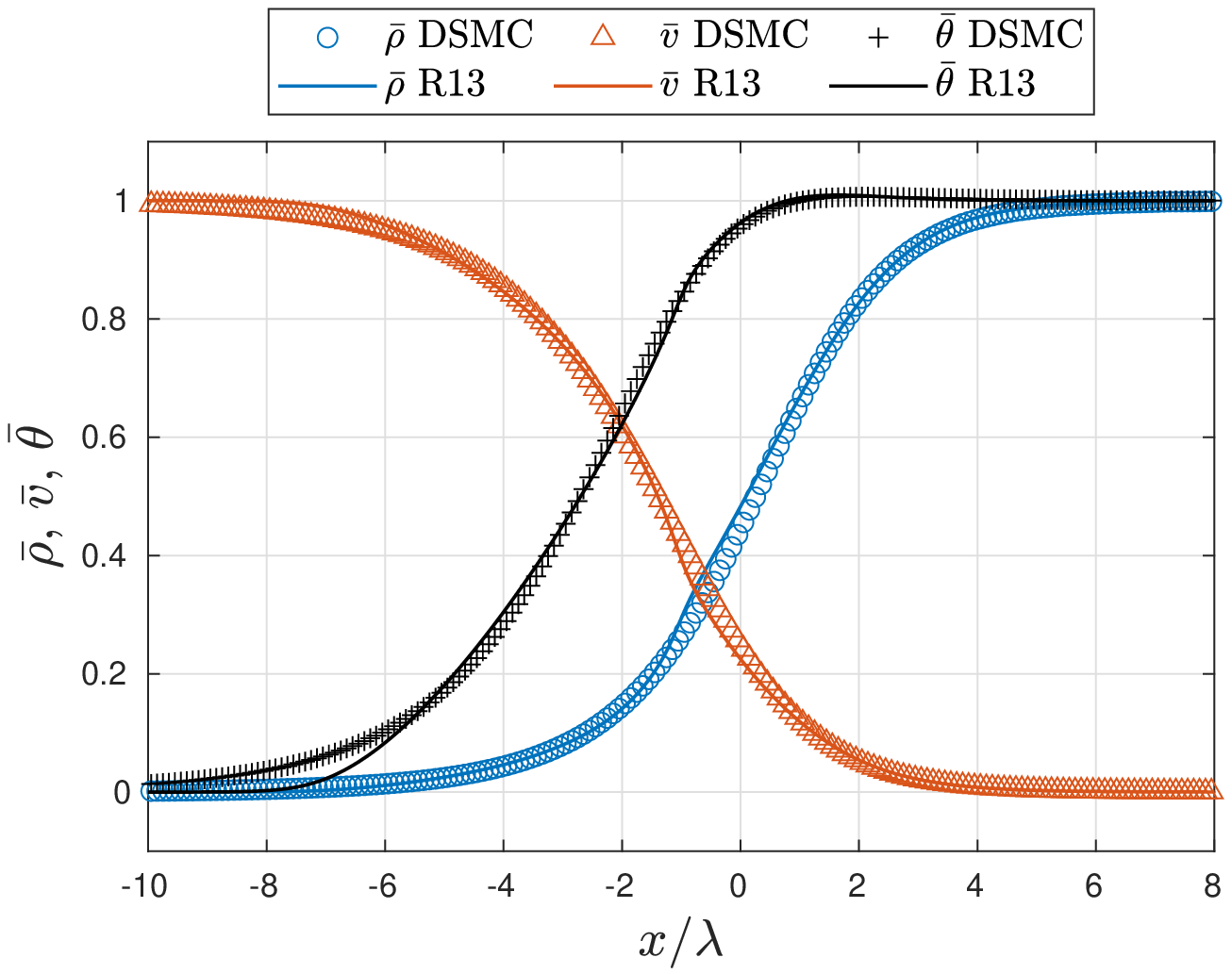}  
  \end{overpic}
  } \quad
\subfigure[$\mMa=9.0$, $\bar{\rho}, \bar{v}, \bar{\theta}$]{%
  \begin{overpic}
  [width=.45\textwidth,clip]{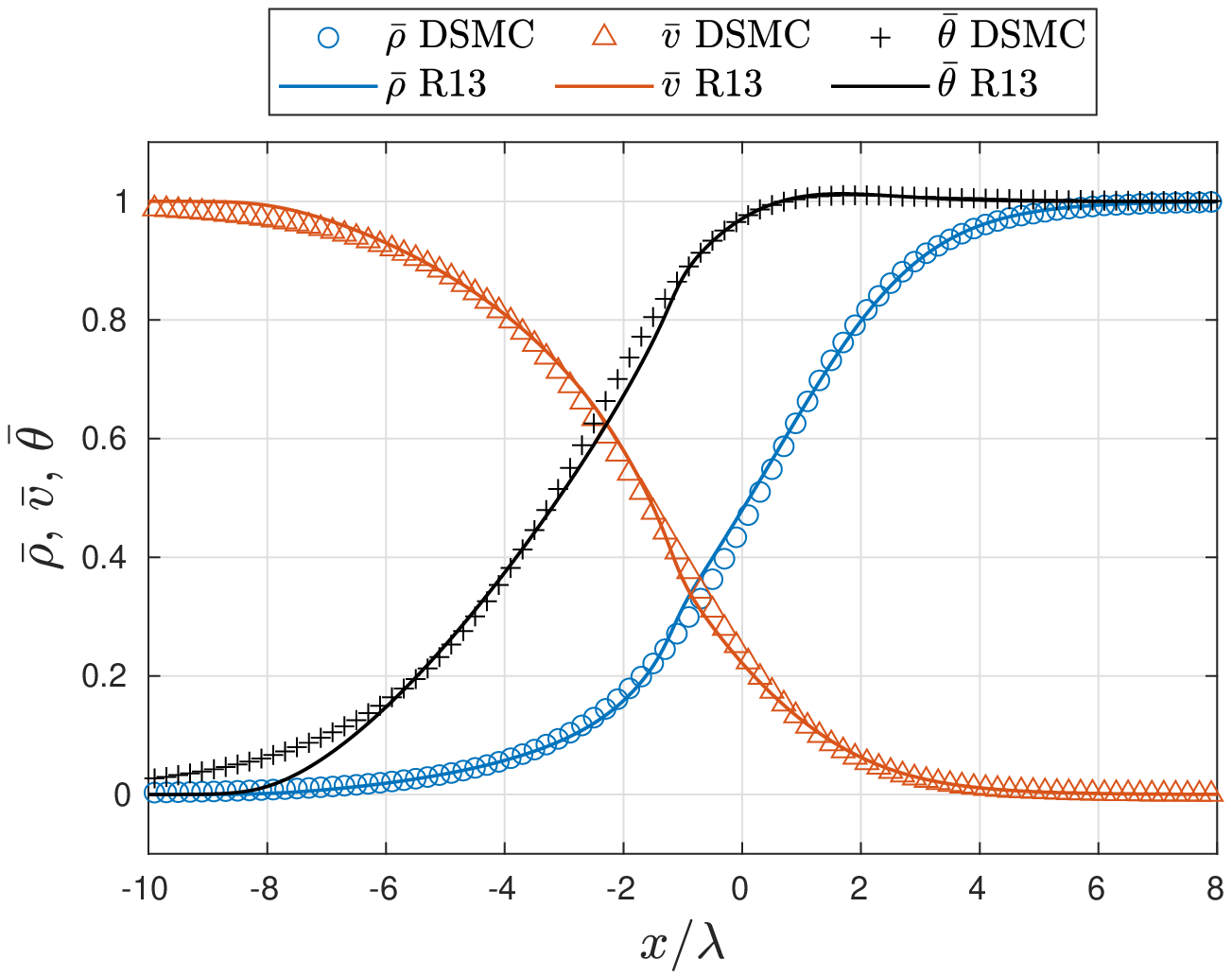}  
  \end{overpic}
  }
  \caption{Normalized density, velocity, and temperature of shock
    structures for the IPL model with $\eta = 10$ and Mach numbers
    $\mMa = 1.55, 3, 6.5, 9$. DSMC solutions for the variable hard
    sphere model are provided as reference results. The horizontal
    axis is $x/\lambda$ with $\lambda$ being the mean free path.}
\label{fig:eta10_shock_rhoutheta}
\end{figure}

\begin{figure}[!ht]
\centering
\subfigure[$\mMa=1.55$, $\sigma, q$]{%
  \begin{overpic}
  [width=.5\textwidth, clip]{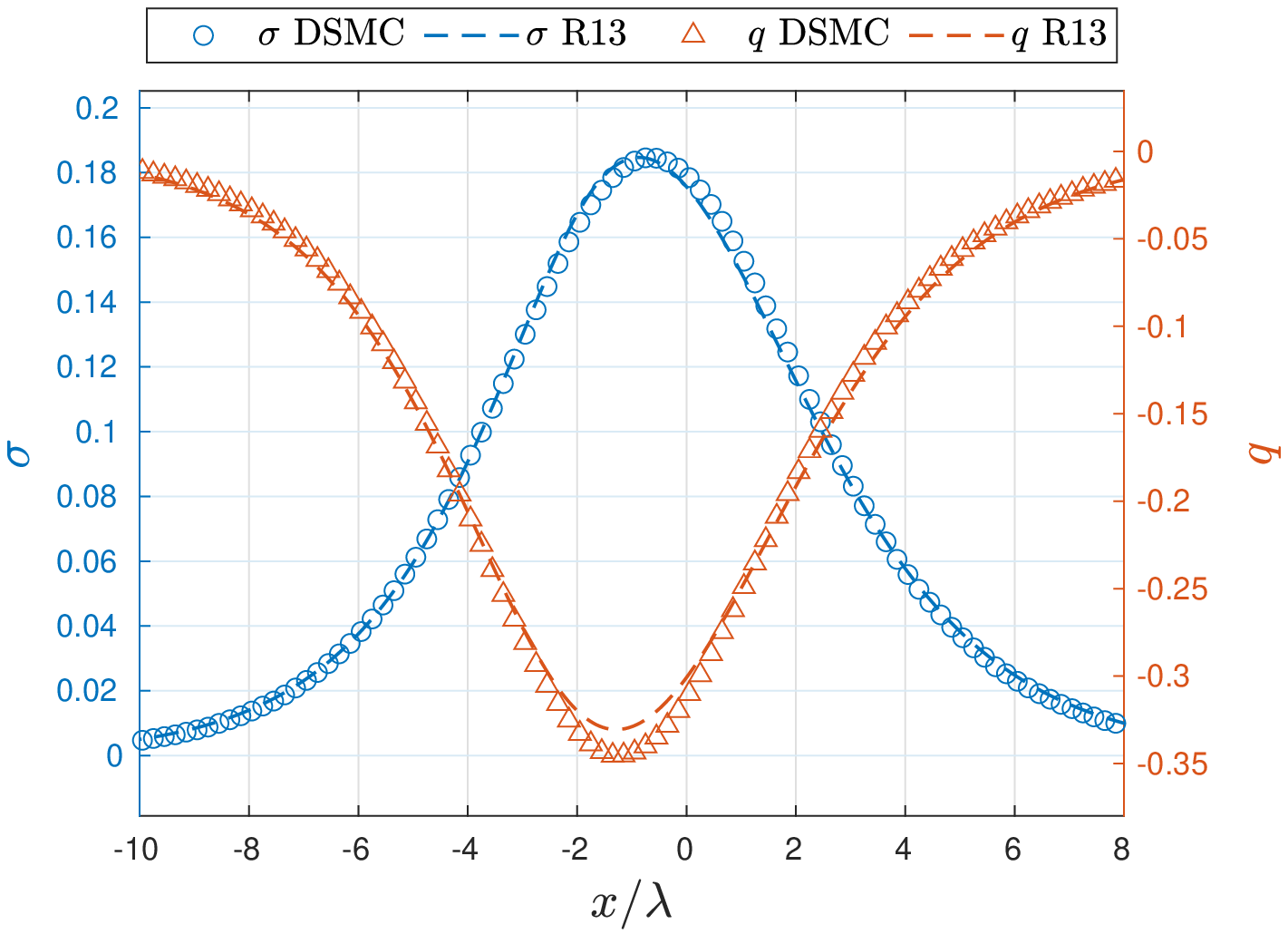}  
  \end{overpic}
  }%
\subfigure[$\mMa=3.0$, $\sigma, q$]{%
  \begin{overpic}
  [width=.5\textwidth, clip]{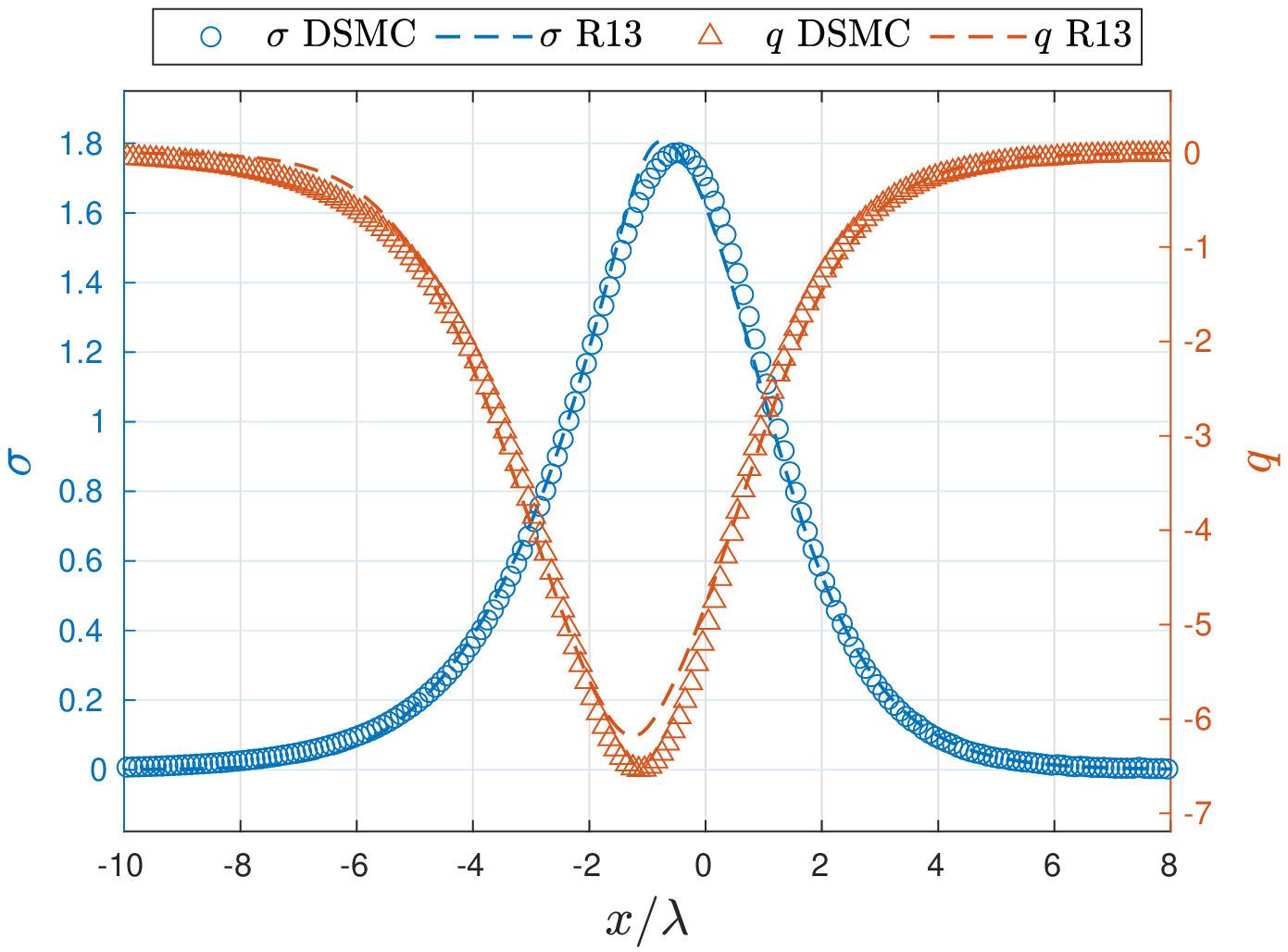}  
  \end{overpic}
  }\\
  \subfigure[$\mMa=6.5$, $\sigma, q$]{%
  \begin{overpic}
  [width=.5\textwidth, clip]{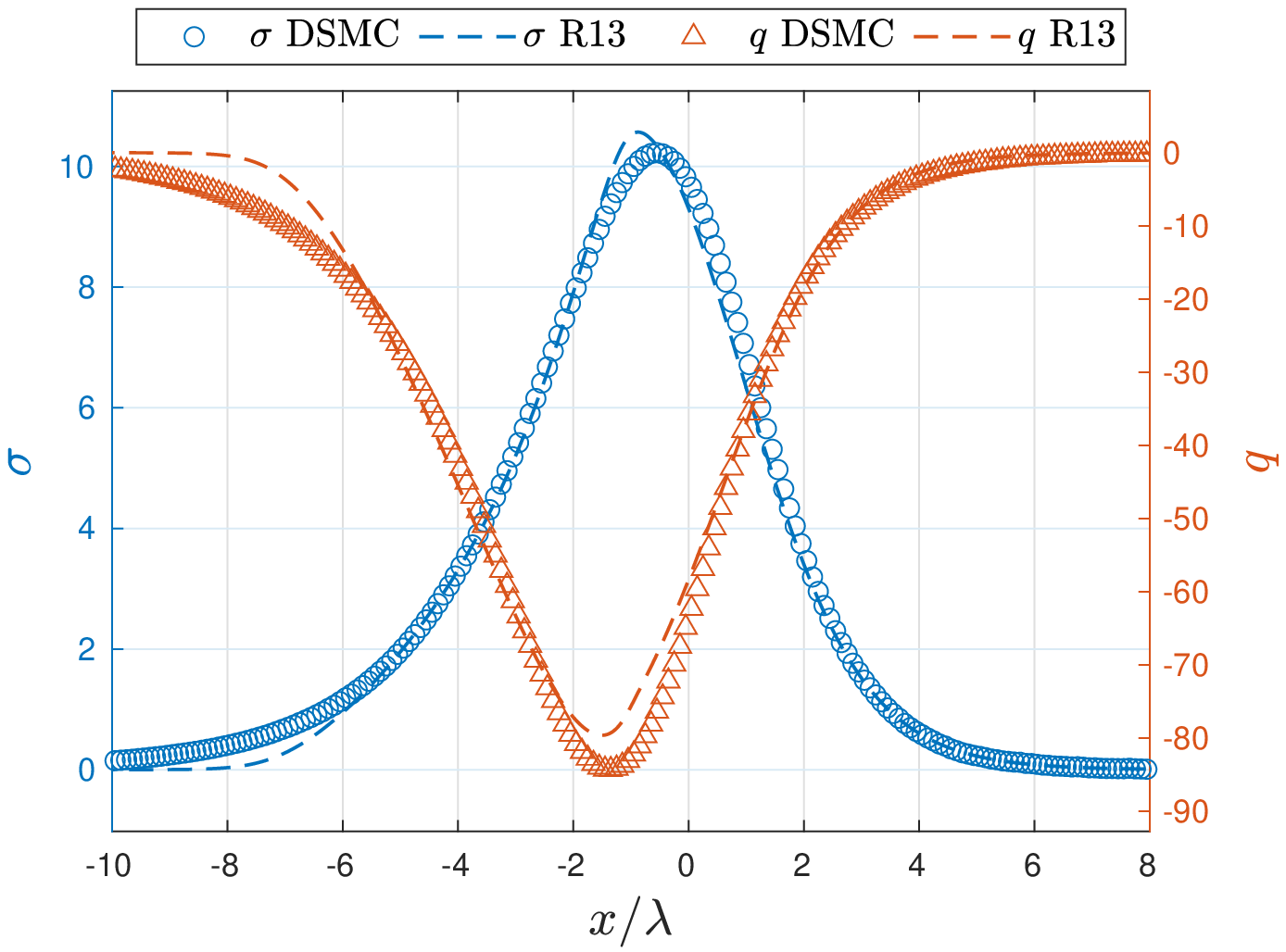}  
  \end{overpic}
  }%
\subfigure[$\mMa=9.0$, $\sigma, q$]{%
  \begin{overpic}
  [width=.5\textwidth, clip]{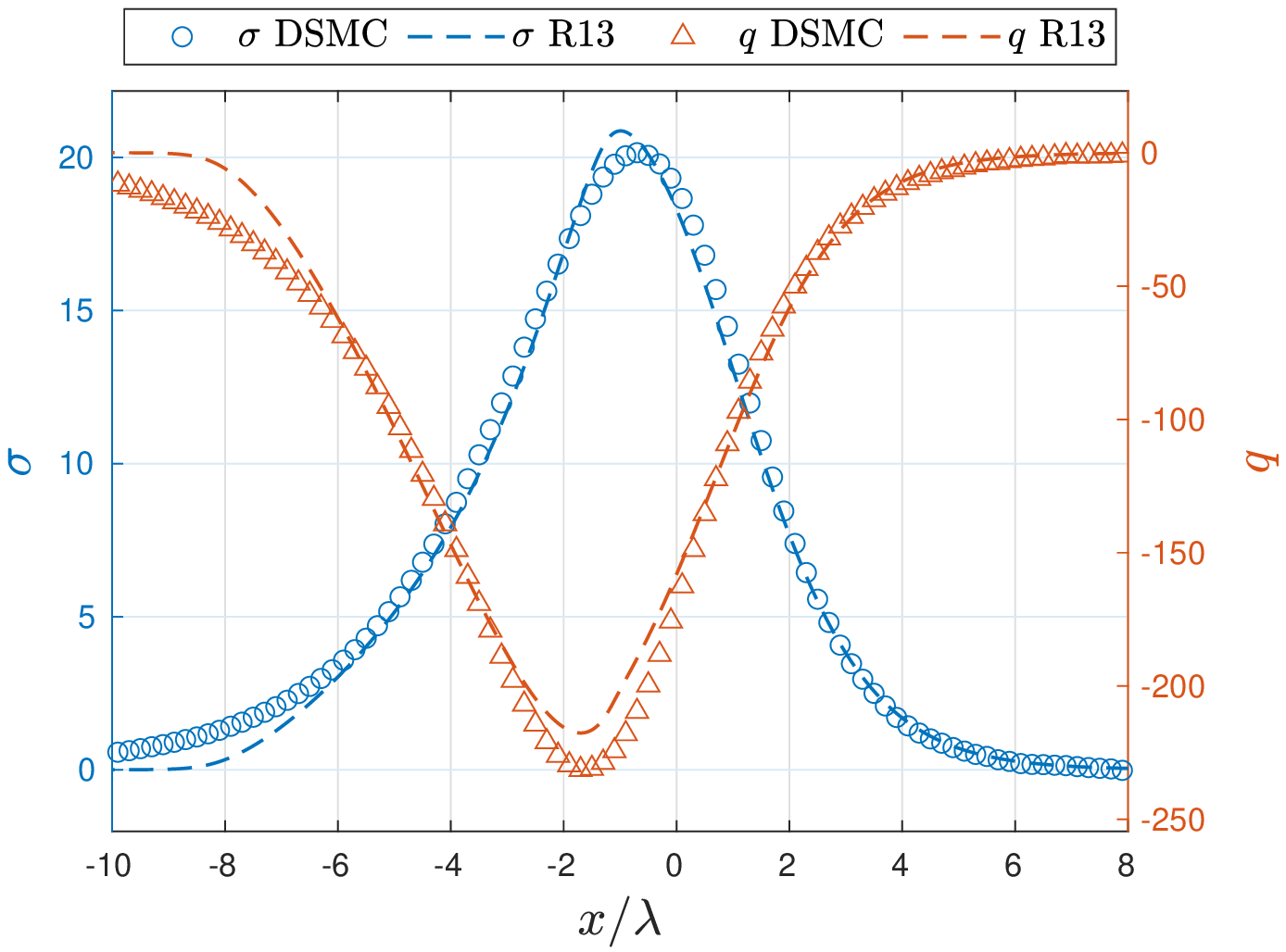}  
  \end{overpic}
  }
  \caption{The stress $\sigma$ and heat flux $q$ of shock structures
    for the IPL model with $\eta = 10$ and Mach numbers $\mMa = 1.55,
    3, 6.5, 9$. DSMC results for the variable hard sphere model are
    provided as references. The horizontal axis is $x/\lambda$ with
    $\lambda$ being the mean free path. The left $y$-axis corresponds
    to the stress and the right $y$-axis corresponds to the heat flux.}
\label{fig:eta10_shock_sigmaq}
\end{figure}

Figure \ref{fig:eta10_shock_rhoutheta} and
\ref{fig:eta10_shock_sigmaq} show the comparison between the R13
results and the DSMC results for $\eta = 10$. The profiles of all the
five quantities have been plotted. Although DSMC uses variable hard
sphere model as an approximation of the inverse power law model, the
simulation results in \cite{Hu2019} show that for the shock structure
problem, the variable hard sphere model and the inverse power law
model show almost identical results for Mach numbers $6.5$ and $9.0$,
which means that it is reliable to use DSMC results to check the
quality of R13 results. For increasing Mach number, it is generally
harder for macroscopic models to accurately capture the nonequilibrium
effects. This can be clearly observed from Figure
\ref{fig:eta10_shock_sigmaq}, which shows that the heat flux is
underestimated in the low density region. Note that the shock
structure in the high density region is well captured for all Knudsen
numbers, since the high density and temperature in this region result
in distribution functions close to the local Maxwellians, which can be
relatively easier to represent using the Chapman-Enskog expansion.

\begin{figure}[!ht]
\centering
\subfigure[$\mMa=1.55$, $\bar{\rho}, \bar{v}, \bar{\theta}$]{%
  \begin{overpic}
    [width = .45\textwidth, clip]{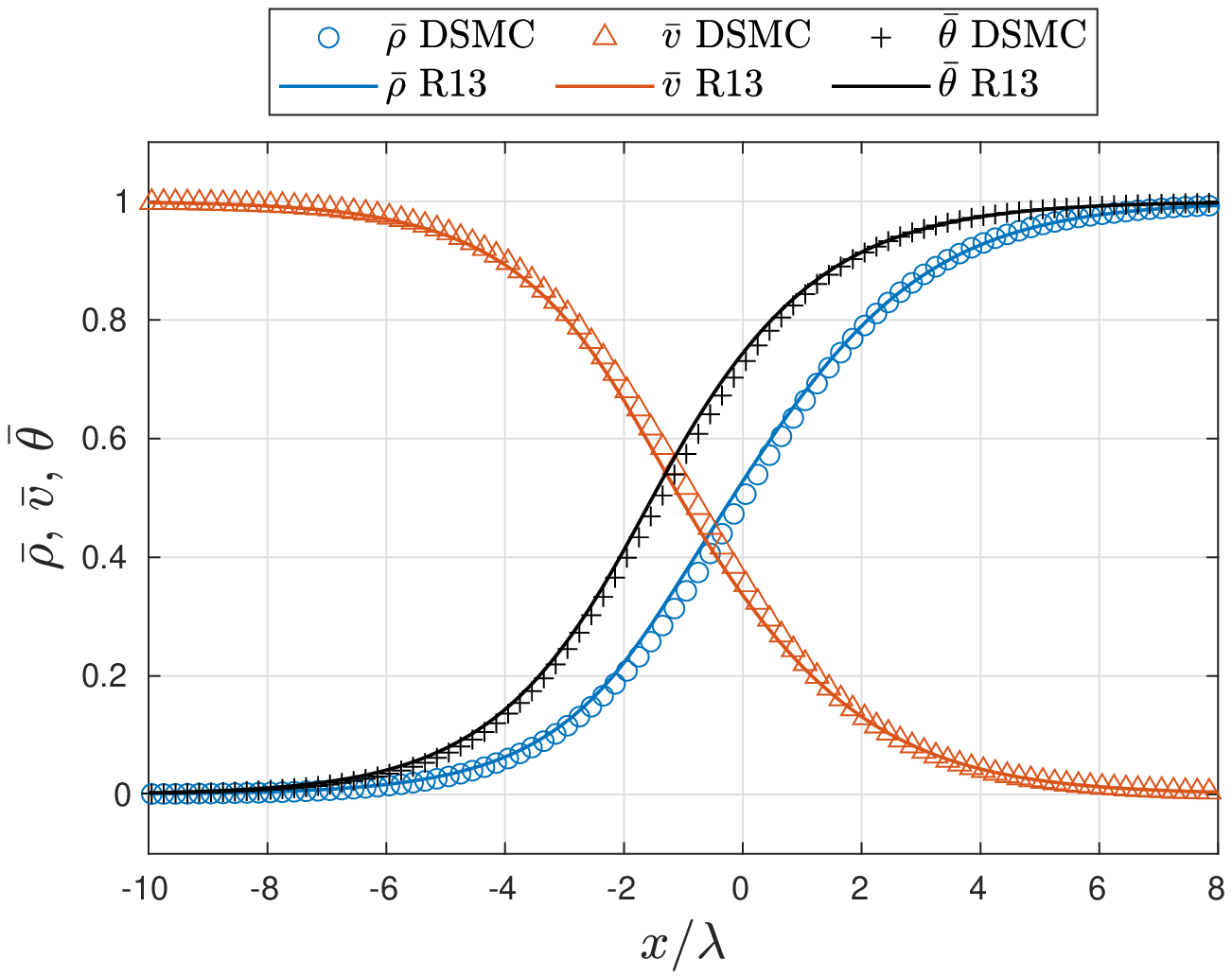}  
  \end{overpic}
  } \quad
\subfigure[$\mMa=3.0$,  $\bar{\rho}, \bar{v}, \bar{\theta}$]{%
  \begin{overpic}
    [width = .45\textwidth, clip]{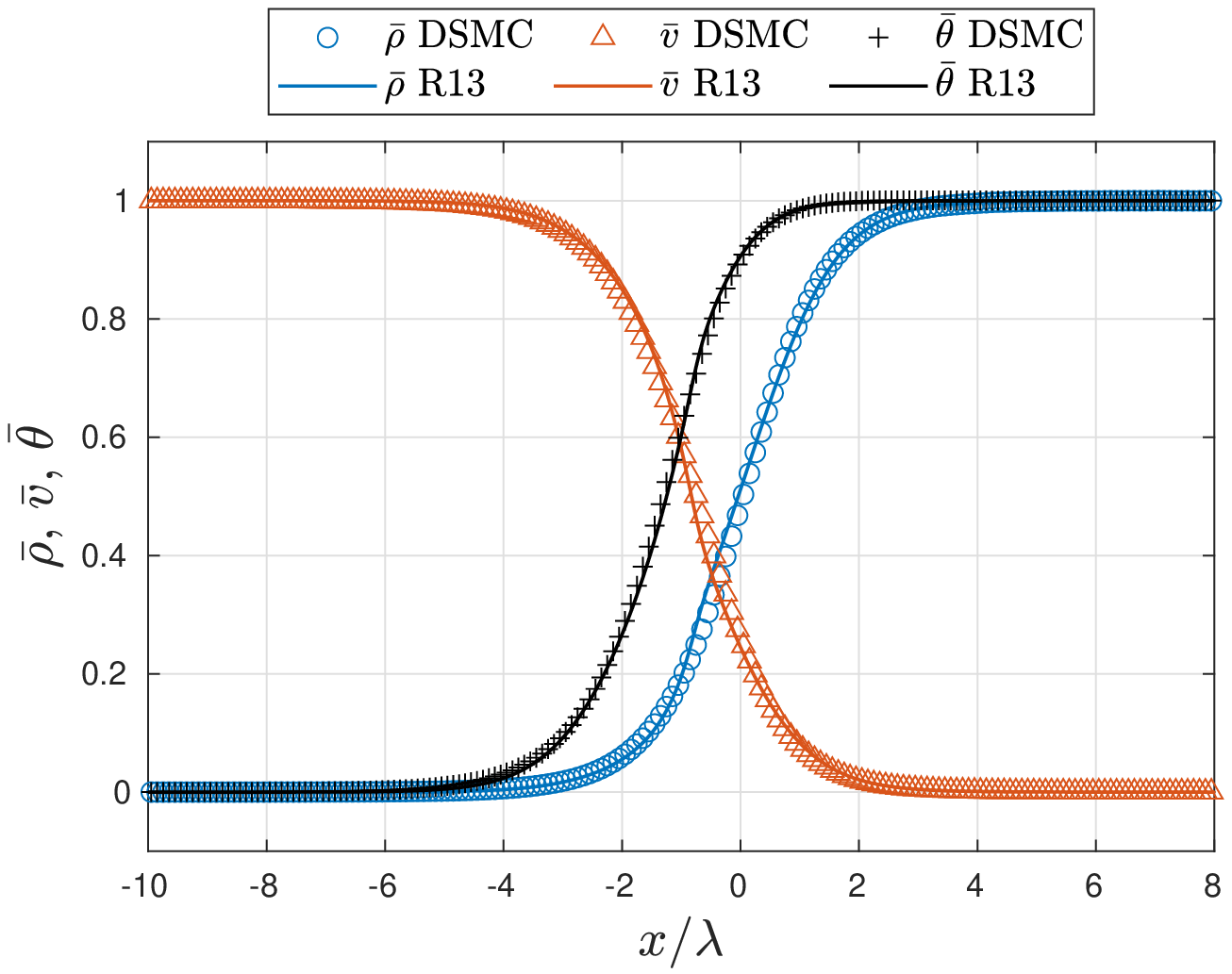}  
  \end{overpic}
  }\\
  \subfigure[$\mMa=6.5$, $\bar{\rho}, \bar{v}, \bar{\theta}$]{%
  \begin{overpic}
  [width = .45\textwidth,clip]{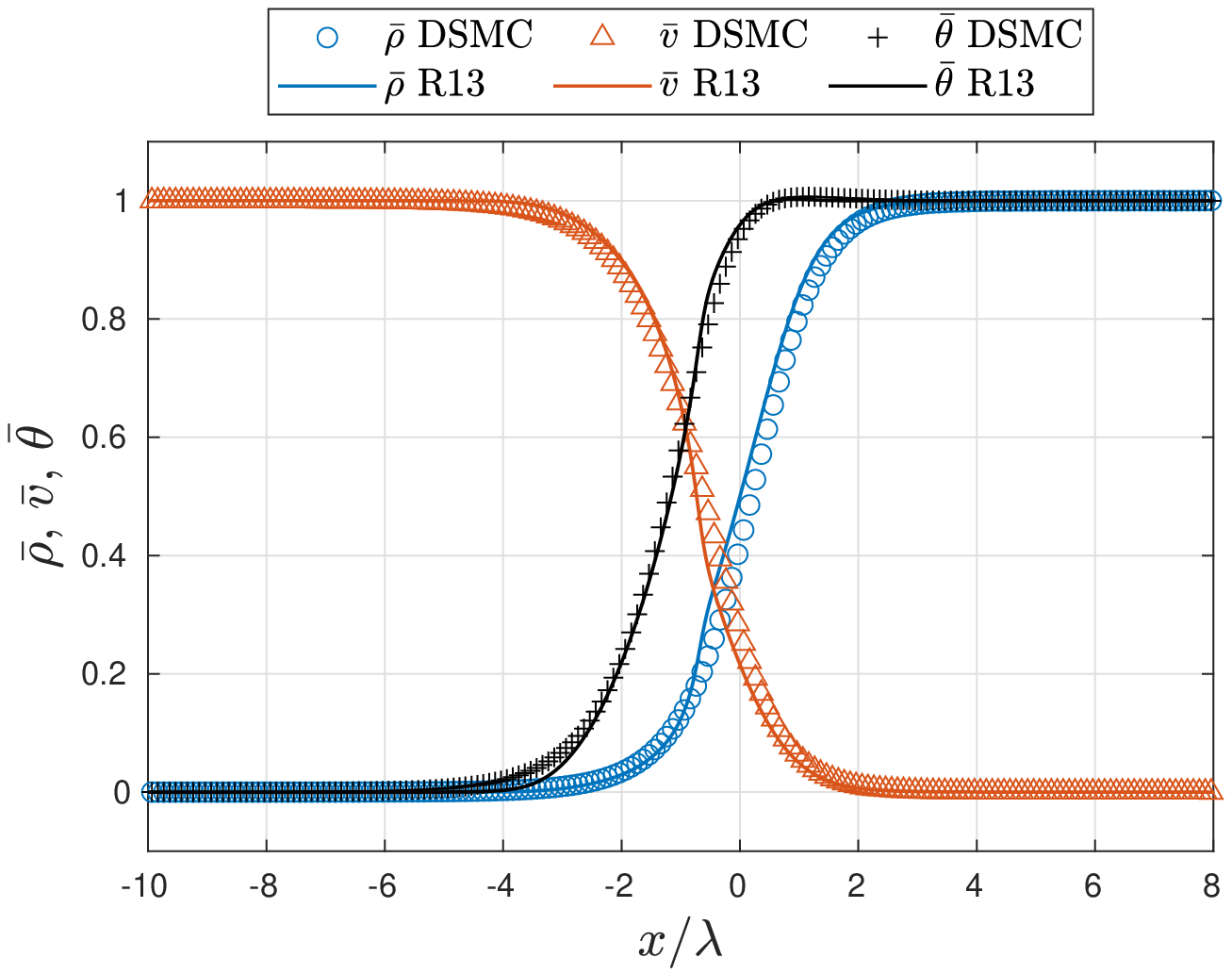}  
  \end{overpic}
  } \quad
  \subfigure[$\mMa=9.0$, $\bar{\rho}, \bar{v}, \bar{\theta}$]{%
  \begin{overpic}
  [width = .45\textwidth,clip]{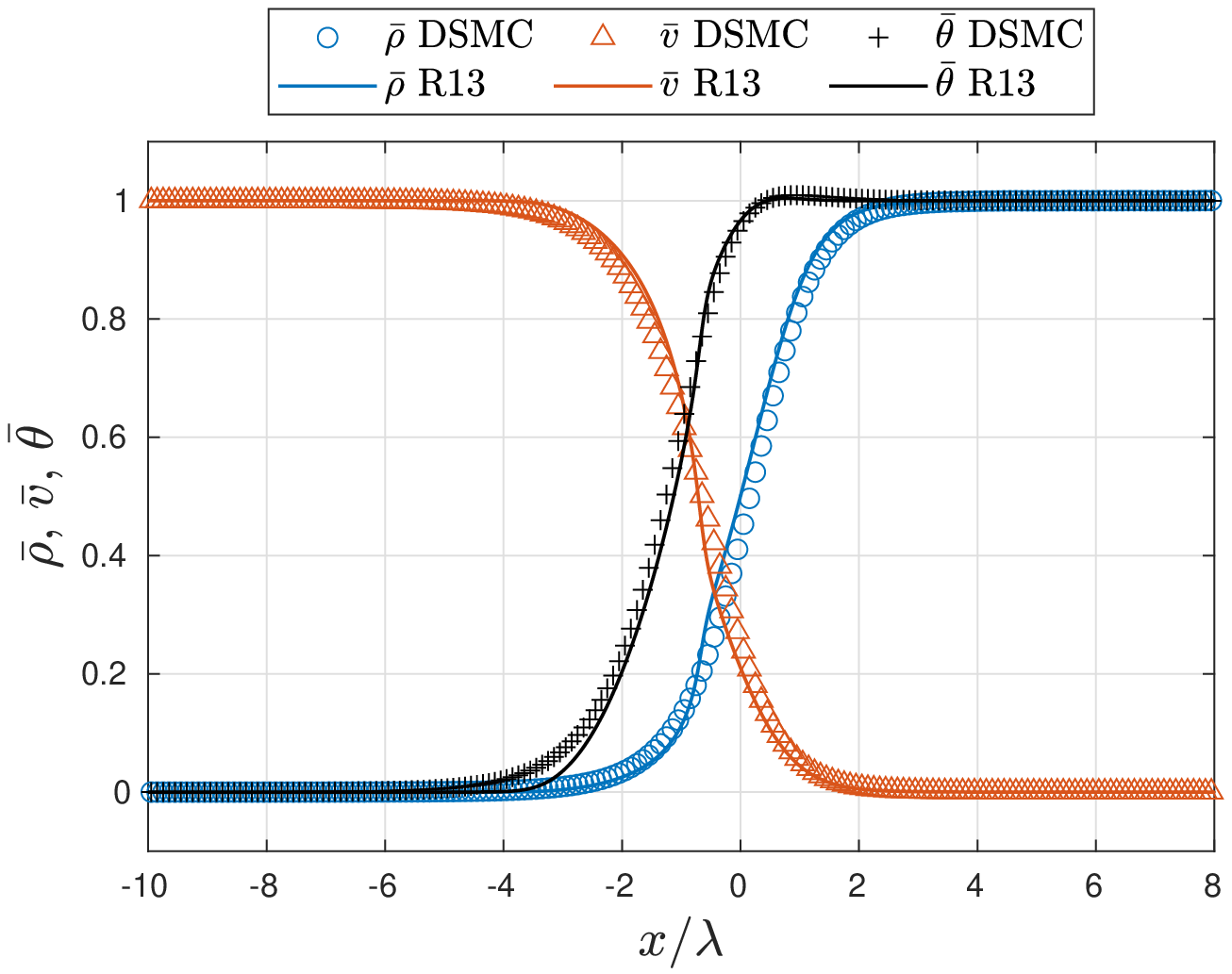}  
  \end{overpic}
  }
  \caption{Normalized density, velocity, and temperature of shock
    structures for the hard-sphere model and Mach numbers $\mMa =
    1.55, 3, 6.5, 9$. DSMC solutions for the variable hard sphere
    model are provided as reference results. The horizontal axis is
    $x/\lambda$ with $\lambda$ being the mean free path.}
\label{fig:hs_shock_rhoutheta}
\end{figure}

\begin{figure}[!ht]
\centering
\subfigure[$\mMa=1.55$, $\sigma, q$]{%
  \begin{overpic}[width = .5\textwidth, clip]{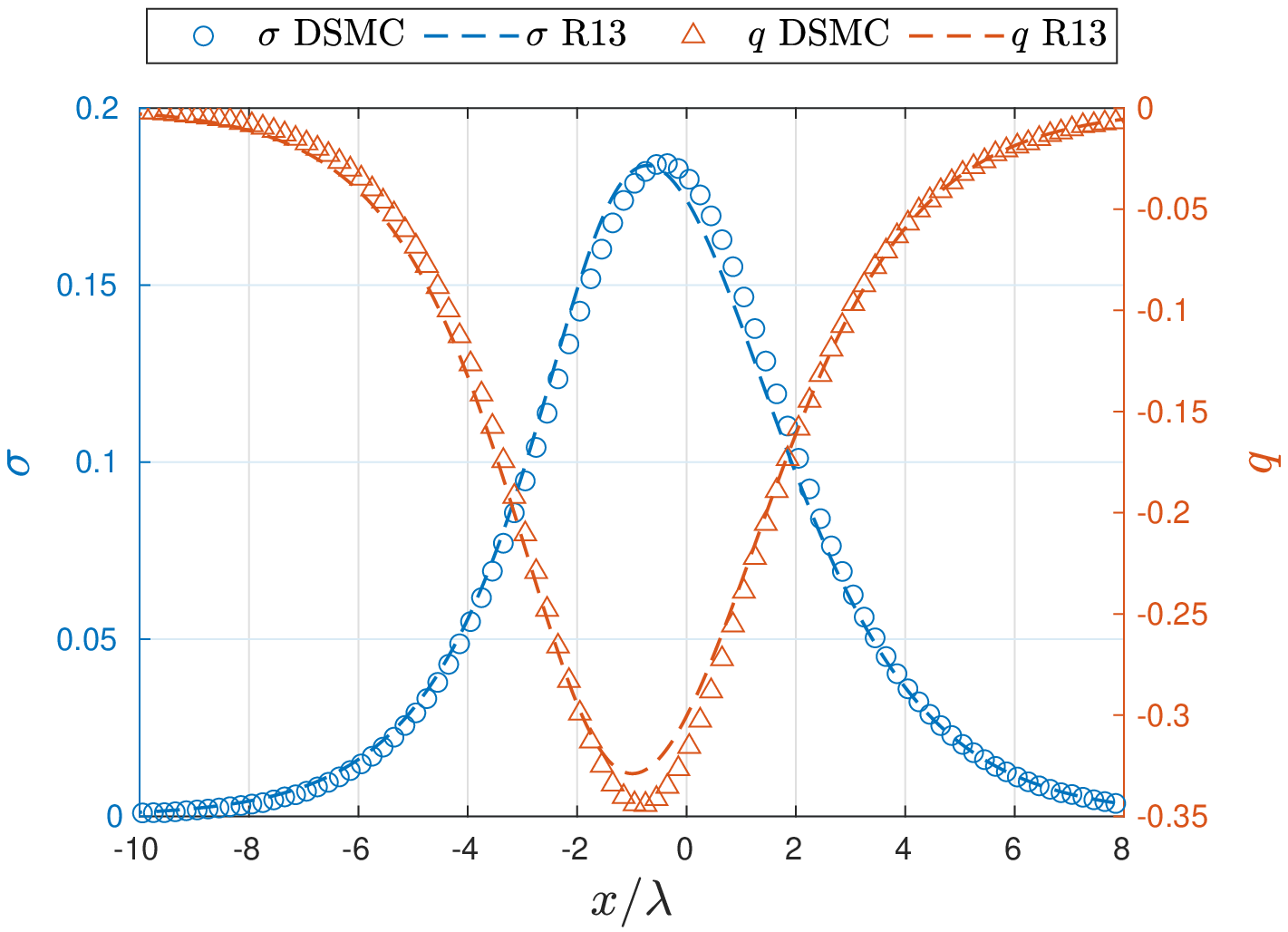}  
  \end{overpic}
  }%
\subfigure[$\mMa=3.0$, $\sigma, q$]{%
  \begin{overpic}
  [width = .5\textwidth,clip]{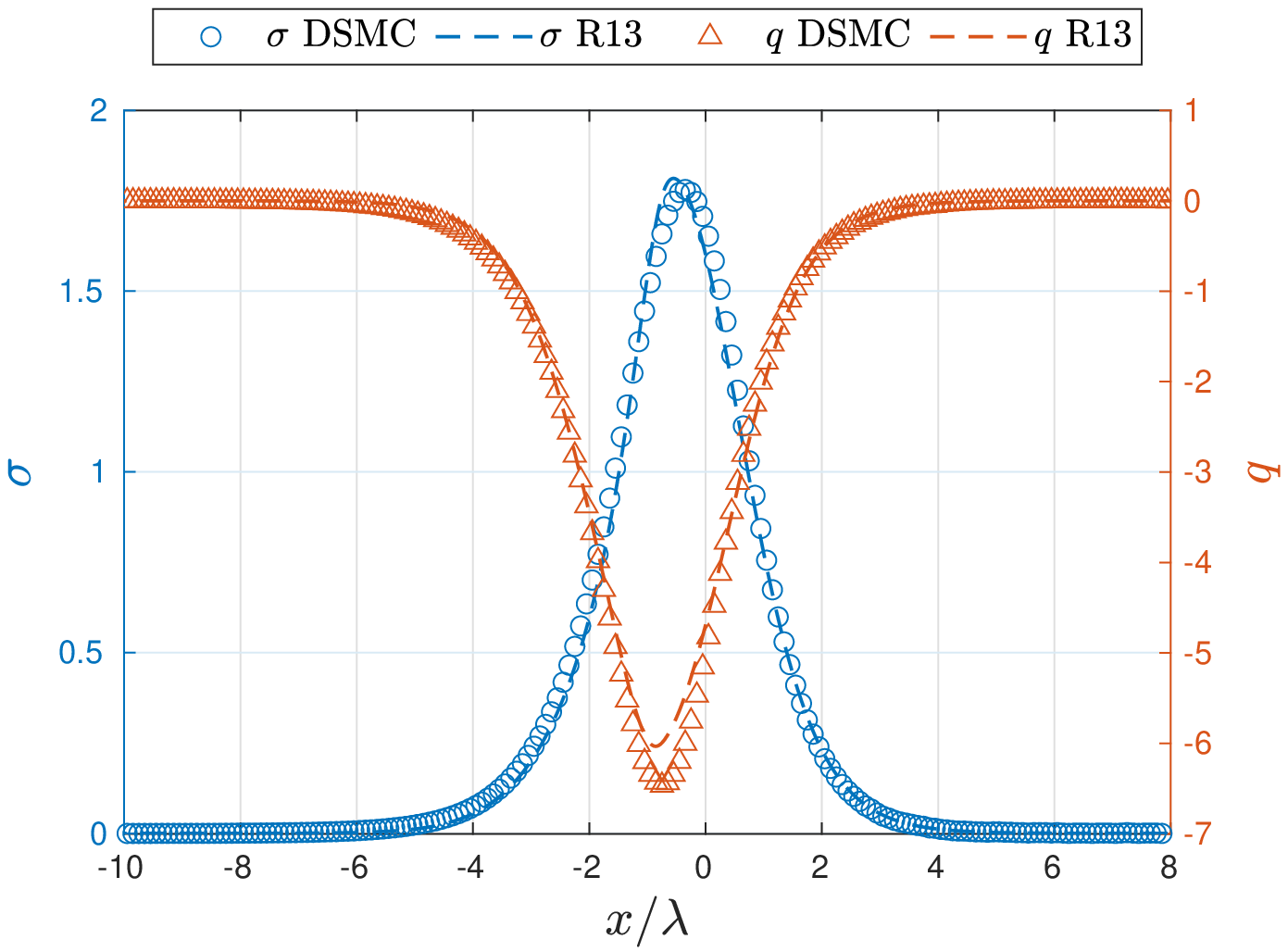}  
  \end{overpic}
  }\\
  \subfigure[$\mMa=6.5$, $\sigma, q$]{%
  \begin{overpic}
  [ width = .5\textwidth,clip]{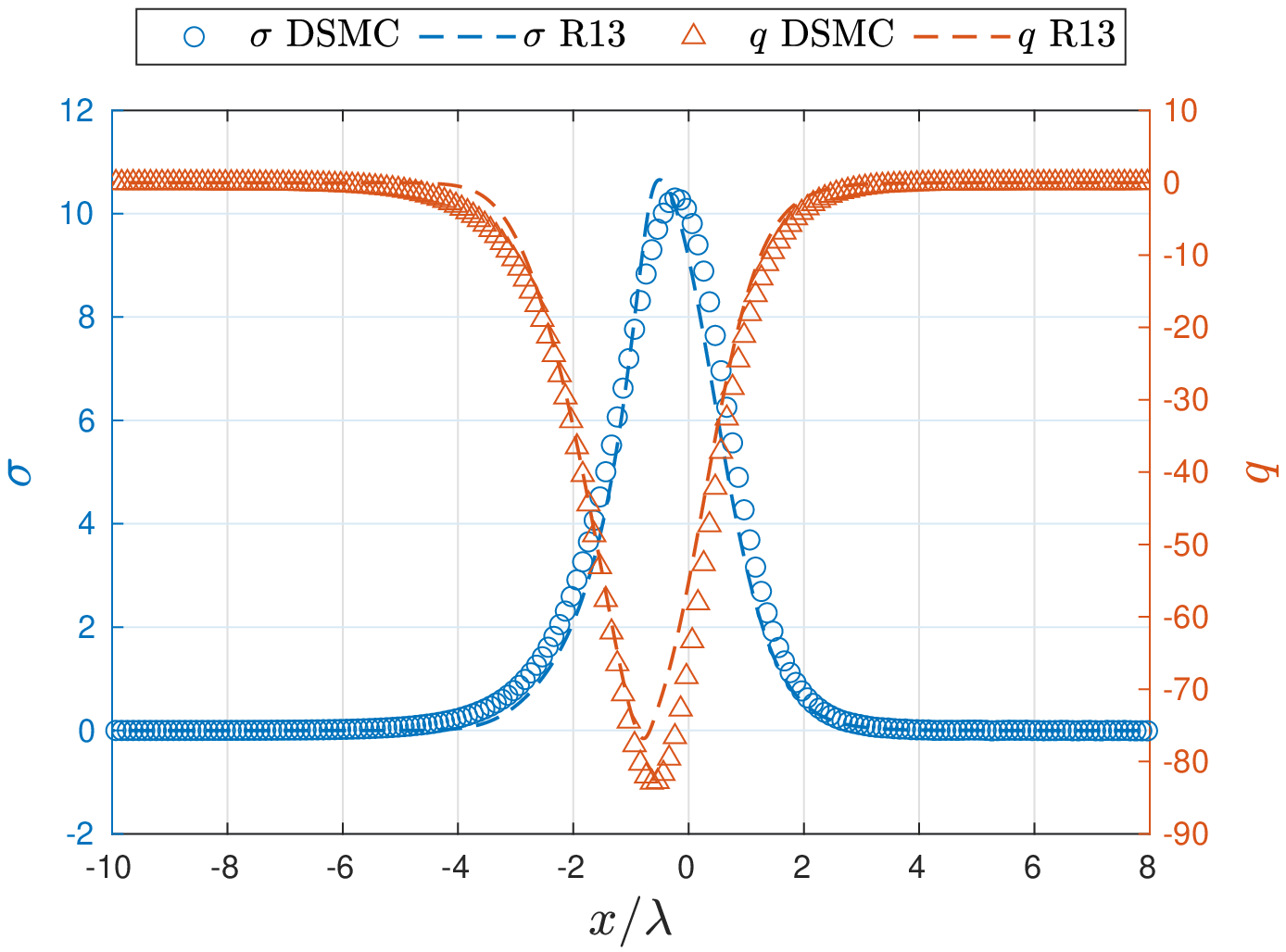}  
  \end{overpic}
  }%
\subfigure[$\mMa=9.0$, $\sigma, q$]{%
  \begin{overpic}
  [ width = .5\textwidth,clip]{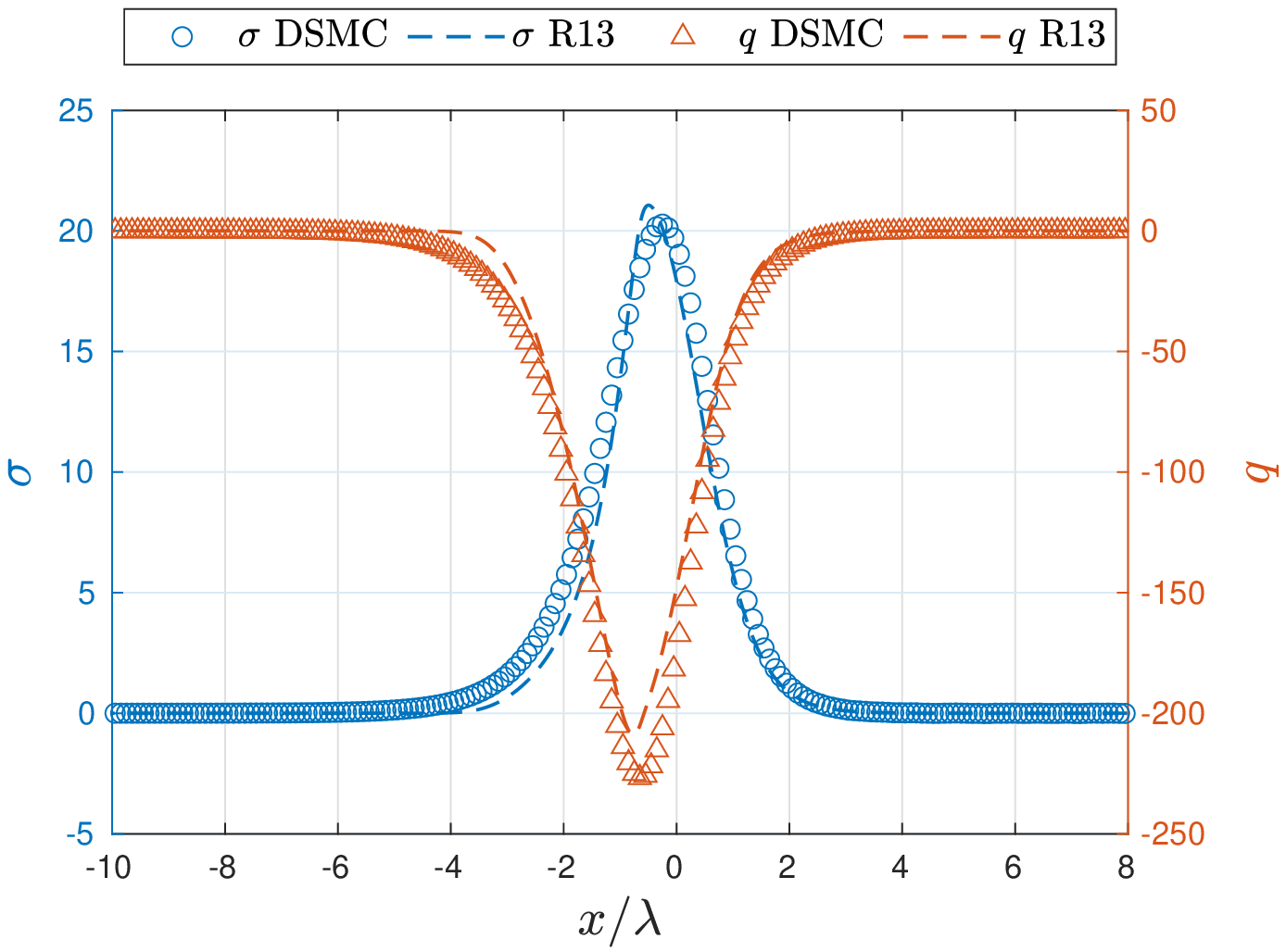}  
  \end{overpic}
  }
  \caption{The stress $\sigma$ and heat flux $q$ of shock structures
    for the hard-sphere model and Mach numbers $\mMa = 1.55, 3, 6.5,
    9$. DSMC results for the variable hard sphere model are provided
    as references. The horizontal axis is $x/\lambda$ with $\lambda$
    being the mean free path. The left $y$-axis corresponds to the
    stress and the right $y$-axis corresponds to the heat flux.}
\label{fig:hs_shock_sigmaq}
\end{figure}

Figure \ref{fig:hs_shock_rhoutheta} and \ref{fig:hs_shock_sigmaq} show
the shock structures for the same Mach numbers for the hard sphere
model.  Similarly, the normalized density $\bar{\rho}$, velocity
$\bar{v}$, and temperature $\bar{\theta}$ are plotted in Figure
\ref{fig:hs_shock_rhoutheta}. It is interesting that when the Mach
number increases from $3.0$ to $9.0$, there is no significant
decrement of the general quality of R13 approximation. In
\cite{Torrilhon2004}, the authors calculated shock structure for the
hard-sphere model using the R13 equations for Maxwell molecules with
its expression of viscosity changed to match the hard-sphere model. At
Mach number $3.0$, such a method already shows significant deviation
in the profile of heat flux. After switching to the ``true'' R13
equations for hard spheres, much better agreement can be
obtained. Note that the peak of the heat flux is \add{again}
underestimated in all results (including the model with $\eta = 10$).
In general, up to Mach number $9.0$, R13 results show
quite satisfactory agreement with the reference solutions for both
models.

\subsection{Shock structures for different indices $\eta$}
\begin{figure}[!ht]
\centering
\subfigure[$\eta= 7$, $\bar{\rho}, \bar{v}, \bar{\theta}$]{%
  \begin{overpic}
  [width = .45\textwidth,clip]{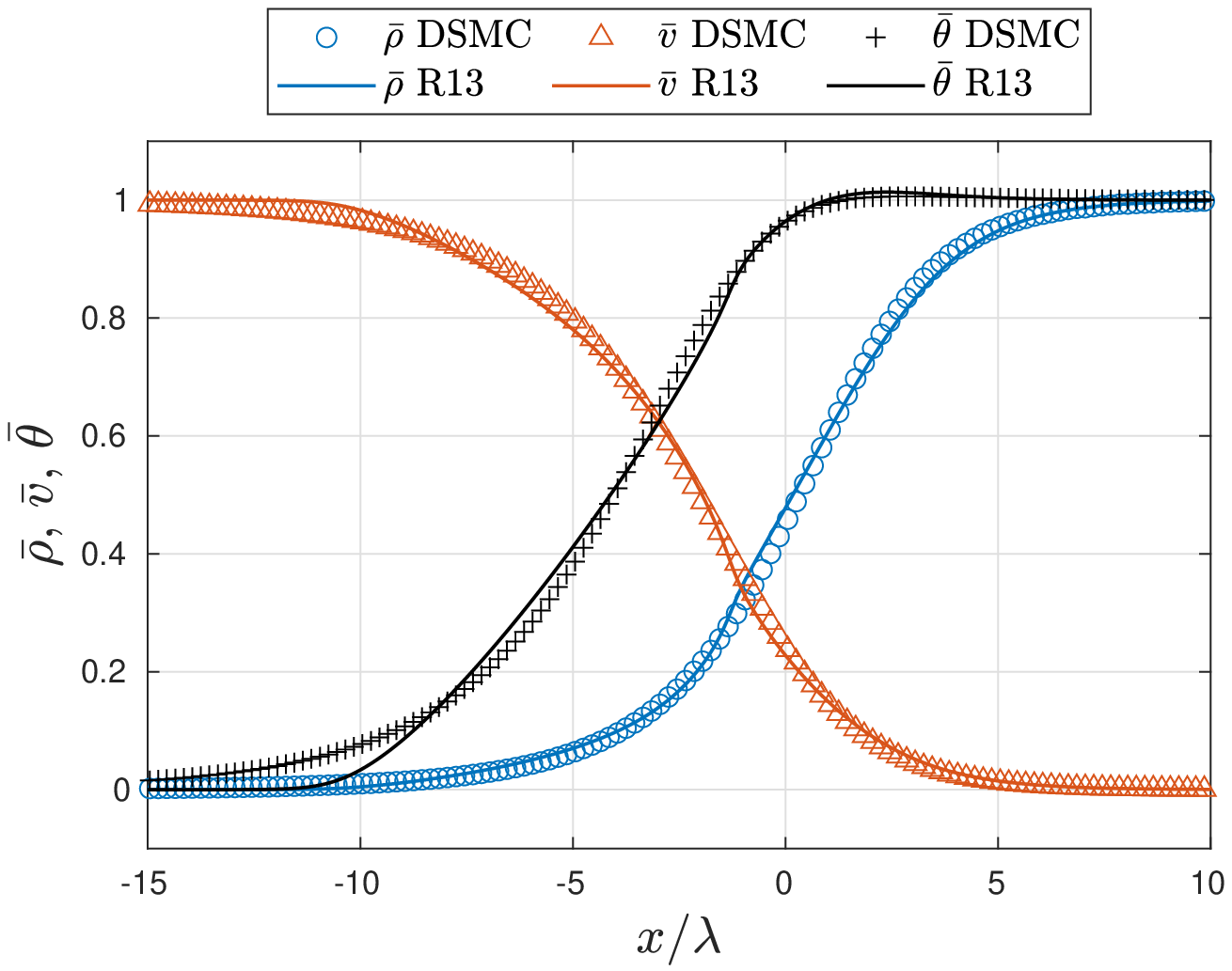}  
  \end{overpic}
  } \quad
  \subfigure[$\eta = 10$, $\bar{\rho}, \bar{v}, \bar{\theta}$]{%
  \begin{overpic}
  [width = .45\textwidth, clip]{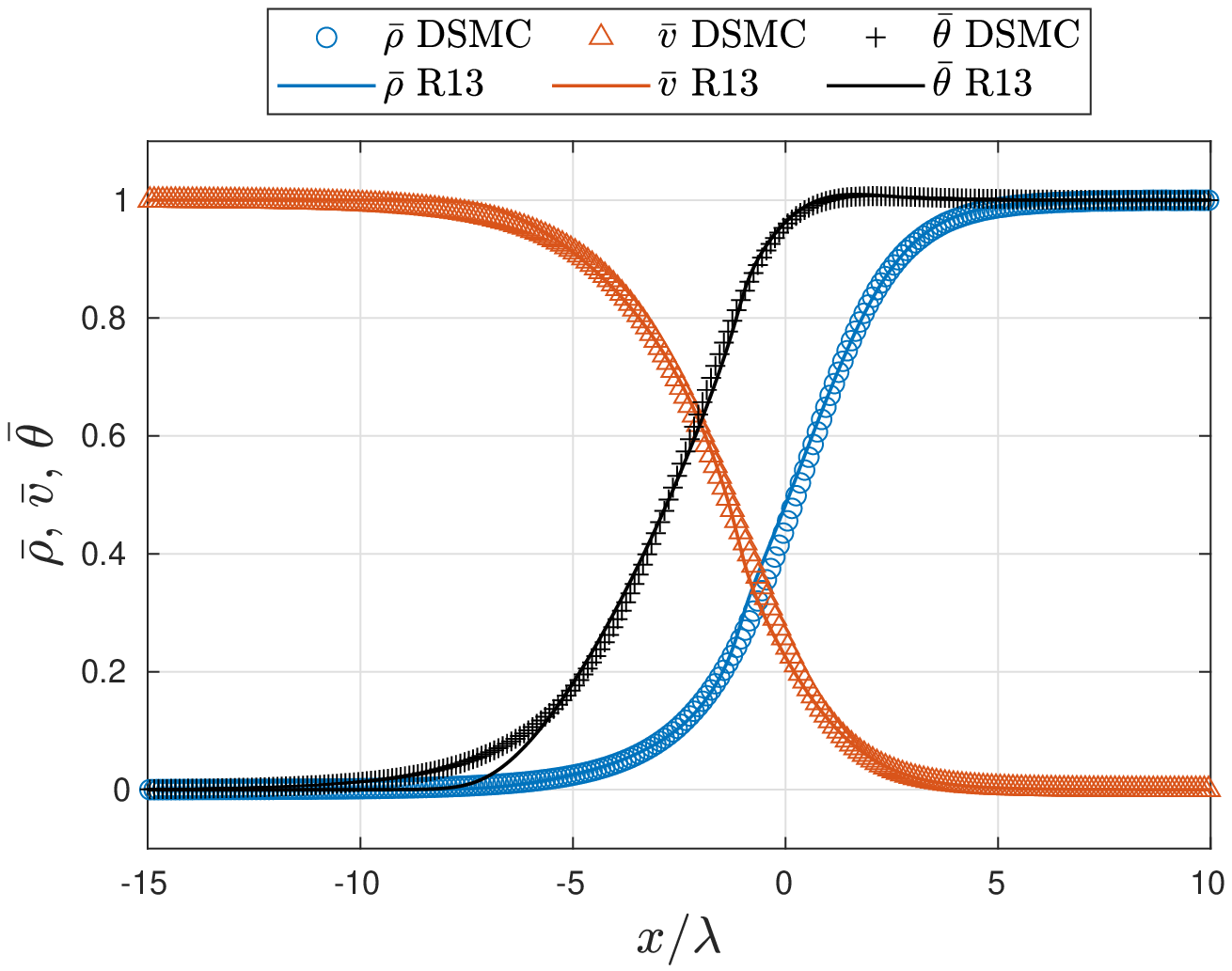}  
  \end{overpic}
  }\\
  \subfigure[$\eta=17$, $\bar{\rho}, \bar{v}, \bar{\theta}$]{%
  \begin{overpic}
  [width = .45\textwidth,clip]{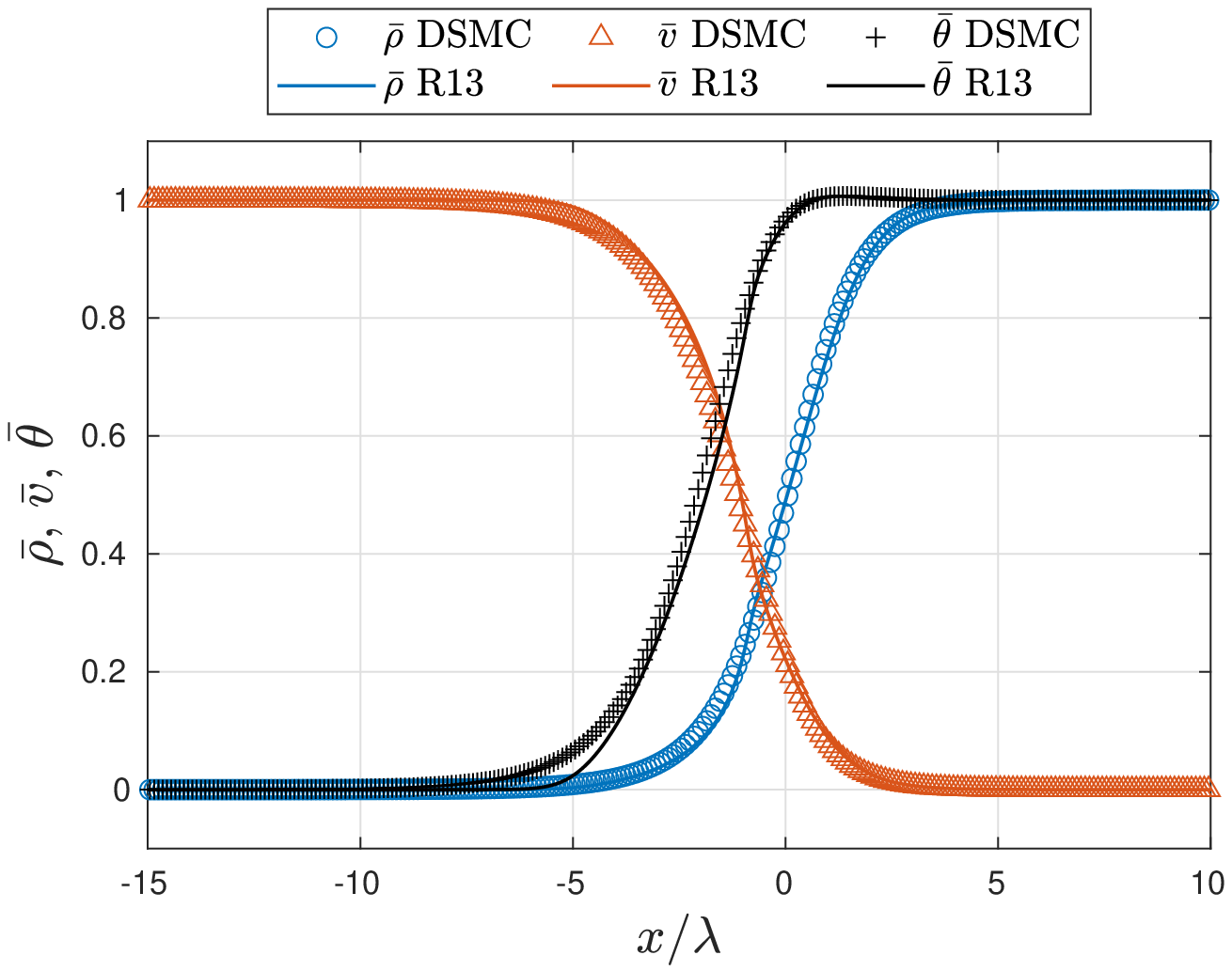}  
  \end{overpic}
  } \quad
  \subfigure[$\eta=\infty$, $\bar{\rho}, \bar{v}, \bar{\theta}$]{%
  \begin{overpic}
  [width = .45\textwidth, clip]{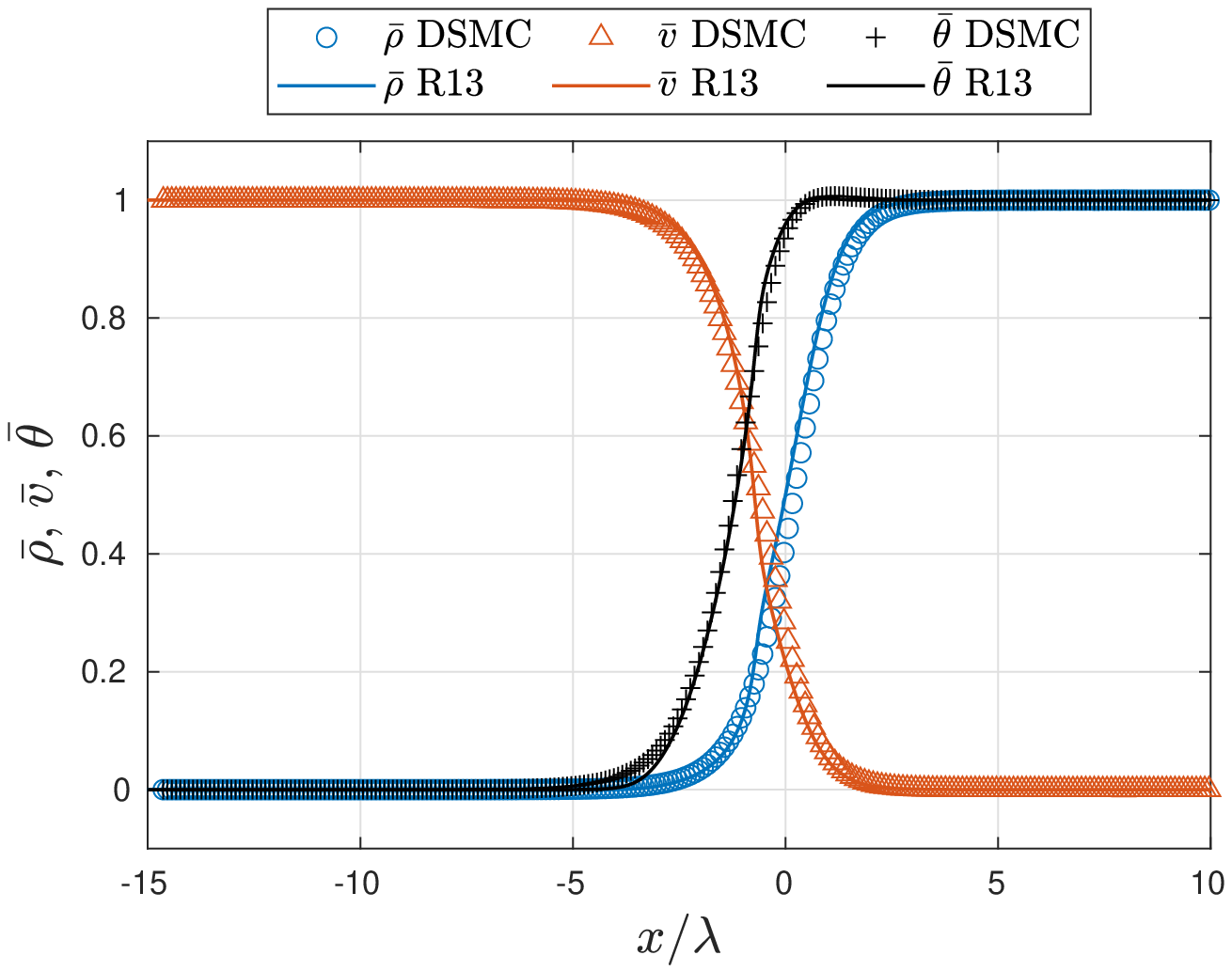}  
  \end{overpic}
  }
  \caption{Normalized density, velocity, and temperature of shock
    structures for IPL models with $\eta = 7, 10, 17, \infty$
    (hard-sphere) and Mach numbers $\mMa = 6.5$. DSMC solutions for
    the variable hard sphere models are provided as reference results.
    The horizontal axis is $x/\lambda$ with $\lambda$ being the mean
    free path.}
\label{fig:ma6.5_shock_rhoutheta}
\end{figure}

\begin{figure}[!ht]
\centering
\subfigure[$\eta = 7$, $\sigma, q$]{%
  \begin{overpic}
  [width = .45\textwidth, clip]{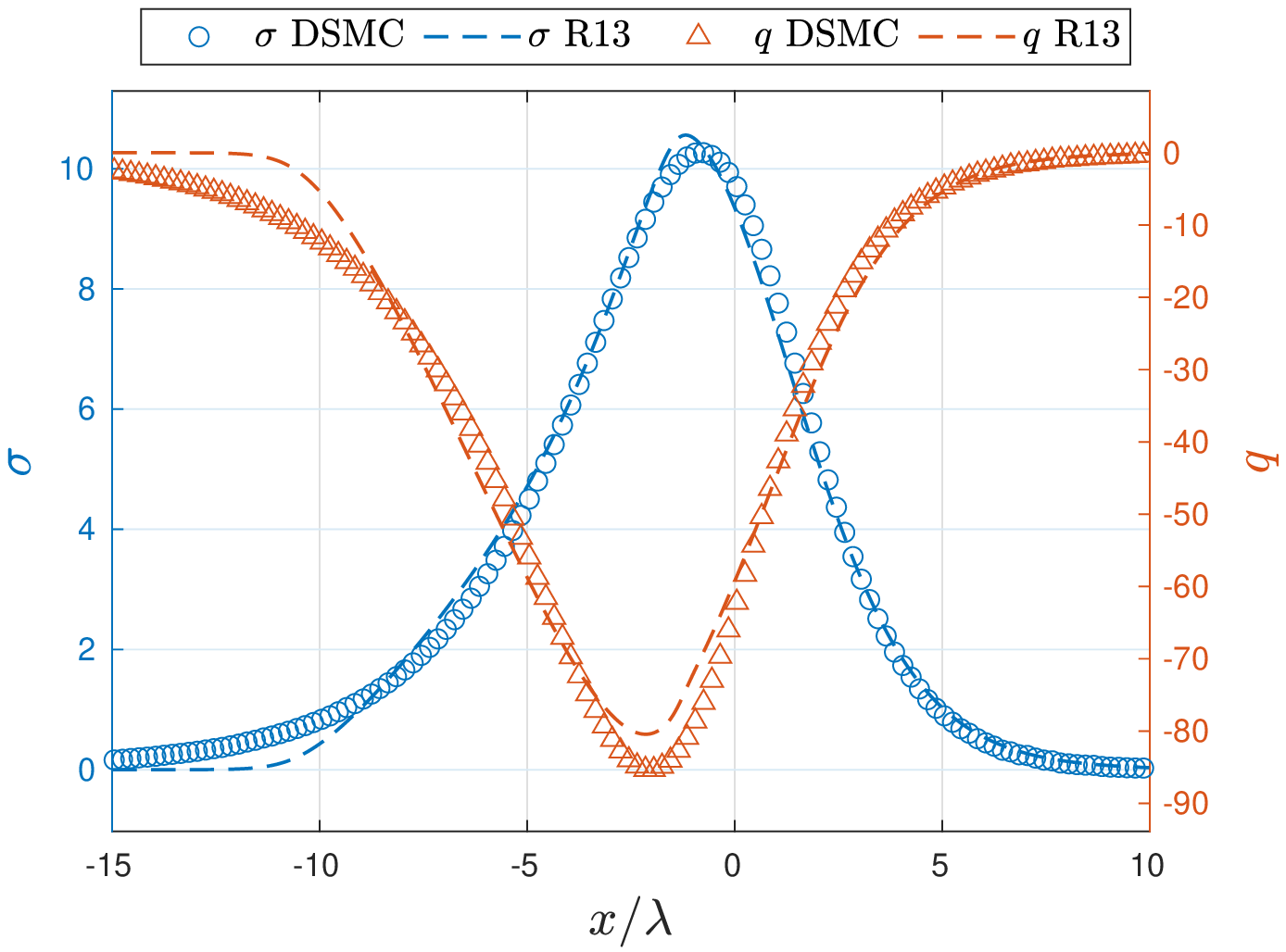}  
  \end{overpic}
  } \quad
\subfigure[$\eta = 10$, $\sigma, q$]{%
  \begin{overpic}
    [width = .45\textwidth, clip]{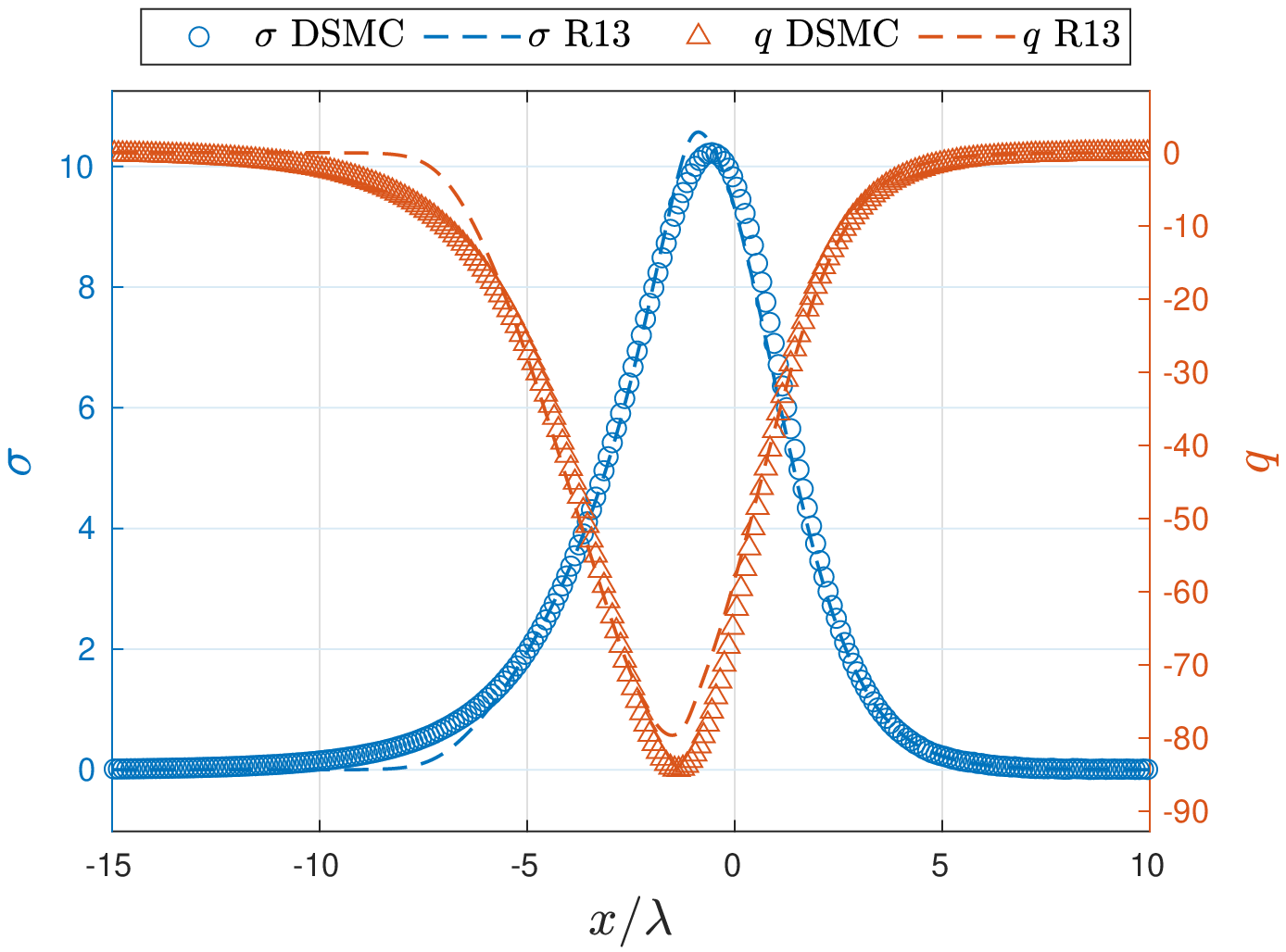}  
  \end{overpic}
  }\\
  \subfigure[$\eta = 17$, $\sigma, q$]{%
  \begin{overpic}
  [width = .45\textwidth, clip]{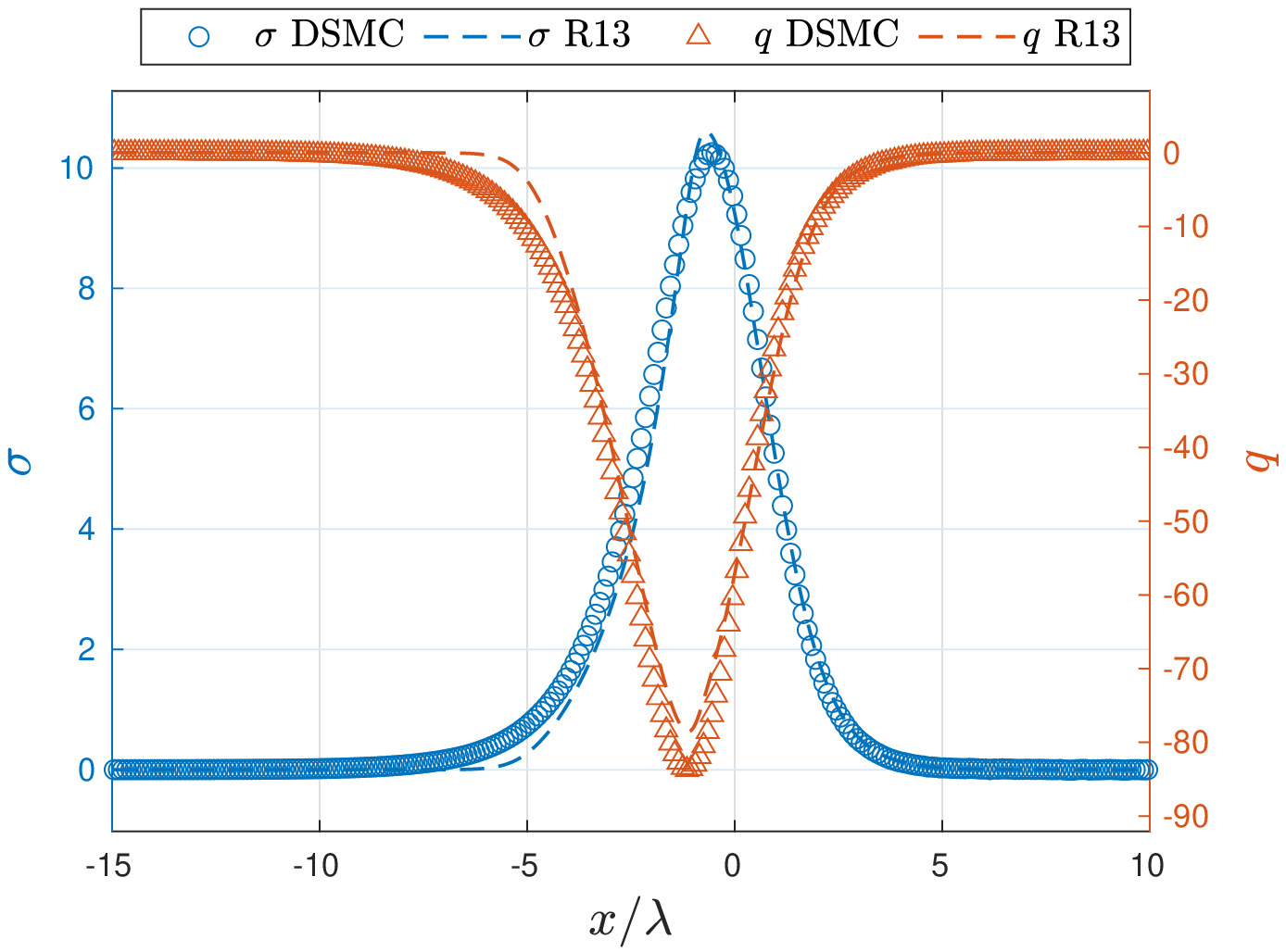}  
  \end{overpic}
  } \quad
\subfigure[$\eta = \infty$, $\sigma, q$]{%
  \begin{overpic}
  [width = .45\textwidth, clip]{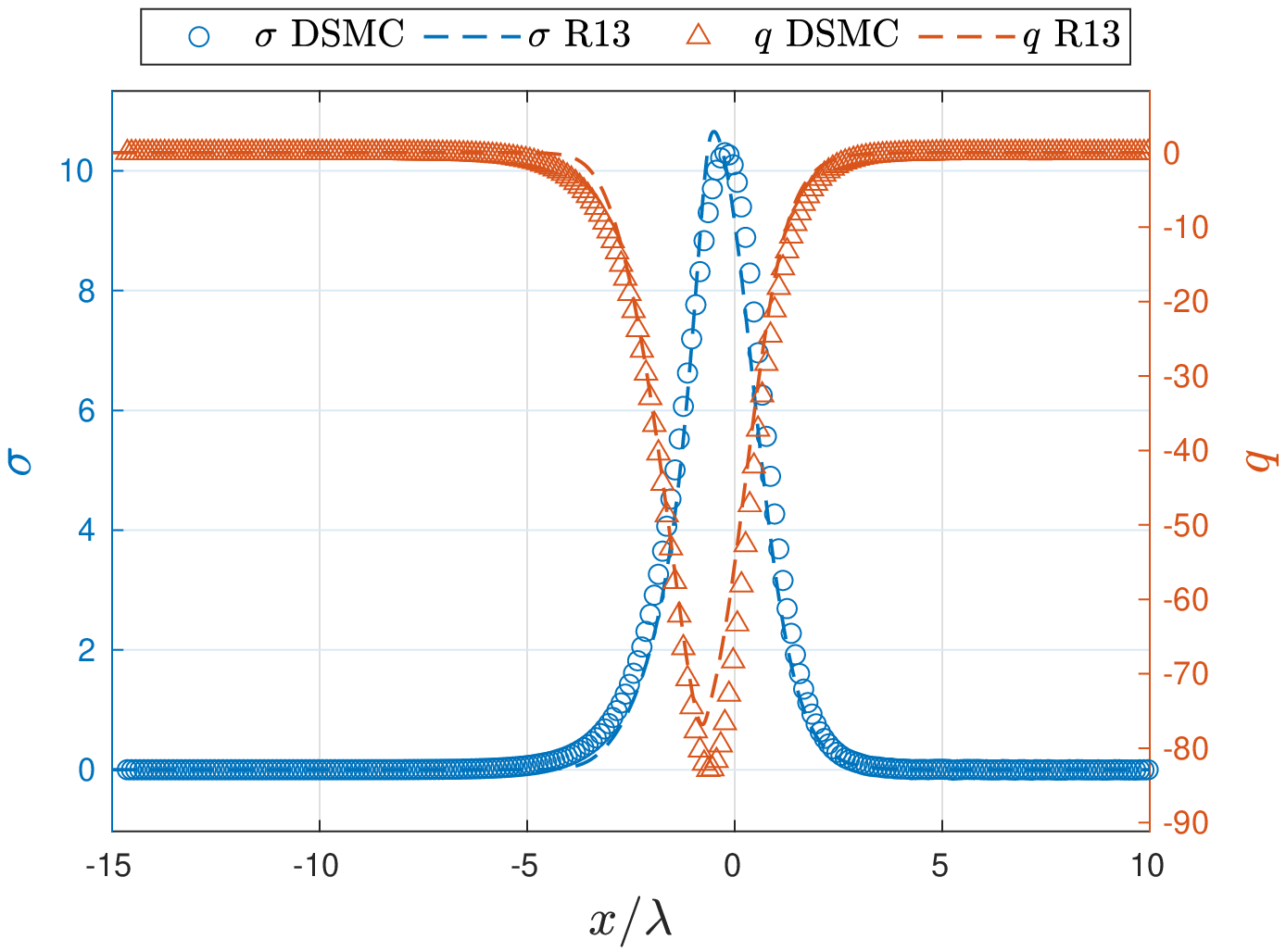}  
  \end{overpic}
  }
  \caption{The stress $\sigma$ and heat flux $q$ of shock structures
    for IPL models with $\eta = 7, 10, 17, \infty$ (hard-sphere) and
    Mach numbers $\mMa = 6.5$. DSMC results for the variable hard
    sphere models are provided as references. The horizontal axis is
    $x/\lambda$ with $\lambda$ being the mean free path. The left
    $y$-axis corresponds to the stress and the right $y$-axis
    corresponds to the heat flux.}
\label{fig:ma6.5_shock_sigmaq}
\end{figure}

Now we perform the tests by fixing the Mach number as $\mMa = 6.5$ and
changing the parameter $\eta$. Here we focus only on hard potentials
with $\eta = 7, 10, 17$ and $\infty$ (hard-sphere model). The results
for all the five quantities are plotted in Figure
\ref{fig:ma6.5_shock_rhoutheta} and \ref{fig:ma6.5_shock_sigmaq}.
Similarly, the normalized density $\bar{\rho}$, velocity $\bar{v}$,
and temperature $\bar{\theta}$ are plotted in Figure
\ref{fig:ma6.5_shock_rhoutheta}. Both R13 results and DSMC results
show that the shock structure differs for different collision models,
and it can be observed that better agreement between two results can
be achieved for larger $\eta$.  The possible reason is that larger
$\eta$ gives smaller viscosity index, which brings the distribution
function closer to the local Maxwellian.

Again, the most obvious deviation between R13 and DSMC results appears
in the plots of heat fluxes in the low density region. In general, the
distribution function inside a shock wave is similar to the
superposition of two Maxwellians: a narrow one coming from the front
of the shock wave and a wide one from the back of the shock wave
\cite{Valentini}. In the low density region, the portion of the wide
Maxwellian is quite small. However, when evaluating high-order
moments, the contribution of this small portion of wide Maxwellian
becomes obvious due to its slow decay at infinity. For the 13-moment
approximation, it can be expected that the contribution of the tail
may be underestimated, since the decay rate of the distribution
function in the Chapman-Enskog expansion is mainly set by the local
temperature, which is significantly faster than the wide Maxwellian in
the low density region.

As a summary, we observe that R13 models predicts reasonable shock
structures both qualitatively and quantitatively, although the
derivation of the models does not involve any special consideration
for this specific problem. This indicates the potential use of such a
model not only for the low Knudsen number case, but also for high
speed rarefied gas flows.

\section{Conclusion} \label{sec:conclusion}
In this work, we have derived the regularized 13-moment equations for
all inverse power law models and the hard-sphere model. This work can
be considered as a generalization of \cite{Struchtrup2003} to a much
more general class of gas molecules. It also generalizes the
methodology of \cite{Struchtrup2005}, which proposed the derivation of
GG13 equations for general collision models, to one more order of
accuracy. The derivation follows a systematic routine which can be in
principle applied to all collision models. In the numerical experiment
for shock structures, these new models show good agreement with the
kinetic model in strong nonequlibrium regimes. \add{To better
understand how R13 equations describe the distribution functions, one
may plot the distribution function predicted by R13 models inside the
shock structure, which is to be considered in the future work.}

A significant drawback of these models is the high complexity of
collision terms, which are given in Appendix \ref{sec:SQ_2}. This may
cause difficulties in both understanding the models and designing the
numerical methods. \add{One possible way to simplify the equations is
to linearize the regularization terms as in \cite{Torrilhon2006,
Cai2012n}, which still needs further justification.} We are currently
also working on the derivation of regularized 13-moment equations for
Boltzmann equation with binary collision terms.

\section*{Acknowledgements}
Zhenning Cai is supported by National University of Singapore Startup
Fund under Grant No. R-146-000-241-133. Yanli Wang is supported by the
National Natural Scientific Foundation of China (Grant No. 91630310
and U1930402).

\appendix
\section{Introduction to the Boltzmann equation and the linearized IPL model}
\label{sec:linearized_Bol}
As introduced in the beginning of Section \ref{sec:Moment}, both GG13
equations and R13 equations are derived from the kinetic equation,
which governs the distribution function $f(\bx,\bxi,t)$.  The relation
between the distribution function and the moments has been
demonstrated in equations \eqref{eq:basis_fun} and
\eqref{eq:macro_coe}. An equivalent but more straightforward way to 
write down the relationship is the follows:
\begin{equation}
  \label{eq:macro}
  \begin{aligned}
  \rho(\bx,t) &= \mm \int_{\bbR^3} f(\bx,\bxi,t) \dd \bxi, \\
  \bv(\bx,t) &=
    \frac{\mm}{\rho(\bx,t)}\int_{\bbR^3} \bxi f(\bx,\bxi,t) \dd \bxi, \\
  \theta(\bx,t) &= \frac{\mm}{3 \rho(\bx,t)}
    \int_{\bbR^3}|\bxi - \bv(\bx,t)|^2 f(\bx,\bxi,t) \dd \bxi, \\
  q_i(\bx,t) &= \frac{\mm}{2}\int_{\bbR^3}|\bxi -\bv(\bx,t)|^2(\xi_i-v_i(\bx,t))
    f(\bx,\bxi,t) \dd \bxi, \qquad i = 1, 2, 3, \\
  \sigma_{ij}(\bx,t) &= \mm \int_{\bbR^3}
    \left( (\xi_i-v_i(\bx,t))(\xi_j -v_j(\bx,t)) -
      \frac{1}{3} \delta_{ij} |\bxi-\bv(\bx,t)|^2 \right)
      f(\bx,\bxi,t) \dd \bxi, \quad i,j = 1,2,3, 
  \end{aligned}
\end{equation}
where $\mm$ is the mass of a single molecule. For monatomic gases,
the governing equation of the distribution function is the Boltzmann
equation, which reads
\begin{equation}
  \label{eq:Bol}
  \pd{f}{t} + \bxi \cdot \nabla_{\bx} f = C[f], 
\end{equation}
where $C(f)$ is the collision term. Here we only focus on the
linearized collision term, whose expression is \cite{Harris}
\begin{equation}
  \label{eq:linear}
  \begin{split}
  C[f](\bx,\bxi,t) &= \int_{\bbR^3}\int_{\bn \perp \bg}
    B(\bg,\chi) \mM(\bx,\bxi,t) \mM(\bx,\bxi_1,t) \times {}\\
  & \qquad \left[
    \frac{f(\bx,\bxi_1',t)}{\mM(\bx,\bxi_1',t)} +
    \frac{f(\bx,\bxi',t)}{\mM(\bx,\bxi',t)} -
    \frac{f(\bx,\bxi_1,t)}{\mM(\bx,\bxi_1,t)} -
    \frac{f(\bx,\bxi,t)}{\mM(\bx,\bxi,t)} \right]
  \dd \chi \dd \bn \dd \bxi_1,
  \end{split}
\end{equation}
where $\mM(\bx,\bxi,t)$ is the local Maxwellian:
\begin{equation}
  \label{eq:Maxwellian}
  \mM(\bx,\bxi,t) = \frac{\rho(\bx,t)}{\mm (2\pi \theta(\bx,t))^{3/2}}
    \exp\left(-\frac{|\bxi-\bv(\bx,t)|^2}{2\theta(\bx,t)}\right),
\end{equation}
which satisfies $C[\mM] = 0$. In \eqref{eq:linear}, $\bg = \bxi -
\bxi_1$ and $\bn$ is a unit vector. The post-collisional velocities
$\bxi'$ and $\bxi_1'$ are
\begin{equation} \label{eq:post_vel}
\begin{aligned}
\bxi' &= \cos^2(\chi/2) \bxi + \sin^2(\chi/2) \bxi_1
  - |\bg| \cos(\chi/2) \sin(\chi/2) \bn, \\
\bxi_1' &= \cos^2(\chi/2) \bxi_1 + \sin^2(\chi/2) \bxi
  + |\bg| \cos(\chi/2) \sin(\chi/2) \bn.
\end{aligned}
\end{equation}

By now, the only unexplained term in the collision term
\eqref{eq:linear} is the collision kernel $B(|\bg|,\chi)$, which is 
a non-negative function determined by the force between gas molecules.
For the IPL model, it has the form \cite{Bird}
\begin{equation}
  \label{eq:IPL}
  B(|\bg|, \chi) =
    \left( \frac{2\kappa}{\mm} \right)^{\frac{2}{\eta-1}}
    |\bg|^{\frac{\eta - 5}{\eta-1}} W_0
    \left| \frac{\mathrm{d} W_0}{\mathrm{d} \chi} \right|,
\end{equation}
where $\eta$ is the same as the parameter used throughout this paper.
The dimensionless impact parameter $W_0$ is related to the angle
$\chi$ by the following two equations:
\begin{displaymath} 
  \chi = \pi - 2 \int_0^{W_1} \left[ 1 -
    W^2 - \frac{2}{\eta - 1} \left( \frac{W}{W_0} \right)^{\eta-1}
  \right]^{-1/2} \dd W, \quad 1-W_1^2 - \frac{2}{\eta - 1} \left(
    \frac{W_1}{W_0} \right)^{\eta-1} = 0.
\end{displaymath}
We also refer the readers to \cite{Struchtrup} for more information on
the Boltzmann equation and the collision models.

\section{Basis functions and moment equations}
\label{sec:burnett}
In this appendix, we are going to explain the basis functions used in
the expansion \eqref{eq:basis_fun} and the terms in the moment
equations \eqref{eq:moment_eq}. Here we follow \cite{Cai2015} to
define the basis function $\psi_{lmn}(\bx,\bxi,t)$ as
\begin{equation}
  \psi_{lmn}(\bx,\bxi,t) = [\theta(\bx,t)]^{-\frac{2n+l}{2}}
    p_{lmn} \left( \frac{\bxi - \bv(\bx,t)}{\sqrt{\theta(\bx,t)}} \right)
    \cdot [2\pi \theta(\bx,t)]^{-3/2} \exp \left(
      -\frac{|\bxi - \bv(\bx,t)|^2}{2\theta(\bx,t)}
    \right),
\end{equation}
where $p_{lmn}(\cdot)$ is an orthogonal polynomial in $\mathbb{R}^3$:
\begin{displaymath}
    p_{lmn}(\bxi) = \sqrt{\frac{2^{1-l} \pi^{3/2} n!}{\Gamma(n+l+3/2)}}
  L_n^{(l+1/2)} \left( \frac{|\bxi|^2}{2} \right) |\bxi|^l
  Y_l^m \left( \frac{\bxi}{|\bxi|} \right), \qquad
l,n \in \bbN, \quad m = -l,\cdots,l,
\end{displaymath}
where we have used Laguerre polynomials
\begin{equation} \label{eq:Laguerre}
L_n^{(\alpha)}(x) = \frac{x^{-\alpha} \exp(x)}{n!}
  \frac{\mathrm{d}^n}{\mathrm{d}x^n}
  \left[ x^{n+\alpha} \exp(-x) \right],
\end{equation}
and spherical harmonics
\begin{displaymath}
Y_l^m(\bn) = \sqrt{\frac{2l+1}{4\pi} \frac{(l-m)!}{(l+m)!}}
  P_l^m(\cos \theta) \exp(\mathrm{i} m \phi), \qquad
\bn = (\sin \theta \cos \phi, \sin \theta \sin \phi, \cos \theta)^T
\end{displaymath}
with $P_l^m$ being the associate Legendre polynomial:
\begin{displaymath}
P_l^m(x) = \frac{(-1)^m}{2^l l!} (1-x^2)^{m/2}
  \frac{\mathrm{d}^{l+m}}{\mathrm{d}x^{l+m}} \left[ (x^2-1)^l \right].
\end{displaymath}
The orthogonality of the polynomials $p_{lmn}$ is
\begin{displaymath}
\frac{1}{(2\pi)^{3/2}}
  \int_{\bbR^3} \overline{p_{l_1 m_1 n_1}(\bxi)} p_{l_2 m_2 n_2}(\bxi)
  \exp\left(-\frac{|\bxi|^2}{2}\right) \dd\bxi =
    \delta_{l_1 l_2} \delta_{m_1 m_2} \delta_{n_1 n_2}.
\end{displaymath}

Next, we will show the expressions of $S_{lmn}$ and $T_{lmn}$ in
\eqref{eq:moment_eq}, which have been obtained in \cite{Cai2018N}.
For simplicity, we introduce the following velocities:
\begin{equation}
  \label{eq:media_var}
  V_{-1} = \frac{1}{2}(v_1 - \imag v_2), \quad V_0 = v_3, \quad
  V_1 = -\frac{1}{2}(v_1 + \imag v_2).
\end{equation}
Then we have
\begin{equation}
  \label{eq:Slmn}
  \begin{aligned}
  S_{lmn}& = - \sqrt{n(n+l+1/2)}\pd{\theta}{t}f_{l,m,n-1} \\
  & +\sqrt{2} \sum_{\mu = -1}^{1}\pd{V_{\mu}}{t}\left[
    (-1)^{\mu}\sqrt{n + l + 1/2}\gamma_{l,
      m+\mu}^{-\mu}f_{l-1,m+\mu,n} - \sqrt{n}\gamma_{-l-1,
      m+\mu}^{-\mu}f_{l+1,m+\mu,n-1} \right],
  \end{aligned}
\end{equation}
where $\gamma_{lm}^{\mu}$ are constants defined by
\begin{equation}
  \label{eq:gamma}
  \gamma_{lm}^{\mu} = \sqrt{\frac{[l + (2\delta_{1,\mu} - 1)m +
    \delta_{1,\mu}][l - (2\delta_{-1,\mu} -1)m +
    \delta_{-1,\mu}]}{(2l-1)(2l+1)}}. 
\end{equation}
To introduce $T_{lmn}$, we first define the following operators:
\begin{equation}
\frac{\partial}{\partial X_{-1}} =
  \frac{\partial}{\partial x_1} + \imag \frac{\partial}{\partial x_2},
\qquad \frac{\partial}{\partial X_0} = \frac{\partial}{\partial x_3},
\qquad \frac{\partial}{\partial X_1} =
  -\frac{\partial}{\partial x_1} + \imag \frac{\partial}{\partial x_2},
\end{equation}
using which we can write down $T_{lmn}$ as
\begin{equation}
  \label{eq:cove_coe}
  \begin{aligned}
    & T_{lmn} = \sum_{\mu=-1}^{1} \bigg(V_{\mu}F_{lmn\mu} +
      \frac{1}{2^{|\mu|}} \Big[ \sqrt{2(n+l) + 1}
        \gamma_{l,m-\mu}^{\mu} \theta F_{l-1,m-\mu,n,\mu}  
        -\sqrt{2(n+1)} \gamma_{l,m-\mu}^{\mu} F_{l-1,m-\mu,n+1,\mu} \\
    & \qquad + (-1)^{\mu}\gamma_{-l-1,m-\mu}^{\mu}
      (\sqrt{2(n+l)+3}F_{l+1,m-\mu,n,\mu} -\sqrt{2n}\theta
      F_{l+1,m-\mu,n-1,\mu}) \Big] \bigg),
  \end{aligned}
\end{equation}
where
\begin{equation}
  \begin{split}
  F_{lmn\mu} & = \pd{f_{lmn}}{X_{\mu}}
    - \sqrt{n(n+l+1/2)}\pd{\theta}{X_{\mu}}f_{l,m,n-1} \\
  & + \sqrt{2} \sum_{\nu = -1}^{1}\pd{V_{\nu}}{X_{\mu}}\left[
    (-1)^{\nu}\sqrt{n + l + 1/2}\gamma_{l,m+\nu}^{-\nu}f_{l-1,m+\nu,n}
    - \sqrt{n}\gamma_{-l-1,m+\nu}^{-\nu}f_{l+1,m+\nu,n-1} \right].
  \end{split}
\end{equation}

\section{R13 collision terms}
\label{sec:SQ_2}
In this section, we provide the explicit forms of $\Sigma_{ij}^{(\eta,
2)}$ and $Q_i^{(\eta, 2)}$ and tabulate some values of the
coefficients for some choices of $\eta$. In the expressions
\eqref{eq:S_2} and \eqref{eq:Q_2}, $\Xi_k^{(\eta)}$ and $\Lambda_k^{(\eta)}$
are constants depending only on $\eta$, whose values are given in
Table \ref{tab:S2} and \ref{tab:Q2}. These tables are to be read
horizontally. For example, in Table \ref{tab:S2}, the row below
``$\eta = 10$'' gives the values of $\Xi_0^{(10)}, \Xi_1^{(10)}, \cdots,
\Xi_{9}^{(10)}$, and the next row gives the values of $\Xi_{10}^{(10)},
\Xi_{11}^{(10)}, \cdots, \Xi_{19}^{(10)}$.

{\small
\begin{equation} \label{eq:S_2}
\begin{split}
  \Sigma_{ij}^{(\eta,2)} &= \Xi_{0}^{(\eta)} \frac{\rho \theta}{\mu}
  \sigma_{ij} + \Xi_1^{(\eta)} \left( \pd{v_{\li}}{x_k} \sigma_{\jl k} +
    \pd{v_k}{x_{\li}} \sigma_{\jl k} \right) + \Xi_2^{(\eta)}
  \pd{v_k}{x_k} \sigma_{ij} + \Xi_3^{(\eta)}\frac{\mu}{\rho }
  \pd{\ln\theta}{x_k} \pd{\ln\theta}{x_{\li}}
  \sigma_{\jl k} \\
  & + \Xi_4^{(\eta)} \pd{\ln\theta}{x_k} \pd{\ln\theta}{x_k} \sigma_{ij}
  + \Xi_5^{(\eta)}\frac{\mu}{\rho } \pd{\ln\rho}{x_k}
  \pd{\ln\theta}{x_{\li}} \sigma_{\jl k} +
  \Xi_6^{(\eta)}\pd{\ln\rho}{x_k} \pd{\ln\theta}{x_k} \sigma_{ij} +
  \Xi_7^{(\eta)}\frac{\mu}{\rho }
  \pd{\ln\theta}{x_k} \pd{\ln\rho}{x_{\li}} \sigma_{\jl k} \\
  & + \Xi_8^{(\eta)}\frac{\mu}{\rho } \pd{\ln\rho}{x_k}
  \pd{\ln\rho}{x_{\li}} \sigma_{\jl k} + \Xi_9^{(\eta)}\pd{\ln\rho}{x_k}
  \pd{\ln\rho}{x_k} \sigma_{ij} + \Xi_{10}^{(\eta)} \frac{\mu}{\rho
    \theta}\pd{v_k}{x_l} \pd{v_k}{x_{\li}}\sigma_{\jl l} +
  \Xi_{11}^{(\eta)} \frac{\mu}{\rho \theta}\pd{v_k}{x_l}
  \pd{v_k}{x_l}\sigma_{ij}  \\
  & + \Xi_{12}^{(\eta)} \frac{\mu}{\rho \theta}\pd{v_k}{x_l}
  \pd{v_l}{x_k} \sigma_{ij} + \Xi_{13}^{(\eta)} \frac{\mu}{\rho
    \theta}\pd{v_k}{x_k} \pd{v_l}{x_l} \sigma_{ij} +
  \Xi_{14}^{(\eta)}\frac{\mu}{\rho \theta} \left(\pd{v_k}{x_l}
    \pd{v_{\li}}{x_k} + \pd{v_l}{x_k}
    \pd{v_k}{x_{\li}}\right)\sigma_{\jl l} \\
  & + \Xi_{15}^{(\eta)}\frac{\mu}{\rho \theta}\pd{v_k}{x_k}
  \left(\pd{v_{\li}}{x_l} + \pd{v_l}{x_{\li}} \right)\sigma_{\jl l} +
  \Xi_{16}^{(\eta)}\frac{\mu}{\rho \theta}\left(\pd{v_k}{x_{\li}}
    \pd{v_l}{x_{\jl}} + \pd{v_{\li}}{x_k}
    \pd{v_{\jl}}{x_{l}}\right)\sigma_{kl} \\
  & + \Xi_{17}^{(\eta)} \frac{\mu}{\rho \theta}\pd{v_l}{x_{\li}}
  \pd{v_{\jl}}{x_k} \sigma_{kl} + \Xi_{18}^{(\eta)}
  \pd{v_l}{x_k}\pd{v_{\li}}{x_k}\sigma_{\jl l} + \Xi_{19}^{(\eta)}
  \frac{\mu}{\rho \theta}\frac{\partial^2 \theta}{\partial
    x_k \partial x_k} \sigma_{ij} + \Xi_{20}^{(\eta)} \frac{\mu}{\rho
    \theta}\frac{\partial^2 \theta}{\partial x_k \partial x_{\li}}
  \sigma_{\jl k} \\
  & + \Xi_{21}^{(\eta)}\frac{\mu}{\rho}
  \pd{\sigma_{ij}}{x_k}\pd{\ln\theta}{x_k} + \Xi_{22}^{(\eta)}
  \frac{\mu}{\rho}\pd{\sigma_{k\li}}{x_k} \pd{\ln\theta}{x_{\jl}} +
  \Xi_{23}^{(\eta)} \frac{\mu}{\rho} \pd{\sigma_{k\li}}{x_{\jl}}
  \pd{\ln\theta}{x_k} + \Xi_{24}^{(\eta)} \frac{\mu}{\rho^2}
  \frac{\partial^2
    \rho}{\partial x_k \partial x_k} \sigma_{ij} \\
  & + \Xi_{25}^{(\eta)}\frac{\mu}{\rho^2} \frac{\partial^2
    \rho}{\partial x_k \partial x_{\li}} \sigma_{\jl k} +
  \Xi_{26}^{(\eta)}\frac{\mu}{\rho} \pd{\sigma_{ij}}{x_k}
  \pd{\ln\rho}{x_k} +
  \Xi_{27}^{(\eta)}\frac{\mu}{\rho}\pd{\sigma_{k\li}}{x_k}\pd{\ln\rho}{x_{\jl}}
  +
  \Xi_{28}^{(\eta)}\frac{\mu}{\rho} \pd{\sigma_{k\li}}{x_{\jl}} \pd{\ln\rho}{x_k} \\
  & + \Xi_{29}^{(\eta)} \frac{\mu}{\rho } \frac{\partial^2
    \sigma_{ij}}{\partial x_k \partial x_k} +
  \Xi_{30}^{(\eta)}\frac{\mu}{\rho } \frac{\partial^2
    \sigma_{k\li}}{\partial x_{\jl} \partial x_k}+ \Xi_{31}^{(\eta)}\mu
  \pd{v_{\li}}{x_k} \pd{v_{\jl}}{x_k} + \Xi_{32}^{(\eta)}
  \mu\pd{v_k}{x_{\li}} \pd{v_k}{x_{\jl}} + \Xi_{33}^{(\eta)}\mu
  \pd{v_{\li}}{x_k} \pd{v_k}{x_{\jl}}
  \\
  & + \Xi_{34}^{(\eta)} \mu\pd{v_k}{x_k} \pd{v_{\li}}{x_{\jl}} +
  \Xi_{35}^{(\eta)} \rho \theta \pd{v_{\li}}{x_{\jl}} +
  \Xi_{36}^{(\eta)}\mu\theta \pd{\ln\theta}{x_{\li}}
  \pd{\ln\theta}{x_{\jl}} + \Xi_{37}^{(\eta)}\mu\theta
  \pd{\ln\rho}{x_{\li}} \pd{\ln\rho}{x_{\jl}} +
  \Xi_{38}^{(\eta)}\mu\theta
  \pd{\ln\rho}{x_{\li}} \pd{\ln\theta}{x_{\jl}} \\
  & + \Xi_{39}^{(\eta)} \frac{\mu}{\rho \theta} \pd{v_k}{x_k}
  \pd{\ln\theta}{x_{\li}}q_{\jl} + \Xi_{40}^{(\eta)} \frac{\mu}{\rho
    \theta}\pd{v_{\li}}{x_k} \pd{\ln\theta}{x_{\jl}} q_k +
  \Xi_{41}^{(\eta)} \frac{\mu}{\rho \theta} \pd{\ln\theta}{x_k}
  \pd{v_{\li}}{x_k}q_{\jl} + \Xi_{42}^{(\eta)} \frac{\mu}{\rho
    \theta}\pd{\ln\theta}{x_k}
  \pd{v_k}{x_{\li}} q_{\jl} \\
  & + \Xi_{43}^{(\eta)}\frac{\mu}{\rho \theta} \pd{\ln\theta}{x_{\li}}
  \pd{v_k}{x_{\jl}}q_k + \Xi_{44}^{(\eta)}\frac{\mu}{\rho \theta}
  \pd{\ln\theta}{x_k} \pd{v_{\li}}{x_{\jl}} q_k + \Xi_{45}^{(\eta)}
  \frac{\mu}{\rho \theta} \pd{v_k}{x_k} \pd{\ln\rho}{x_{\li}}q_{\jl} +
  \Xi_{46}^{(\eta)} \frac{\mu}{\rho
    \theta}\pd{v_{\li}}{x_k} \pd{\ln\rho}{x_{\jl}}q_k  \\
  & + \Xi_{47}^{(\eta)}\frac{\mu}{\rho \theta} \pd{\ln\rho}{x_k}
  \pd{v_{\li}}{x_k}q_{\jl} + \Xi_{48}^{(\eta)}\frac{\mu}{\rho
    \theta}\pd{\ln\rho}{x_k} \pd{v_k}{x_{\li}}q_{\jl} +
  \Xi_{49}^{(\eta)}\frac{\mu}{\rho \theta} \pd{v_{\li}}{x_k}
  \pd{\ln\rho}{x_{\jl}} q_k + \Xi_{50}^{(\eta)}\frac{\mu}{\rho \theta}
  \pd{\ln\rho}{x_k} \pd{v_{\li}}{x_{\jl}}q_k \\
  & + \Xi_{51}^{(\eta)}\frac{\mu}{\rho \theta}\pd{q_{\li}}{x_k}
  \pd{v_{\jl}}{x_k} + \Xi_{52}^{(\eta)} \frac{\mu}{\rho
    \theta}\pd{q_{\li}}{x_k} \pd{v_k}{x_{\jl}} + \Xi_{53}^{(\eta)}
  \frac{\mu}{\rho \theta}\left( \pd{q_k}{x_{\li}} \pd{v_k}{x_{\jl}} +
    \pd{q_k}{x_{\li}} \pd{v_{\jl}}{x_k} \right) + \Xi_{54}^{(\eta)}
  \frac{\mu}{\rho \theta}\pd{q_k}{x_k}
  \pd{v_{\li}}{x_{\jl}} \\
  & + \Xi_{55}^{(\eta)} \frac{\mu}{\rho \theta}\pd{q_{\li}}{x_{\jl}}
  \pd{v_k}{x_k} + \Xi_{56}^{(\eta)} \frac{\mu}{\rho
    \theta}\frac{\partial^2 v_{\li}}{\partial x_k \partial x_k}q_{\jl}
  + \Xi_{57}^{(\eta)} \frac{\mu}{\rho \theta}\frac{\partial^2
    v_k}{\partial x_k \partial x_{\li}} q_{\jl} +
  \Xi_{58}^{(\eta)}\frac{\mu}{\rho \theta} \frac{\partial^2
    v_k}{\partial x_{\li} \partial x_{\jl}} q_k \\
  & + \Xi_{59}^{(\eta)} \frac{\mu}{\rho \theta}\frac{\partial^2
    v_{\li}}{\partial x_{\jl} \partial x_k} q_k + \Xi_{60}^{(\eta)}
  q_{\li} \pd{\ln\theta}{x_{\jl}} + \Xi_{61}^{(\eta)} \frac{\partial^2
    v_{\li}}{\partial x_{\jl} \partial x_k} q_k +
  \Xi_{62}^{(\eta)}\frac{ \mu \partial^2 \theta}{\partial
    x_{\li} \partial x_{\jl}} + \Xi_{63}^{(\eta)} \frac{\partial
    q_{\li}}{\partial x_{\jl}} +
  \Xi_{64}^{(\eta)}\frac{\mu\theta}{\rho}\frac{\partial^2 \rho}{\partial
    x_{\li} \partial x_{\jl}} \\
  & + \frac{1}{\rho \theta}\left(
    \Xi_{65}^{(\eta)}\sigma_{ij} \pd{q_{k}}{x_k} + \Xi_{66}^{(\eta)}\sigma_{ij}\pd{v_k}{x_l}\sigma_{kl}
  \right).
\end{split}
\end{equation}

\begin{equation} \label{eq:Q_2}
  \begin{split}
    Q_i^{(\eta, 2)} & = \Lambda_0^{(\eta)} \frac{\theta\rho}{\mu}q_i +
    \Lambda_1^{(\eta)}\sigma_{ik}\pd{\theta}{x_k} +
    \Lambda_2^{(\eta)}\frac{\mu}{\rho}q_i \pd{\ln \theta}{x_k}\pd{\ln
      \theta}{x_k} + \Lambda_3^{(\eta)}\frac{\mu}{\rho}q_k \pd{\ln
      \theta}{x_k}\pd{\ln \theta}{x_i} + \Lambda_4^{(\eta)}\theta
    \sigma_{ik}\pd{\ln \rho}{x_k} + \Lambda_5^{(\eta)} \frac{\mu}{\rho} q_k
    \pd{\ln
      \rho}{x_k}\pd{\ln \theta}{x_i} \\
    & + \Lambda_6^{(\eta)} \frac{\mu}{\rho}q_k \pd{\ln \rho}{x_i}\pd{\ln
      \theta}{x_k} + \Lambda_7^{(\eta)} \frac{\mu}{\rho}q_i \pd{\ln
      \rho}{x_k}\pd{\ln \theta}{x_k} + \Lambda_8^{(\eta)}
    \frac{\mu}{\rho}q_i \pd{\ln \rho}{x_k}\pd{\ln \rho}{x_k} +
    \Lambda_9^{(\eta)} \frac{\mu}{\rho}q_k \pd{\ln \rho}{x_k}\pd{\ln
      \rho}{x_i} + \Lambda_{10}^{(\eta)} \frac{\mu}{\rho}\pd{\ln
      \theta}{x_i}\pd{q_k}{x_k} \\
    & + \Lambda_{11}^{(\eta)} \frac{\mu}{\rho}\pd{\ln
      \theta}{x_k}\pd{q_i}{x_k} + \Lambda_{12}^{(\eta)}
    \frac{\mu}{\rho}\pd{\ln \theta}{x_k}\pd{q_k}{x_i} +
    \Lambda_{13}^{(\eta)} \frac{\mu}{\rho}\pd{\rho}{x_i}\pd{q_k}{x_k} +
    \Lambda_{14}^{(\eta)} \frac{\mu}{\rho}\pd{\rho}{x_k}\pd{q_i}{x_k} +
    \Lambda_{15}^{(\eta)}
    \frac{\mu}{\rho}\pd{\rho}{x_k}\pd{q_k}{x_i}  \\
    & + \Lambda_{16}^{(\eta)}q_k\left(\pd{v_k}{x_i} + \pd{v_i}{x_k}\right) +
    \Lambda_{17}^{(\eta)}q_i\pd{v_k}{x_k} + \Lambda_{18}^{(\eta)}\mu
    \pd{\theta}{x_k}\pd{v_k}{x_i} + \Lambda_{19}^{(\eta)}\mu
    \pd{\theta}{x_k}\pd{v_i}{x_k} + \Lambda_{20}^{(\eta)}\mu
    \pd{\theta}{x_i}\pd{v_k}{x_k}  \\
    & + \Lambda_{21}^{(\eta)}\frac{\mu}{\rho}\sigma_{kl}\left(\pd{\ln
        \theta}{x_k}\pd{v_i}{x_l} +\pd{\ln \theta}{x_l}\pd{v_i}{x_k} +
      \pd{\ln \theta}{x_k}\pd{v_k}{x_i} \right) +
    \Lambda_{22}^{(\eta)}\frac{\mu}{\rho}\sigma_{kl}\left(\pd{\ln
        \theta}{x_i}\pd{v_l}{x_k} +\pd{\ln
        \theta}{x_i}\pd{v_k}{x_l}\right) \\
    & + \Lambda_{23}^{(\eta)}\frac{\mu}{\rho}\sigma_{il}\pd{\ln
      \theta}{x_k}\pd{v_l}{x_k} +
    \Lambda_{24}^{(\eta)}\frac{\mu}{\rho}\sigma_{il}\pd{\ln
      \theta}{x_k}\pd{v_k}{x_l} +
    \Lambda_{25}^{(\eta)}\frac{\mu}{\rho}\sigma_{il}\pd{\ln
      \theta}{x_l}\pd{v_k}{x_k} + \Lambda_{26}^{(\eta)}\mu\theta \pd{\ln
      \rho}{x_i}\pd{v_k}{x_k} \\
    & + \Lambda_{27}^{(\eta)}\mu\theta\left(\pd{\ln \rho}{x_k}\pd{v_i}{x_k}
      +\pd{\ln \rho}{x_k}\pd{v_k}{x_i} \right) + \Lambda_{28}^{(\eta)}
    \frac{\mu}{\rho} \sigma_{kl}\left(\pd{\ln \rho}{x_k}\pd{v_i}{x_l}
      +\pd{\ln \rho}{x_l}\pd{v_i}{x_k} + \pd{\ln
        \rho}{x_k}\pd{v_l}{x_i} +
      \pd{\ln \rho}{x_l}\pd{v_k}{x_i}   \right)\\
    & + \Lambda_{29}^{(\eta)} \frac{\mu}{\rho} \sigma_{kl}\left(\pd{\ln
        \rho}{x_i}\pd{v_l}{x_k} +\pd{\ln
        \rho}{x_i}\pd{v_k}{x_l}\right) + \Lambda_{30}^{(\eta)}
    \frac{\mu}{\rho} \sigma_{il}\pd{\ln \rho}{x_k}\pd{v_l}{x_k} +
    \Lambda_{31}^{(\eta)} \frac{\mu}{\rho} \sigma_{il}\pd{\ln
      \rho}{x_k}\pd{v_k}{x_l} + \Lambda_{32}^{(\eta)} \frac{\mu}{\rho}
    \sigma_{il}\pd{\ln \rho}{x_l}\pd{v_k}{x_k} \\
    & +
    \Lambda_{33}^{(\eta)}\frac{\mu}{\rho\theta}q_i\pd{v_k}{x_k}\pd{v_l}{x_l}
    +
    \Lambda_{34}^{(\eta)}\frac{\mu}{\rho\theta}q_i\pd{v_k}{x_l}\pd{v_k}{x_l}
    +\Lambda_{35}^{(\eta)}\frac{\mu}{\rho\theta}q_i\pd{v_k}{x_l}\pd{v_l}{x_k}
    + \Lambda_{36}^{(\eta)}\frac{\mu}{\rho\theta}q_k\left(\pd{v_k}{x_i}
      \pd{v_l}{x_l}+
      \pd{v_i}{x_k}\pd{v_l}{x_l} \right) \\
    & +
    \Lambda_{37}^{(\eta)}\frac{\mu}{\rho\theta}q_k\pd{v_k}{x_l}\pd{v_i}{x_l}
    + \Lambda_{38}^{(\eta)}\frac{\mu}{\rho\theta}q_k
    \left(\pd{v_k}{x_l}\pd{v_l}{x_i} + \pd{v_l}{x_k}\pd{v_i}{x_l} +
      \pd{v_l}{x_k}\pd{v_l}{x_i}\right) + \Lambda_{39}^{(\eta)}
    \frac{\mu}{\rho}\pd{v_k}{x_k}\pd{\sigma_{il}}{x_l} +
    \Lambda_{40}^{(\eta)}
    \frac{\mu}{\rho}\pd{v_l}{x_k}\pd{\sigma_{il}}{x_k} \\
    & + \Lambda_{41}^{(\eta)}
    \frac{\mu}{\rho}\pd{v_k}{x_l}\pd{\sigma_{il}}{x_k} +
    \Lambda_{42}^{(\eta)}
    \frac{\mu}{\rho}\left(\pd{v_k}{x_l}\pd{\sigma_{kl}}{x_i} +
      \pd{v_l}{x_k}\pd{\sigma_{kl}}{x_i}\right) + \Lambda_{43}^{(\eta)}
    \frac{\mu}{\rho}\left(\pd{v_i}{x_k}\pd{\sigma_{kl}}{x_l}
      +\pd{v_k}{x_i}\pd{\sigma_{kl}}{x_l}\right) + \Lambda_{44}^{(\eta)}
    \frac{\mu}{\rho}\sigma_{il}\frac{ \partial ^2v_k}{\partial
      x_k \partial x_l}\\
    & + \Lambda_{45}^{(\eta)} \frac{\mu}{\rho}\sigma_{il}\frac{ \partial
      ^2v_l}{\partial x_k \partial x_k} + \Lambda_{46}^{(\eta)}
    \frac{\mu}{\rho}\sigma_{kl}\frac{ \partial ^2v_i}{\partial
      x_k \partial x_l} + \Lambda_{47}^{(\eta)}
    \frac{\mu}{\rho}\sigma_{kl}\frac{ \partial ^2v_k}{\partial
      x_i \partial x_l} + \Lambda_{48}^{(\eta)}\theta \pd{\sigma_{ik}}{x_k}
    + \Lambda_{49}^{(\eta)} \frac{\mu}{\theta \rho}q_k\frac{\partial ^2
      \theta}{\partial
      x_k \partial x_i} \\
    & + \Lambda_{50}^{(\eta)} \frac{\mu}{\theta \rho}q_i\frac{\partial ^2
      \theta}{\partial x_k \partial x_k} +
    \Lambda_{51}^{(\eta)}\frac{\mu}{\rho^2} q_k \frac{\partial^2
      \rho}{\partial x_k \partial x_i} +
    \Lambda_{52}^{(\eta)}\frac{\mu}{\rho^2} q_i \frac{\partial^2
      \rho}{\partial x_k \partial x_k} +
    \Lambda_{53}^{(\eta)}\frac{\mu}{\rho} \frac{\partial^2 q_i}{\partial
      x_k \partial x_k} + \Lambda_{54}^{(\eta)}\frac{\mu}{\rho}
    \frac{\partial^2 q_k}{\partial x_i \partial x_k} \\
    & + \Lambda_{55}^{(\eta)}\mu \theta\frac{\partial^2 v_i}{\partial
      x_k\partial x_k} + \Lambda_{56}^{(\eta)}\mu \theta\frac{\partial^2
      v_k}{\partial x_i\partial x_k} + \Lambda_{57}^{(\eta)}\theta\rho
    \pd{\theta}{x_i}
    +\frac{1}{\rho}\Lambda_{58}^{(\eta)}\sigma_{ik}\pd{\sigma_{lk}}{x_l} +
    \frac{q_i}{\theta\rho}\left( \Lambda_{59}^{(\eta)}\pd{q_k}{x_k} +
      \Lambda_{60}^{(\eta)}\sigma_{lk}\pd{v_l}{x_k} \right).
    \end{split}
\end{equation}
}

\begin{table}[ht!]
  \centering
   \def\arraystrech{1.5}
  {\footnotesize
    \begin{tabular}[ht]{r|r|r|r|r|r|r|r|r|r}
      $\eta = 7$ &  & &  & &  &  & &  & \\ 
      \hline    
      $0.0013$ & $0.0342$ & $-0.0228$ & $-0.0198$ & $-0.0147$ & $0.0096$ & $0.0664$ & $0.0367$ & $-0.0162$ &
                                                                                           $-0.0151$ \\
      $0.0092$ & $0.0101$ & $-0.0023$ & $-0.0100 $& $-0.0125$ & $0.0377$ & $-0.0155$ & $-0.0526$ & $-0.0603$ &
                                                                                             $-0.0721$ \\
      $0.0435$ & $-0.0973$ & $0.0018$ & $-0.0249$ & $-0.0110$ &$ 0.0347$ & $0.0378$ & $-0.0450$ &$ -0.0077$ &
                                                                                            $-0.0120$ \\
      $0.0337$ & $0.0317$ & $0.0344$ & $0.0687$ & $-0.0910$ & $0.0026$ & $0.1587$ &$ -0.0026$ & $-0.0002$ &
                                                                                          $0.0162$ \\
      $0.0065$ & $-0.0586$ & $-0.0252$ & $0.0065$ & $-0.1879$ & $-0.0121$ & $-0.0109$ & $0.1117$ & $0.0124$ &
                                                                                            $-0.0109$ \\
      $0.0968$ & $-0.0927$ & $0.0053$ & $0.0105$ & $-0.1034$ & $0.0003$ & $-0.0500$ & $0.0237$ & $0.0042$ &
                                                                                          $0.0032$\\
      $0.0158$ & $-0.0007$ & $0.1214$ & $0.0326$ & $0.0026$ & $0.0009$ & $0.0009$ & $\times$ &  $\times$& $\times$ \\
      \hline  \hline 
      $\eta = 10$ &  & &  & &  &  & &  &\\ 
      \hline 
       $0.0037$ & $0.0566$ & $-0.0377$ & $-0.0303$ & $-0.0072$ & $0.0212$ & $0.0981$ & $0.0477$ & $-0.0218$ & $-0.0227$ \\
      $0.0217$ & $0.0139$ & $-0.0071$ & $-0.0074$ & $-0.0140$ & $0.0456$ & $-0.0227$ & $-0.0810$ & $-0.0948$ &
                                                                                            $ -0.1173 $\\
      $0.0764$ & $-0.1527$ & $-0.0139$ & $-0.0315$ & $-0.0201$ & $0.0554$ & $0.0603$ & $-0.0773$ & $-0.0154$ &
                                                                                             $-0.0184$ \\
      $0.0579$ & $0.0497$ &$ 0.0571$ & $0.1142$ & $-0.1512$ & $0.0074$ &$ 0.2231$ & $-0.0074$ & $-0.0012$ &
                                                                                          $0.0219 $\\
      $0.0080$ & $-0.0634$ & $-0.0380$ & $0.0080$ & $-0.2779$ & $-0.0088$ & $-0.0198$ &$ 0.1825$ &$ 0.0143$ &
                                                                                            $-0.0198$ \\
      $0.1530$ & $-0.1523$ & $0.0125$ & $0.0171$ & $-0.1582$ & $-0.0075$ & $-0.0808$ & $0.0306$ & $0.0061$ &
                                                                                           $0.0076$     \\
      $0.0216$ & $-0.0019$ & $0.2029$ & $0.0543$ & $0.0074$ & $0.0026$ & $0.0026$ & $\times$ & $\times$ & $\times$ \\
      \hline  \hline
      $\eta = 17$ &  & &  & &  &   &  & &\\ 
      \hline
      $0.0068$ & $0.0761$ & $-0.0507$ & $-0.0403$ &$ 0.0093$ & $0.0358$ & $0.1187$ & $0.0493$ &$ -0.0236$ &
                                                                                           $-0.0278$ \\
      $0.0368$ & $0.0155$ & $-0.0135$ & $0.0009$ &$ -0.0111$ & $0.0417$ & $-0.0271$ & $-0.1022$ & $-0.1221$ &
                                                                                             $-0.1564 $\\
      $0.1076$ & $-0.1969 $& $-0.0385$ & $-0.0315$ &$ -0.0294$ & $0.0722$ & $0.0786$ & $-0.1081$ & $-0.0239$ &
                                                                                             $ -0.0231$ \\
      $0.0808$ & $0.0635$ & $0.0774$ &$ 0.1548$ & $-0.2052$ & $0.0137$ & $0.2536$ & $-0.0139$ & $-0.0032$ &
                                                                                           $0.0276$ \\
      $0.0071$ & $-0.0462 $& $-0.0488$ &$ 0.0071$ & $-0.3379$ & $0.0019$ & $-0.0290$ & $0.2436$ & $0.0116$ &
                                                                                            $-0.0290$ \\
      $0.1978$ & $-0.2042$ &$ 0.0211 $& $0.0227$ & $-0.1987$ & $-0.0198$ & $-0.1069$ & $0.0304$ & $0.0073$ &
                                                                                            $0.0130$ \\
      $0.0234$ & $-0.0036$ & $0.2760$ & $0.0736$ & $0.0139$ & $0.0050$ & $0.0050$ & $\times$ & $\times$
                                      & $\times$ \\
     \hline \hline
      $\eta = \infty$ &  & &  & &  &  & & &\\ 
      \hline    
      $0.0124$ & $0.1016$ & $-0.0677$ & $-0.0574$ & $0.0433$ & $0.0628$ & $0.1364$ & $0.0399$ & $-0.0213$ &
                                                                                           $-0.0322$\\
      $0.0627$ & $0.0152$ & $-0.0250$ & $0.0202$ & $-0.0015$ & $0.0213$ & $-0.0305$ & $-0.1253$ & $-0.1542$ &
                                                                                             $-0.2079$ \\
      $0.1511$ & $-0.2502$ & $-0.0861$ & $-0.0239$ & $-0.0438$ & $0.0927$ & $0.1006$ & $-0.1524$ & $-0.0372$ &
                                                                                              $-0.0280 $\\
     $ 0.1134 $& $0.0788$ & $0.1048$ & $0.2096$ & $-0.2794$ &$0.0252$ & $0.2561$ & $-0.0260$ & $-0.0086$ &
                                                                                            $0.0414 $\\
      $0.0025$ & $0.0058$ & $-0.0647$ & $0.0025$ & $-0.3929$ & $0.0277$ & $-0.0435$ & $0.3228$ & $0.0020$ &
                                                                                           $-0.0435 $\\
      $ 0.2502$ & $-0.2727$ & $0.0353$ & $0.0302$ & $-0.2419$ & $-0.0437$ & $-0.1405$ & $0.0203$ & $0.0084$&
                                                                                             $0.0220$ \\
      $0.0208$ & $-0.0067$ &$ 0.3748$ & $0.0988$ & $0.0260$ & $0.0096$ & $0.0096$ & $\times$ & $\times$
                                      & $\times$ \\
    \end{tabular}
  }
  \caption{Coefficients $\Xi_{i}^{(\eta)}$ in \eqref{eq:S_2} for different $\eta$.}
  \label{tab:S2}
\end{table}

\begin{table}[ht!]
  \centering
   \def\arraystrech{1.5}
  {\footnotesize
    \begin{tabular}[ht]{r|r|r|r|r|r|r|r|r|r|r}
      $\eta = 7$ &  & &  & &  &  & &  & &\\ 
      \hline     
      $0.0008$ & $0.1037$ & $-0.1049$ & $-0.3554$ & $0.0014$ & $0.3359$ & $0.6875$ & $0.2591$ & $-0.0778$ &
                                                                                          $-0.4526 $& $-0.2974$ \\
      $-0.2381$ & $-0.3980$ & $0.1920$ & $0.1068$ & $0.1569$ & $0.0292$ & $-0.0194$ & $0.2124$ & $0.2032$ &
                                                                                          $-0.1406$ & $-0.0080$ \\
      $-0.1119$ & $-0.1658$ & $0.0497$ & $0.0015$ & $-0.0017$ & $0.0026$ & $-0.0005$ & $0.0762$ & $0.0694$ &
                                                                                           $0.0705$ & $-0.0462$ \\
      $0.0447$ & $-0.0373$ & $-0.0667$ & $-0.0105$ & $-0.0732$ & $0.0153$ & $0.0116$ & $-0.0420$ & $-0.0420$ &
                                                                                             $-0.0415$ & $0.0246$ \\
      $0.0803$ & $-0.0237$ & $-0.0238$ & $-0.0885$ & $-0.0003$ & $-0.3693$ & $-0.1784$ &$ 0.2070$ &$ 0.0048$ &
                                                                                             $-0.0340$ & $-0.1100$ \\
      $-0.0015$ &$ 0.0026$ &$ 0.0031$ &$ -0.0012$ &$ 0.0009$ &$ 0.0009$ & $\times$ & $\times$ &
                                                                      $\times$ & $\times$ & $\times$ \\ 
      \hline \hline
      $\eta = 10$ &  & &  & &  &  & &   & &  \\ 
      \hline 
      $0.0023$ &$ 0.1666$ &$ -0.1119$ &$ -0.4234$ &$ 0.0039$ &$ 0.5006$ &$ 1.0827$ &$ 0.3893$ &$ -0.1230$ &
                                                                                                            $-0.7539$ &$ -0.4336$ \\
      $-0.3671$ &$ -0.6176$ &$ 0.3035$ &$ 0.1712$ &$ 0.2586$ &$ 0.0483$ &$ -0.0322$ &$ 0.3484$ & $ 0.3220$ &
                                                                                          $-0.2303$ & $-0.0089$ \\
      $-0.1619$ &$ -0.2606$ &$ 0.0956$ &$ -0.0254$ &$ -0.0049$ &$ 0.0074$ &$ -0.0021$ &$ 0.1181$ &$ 0.1099$ &
                                                                                                              $0.1130$ &$ -0.0723$ \\
      $0.0779$ &$ -0.0600$ &$ -0.1094$ &$ -0.0257$ &$ -0.1209$ &$ 0.0275$ &$ 0.0156$ &$ -0.0636$ &$ -0.0637$ &
                                                                                                               $-0.0627$ &$ 0.0402$ \\
      $0.1322 $& $-0.0361$ &$ -0.0362$ &$ -0.1341$ &$ -0.0008$ &$ -0.5668$ &$ -0.2818$ &$ 0.3422$ &$ 0.0037$ &
                                                                                                               $-0.0527$ &$ -0.1698$ \\
      $-0.0039$ &$ 0.0075$ & $0.0087$ & $-0.0035$ & $0.0025$ & $0.0025$ & $\times$ & $\times$ & $\times$ & $\times$
               & $\times$ \\
      \hline \hline
       $\eta = 17$ &  & &  & &  &  & &  & &\\ 
      \hline 
      $0.0042$ &$ 0.2193$ &$ -0.0840$ &$ -0.3948$ &$ 0.0073$ &$ 0.6105$ &$ 1.3974$ &$ 0.4799$ &$ -0.1590$ &
                                                                                                            $-1.0247$ &$ -0.5195$ \\
      $-0.4656$ &$ -0.7863$ & $0.3925$ &$ 0.2244$ &$ 0.3483$ &$ 0.0651$ &$ -0.0434$ &$ 0.4683$ &$ 0.4188$ &
                                                                                                            $-0.3101$ &$ -0.0072$ \\
      $-0.1928 $& $-0.3370$ &$ 0.1438$ &$ -0.0662$ &$ -0.0093$ &$ 0.0140$ &$ -0.0045$ &$ 0.1497$ &$ 0.1423$ &
                                                                                                              $0.1483$ &$ -0.0927$ \\
      $0.1099$ & $-0.0791$ &$ -0.1468$ &$ -0.0446$ &$ -0.1632$ &$ 0.0398$ &$ 0.0169$ &$ -0.0787$ &$ -0.0790$ &
                                                                                                               $-0.0776$ & $0.0536$ \\
      $0.1779$ & $-0.0450$ &$ -0.0452$ &$ -0.1665$ &$ -0.0012$ &$ -0.7144$ &$ -0.3660$ &$ 0.4618$ &$ -0.0001$ &
                                                                                                                $-0.0667$ &$ -0.2146$ \\
      $-0.0067$ &$ 0.0143$ &$ 0.0161$ &$ -0.0065$ &$ 0.0048$ &$ 0.0048$ & $\times$ & $\times$ & $\times$ & $\times$
                                        & $\times$ \\
     \hline  \hline 
      $\eta = \infty$ &  & &  & &  &  & &  & & \\ 
      \hline    
      $0.0077$ &$ 0.2861$ &$ -0.0063$ &$ -0.2555$ &$ 0.0139$ &$ 0.7069$ &$ 1.7691$ &$ 0.5681$ &$ -0.2012$ &
                                                                                                            $-1.3933$ &$ -0.5912$ \\
      $-0.5779 $& $-0.9772$ &$ 0.4972$ &$ 0.2897$ &$ 0.4675$ &$ 0.0871$ &$ -0.0581$ &$ 0.6297$ &$ 0.5374$ &
                                                                                                            $-0.4198$ &$ -0.0016$ \\
      $-0.2184$ &$ -0.4310$ &$ 0.2172$ &$ -0.1426$ &$ -0.0176$ &$ 0.0264$ &$ -0.0087$ &$ 0.1839$ &$ 0.1806$ &
                                                                                                              $0.1921$ &$ -0.1165$ \\
      $0.1566$ &$ -0.1037$ &$ -0.1963$ &$ -0.0773$ &$ -0.2199$ &$ 0.0580$ &$ 0.0154$ &$ -0.0933$ & $-0.0940$ &
                                                                                                               $-0.0920$ &$ 0.0709$ \\
      $0.2392$ &$ -0.0539$ &$ -0.0543$ &$ -0.1986$ &$ -0.0015$ &$ -0.8759$ &$ -0.4699$ &$ 0.6216$ &$ -0.0097$ &
                                                                                                                $-0.0820$ &$ -0.2630$ \\
      $-0.0111$ &$ 0.0271$ &$ 0.0295$ &$ -0.0120$ &$ 0.0092$ &$ 0.0092$ & $\times$ & $\times$ & $\times$ & $\times$
             & $\times$ \\
\end{tabular}
  }
  \caption{Coefficients $\Lambda_{i}^{(\eta)}$ in \eqref{eq:Q_2} for different $\eta$.}
  \label{tab:Q2}
\end{table}

\bibliographystyle{plain}
\bibliography{article}

\begin{thebibliography}{10}

\bibitem{Arfken}
G.~B. Arfken, H.~J. Weber, and F.~E. Harris.
\newblock {\em Mathematical Methods for Physicists (Seventh Edition)}.
\newblock Academic Press, 2013.

\bibitem{Bayin}
S.~Bayin.
\newblock {\em Mathematical Methods in Science and Engineering, Second
  Edition}.
\newblock Wiley, 2018.

\bibitem{Bird}
G.~A. Bird.
\newblock {\em Molecular Gas Dynamics and the Direct Simulation of Gas Flows}.
\newblock Oxford: Clarendon Press, 1994.

\bibitem{Bobylev1982}
A.~V. Bobylev.
\newblock The {C}hapman-{E}nskog and {G}rad methods for solving the {B}oltzmann
  equation.
\newblock {\em Sov. Phys. Dokl.}, 27:29--31, 1982.

\bibitem{Bobylev2006}
A.~V. Bobylev.
\newblock Instabilities in the {C}hapman-{E}nskog expansion and hyperbolic
  {B}urnett equations.
\newblock {\em J. Stat. Phys.}, 124(2--4):371--399, 2006.

\bibitem{Bobylev2008}
A.~V. Bobylev.
\newblock Generalized {B}urnett hydrodynamics.
\newblock {\em J. Stat. Phys.}, 132:569--580, 2008.

\bibitem{Burnett1936}
D.~Burnett.
\newblock The distribution of molecular velocities and the mean motion in a
  non-uniform gas.
\newblock {\em Proc. London Math. Soc.}, 40(1):382--435, 1936.

\bibitem{Grad13toR13}
Z.~Cai, Y.~Fan, and R.~Li.
\newblock On hyperbolicity of 13-moment system.
\newblock {\em Kin. Rel. Models}, 7(3):415--432, 2014.

\bibitem{Cai2015a}
Z.~Cai, Y.~Fan, and R.~Li.
\newblock A framework on moment model reduction for kinetic equation.
\newblock {\em SIAM J. Appl. Math.}, 75(5):2001--2023, 2015.

\bibitem{Cai2012n}
Z.~Cai, R.~Li, and Y.~Wang.
\newblock Numerical regularized moment method for high {M}ach number flow.
\newblock {\em Commun. Comput. Phys.}, 11(5):1415--1438, 2012.

\bibitem{Cai2015}
Z.~Cai and M.~Torrilhon.
\newblock Approximation of the linearized {B}oltzmann collision operator for
  hard-sphere and inverse-power-law models.
\newblock {\em J. Comput. Phys.}, 295:617--643, 2015.

\bibitem{Cai2018N}
Z.~Cai and M.~Torrilhon.
\newblock Numerical simulation of microflows using moment methods with
  linearized collision operator.
\newblock {\em J. Sci. Comput.}, 74(1):336--374, 2018.

\bibitem{Chapman}
S.~Chapman.
\newblock On the law of distribution of molecular velocities, and on the theory
  of viscosity and thermal conduction, in a non-uniform simple monatomic gas.
\newblock {\em Phil. Trans. R. Soc. A}, 216(538--548):279--348, 1916.

\bibitem{Cowling}
S.~Chapman and T.~G. Cowling.
\newblock {\em The Mathematical Theory of Non-uniform Gases, Third Edition}.
\newblock Cambridge University Press, 1990.

\bibitem{Dimarco2018}
G.~Dimarco, R.~Loub{\'e}re, J.~Narski, and T.~Rey.
\newblock An efficient numerical method for solving the {B}oltzmann equation in
  multidimensions.
\newblock {\em J. Comput. Phys.}, 353:46--81, 2018.

\bibitem{Dreyer}
W.~Dreyer.
\newblock Maximisation of the entropy in non-equilibrium.
\newblock {\em J. Phys. A: Math. Gen.}, 20(18):6505--6517, 1987.

\bibitem{Enskog}
D.~Enskog.
\newblock The numerical calculation of phenomena in fairly dense gases.
\newblock {\em Arkiv Mat. Astr. Fys.}, 16(1):1--60, 1921.

\bibitem{Grad1949}
H.~Grad.
\newblock On the kinetic theory of rarefied gases.
\newblock {\em Comm. Pure Appl. Math.}, 2(4):331--407, 1949.

\bibitem{Grad1958}
H.~Grad.
\newblock Principles of the kinetic theory of gases.
\newblock {\em Handbuch der Physik}, 12:205--294, 1958.

\bibitem{Gupta2012}
V.~K. Gupta and M.~Torrilhon.
\newblock Automated {B}oltzmann collision integrals for moment equations.
\newblock {\em AIP Conference Proceedings}, 1501(1):67--74, 2012.

\bibitem{Harris}
S.~Harris.
\newblock {\em An Introduction to the Theory of the Boltzmann equation}.
\newblock Dover Publications, 1971.

\bibitem{Hu2019}
Z.~Hu and Z.~Cai.
\newblock Burnett spectral method for high-speed rarefied gas flows, 2019.
\newblock In preparation.

\bibitem{Jin2001}
S.~Jin and M.~Slemrod.
\newblock Regularization of the {B}urnett equations via relaxation.
\newblock {\em J. Stat. Phys.}, 103(5--6):1009--1033, 2001.

\bibitem{Kumar1966}
K.~Kumar.
\newblock Polynomial expansions in kinetic theory of gases.
\newblock {\em Ann. Phys.}, 37(1):113--141, 1966.

\bibitem{Kumar}
K.~Kumar.
\newblock Polynomial expansions in kinetic theory of gases.
\newblock {\em Ann. Phys.}, 37:113--141, 1966.

\bibitem{Levermore}
C.~D. Levermore.
\newblock Moment closure hierarchies for kinetic theories.
\newblock {\em J. Stat. Phys.}, 83(5--6):1021--1065, 1996.

\bibitem{McDonald2013}
J.~McDonald and M.~Torrilhon.
\newblock Affordable robust moment closures for {CFD} based on the
  maximum-entropy hierarchy.
\newblock {\em J. Comput. Phys.}, 251:500--523, 2013.

\bibitem{Meyer1957}
E.~Meyer and G.~Sessler.
\newblock Schallausbreitung in {G}asen bei hohen {F}requenzen und sehr
  niedrigen {D}rucken.
\newblock {\em Z. Phys.}, 149(1):15--39, 1957.

\bibitem{Mouhot2007}
C.~Mouhot and R.~M. Strain.
\newblock Spectral gap and coercivity estimates for linearized {B}oltzmann
  collision operators without angular cutoff.
\newblock {\em J. Math. Pures Appl.}, 87(5):515--535, 2007.

\bibitem{Muller}
I.~M{\"u}ller and T.~Ruggeri.
\newblock {\em Rational Extended Thermodynamics, Second Edition}, volume~37 of
  {\em Springer tracts in natural philosophy}.
\newblock Springer-Verlag, New York, 1998.

\bibitem{Myong1999}
R.~S. Myong.
\newblock Thermodynamically consistent hydrodynamic computational models for
  high-knudsen-number gas flows.
\newblock {\em Phys. Fluids}, 11(9):2788--2802, 1999.

\bibitem{Reinecke}
S.~Reinecke and G.~M. Kremer.
\newblock Method of moments of {G}rad.
\newblock {\em Phys. Rev. A}, 42(2):815--820, 1990.

\bibitem{Shavaliyev}
M.~Sh. Shavaliyev.
\newblock Super-{B}urnett corrections to the stress tensor and the heat flux in
  a gax of {M}axwellian molecules.
\newblock {\em J. Appl. Mahts. Mechs.}, 57(3):573--576, 1993.

\bibitem{Struchtrup2005}
H.~Struchtrup.
\newblock Derivation of 13 moment equations for rarefied gas flow to second
  order accuracy for arbitrary interaction potentials.
\newblock {\em Multiscale Model. Simul.}, 3(1):221--243, 2005.

\bibitem{Struchtrup}
H.~Struchtrup.
\newblock {\em Macroscopic Transport Equations for Rarefied Gas Flows:
  Approximation Methods in Kinetic Theory}.
\newblock Springer, 2005.

\bibitem{Struchtrup2003}
H.~Struchtrup and M.~Torrilhon.
\newblock Regularization of {G}rad's 13 moment equations: Derivation and linear
  analysis.
\newblock {\em Phys. Fluids}, 15(9):2668--2680, 2003.

\bibitem{Struchtrup2013}
H.~Struchtrup and M.~Torrilhon.
\newblock Regularized 13 moment equations for hard sphere molecules: Linear
  bulk equations.
\newblock {\em Phys. Fluids}, 25:052001, 2013.

\bibitem{Struchtrup2007}
H.~Struchtrup and M.~Torrilhon.
\newblock H-theorem, regularization, and boundary conditions for linearized 13
  moment equations.
\newblock {\em Phys. Rev. Lett.}, 99:014502, 2017.

\bibitem{Tallec}
P.~L. Tallec and J.~P. Perlat.
\newblock Numerical analysis of {L}evermore's moment system.
\newblock Rapport de recherche 3124, INRIA Rocquencourt, March 1997.

\bibitem{Timokhin2017}
M.~Yu. Timokhin, H.~Struchtrup, A.~A. Kokhanchik, and Ye.~A. Bondar.
\newblock Different variants of {R}13 moment equations applied to the
  shock-wave structure.
\newblock {\em Phys. Fluids}, 29:037105, 2017.

\bibitem{Torrens}
I.~M. Torrens.
\newblock {\em Interatomic Potentials}.
\newblock Academic Press, 1972.

\bibitem{Torrilhon2006}
M.~Torrilhon.
\newblock Two dimensional bulk microflow simulations based on regularized
  {G}rad's 13-moment equations.
\newblock {\em SIAM Multiscale. Model. Simul.}, 5(3):695--728, 2006.

\bibitem{Torrilhon2012}
M.~Torrilhon.
\newblock H-theorem for nonlinear regularized 13-moment equations in kinetic
  gas theory.
\newblock {\em Kin. Rel. Models}, 5(1):185--201, 2012.

\bibitem{Torrilhon2004}
M.~Torrilhon and H.~Struchtrup.
\newblock Regularized 13-moment equations: shock structure calculations and
  comparison to {B}urnett models.
\newblock {\em J. Fluid Mech.}, 513:171--198, 2004.

\bibitem{Torrilhon2008}
M.~Torrilhon and H.~Struchtrup.
\newblock Boundary conditions for regularized 13-moment-equations for
  micro-channel-flows.
\newblock {\em J. Comput. Phys.}, 227(3):1982--2011, 2008.

\bibitem{Valentini}
P.~Valentini and T.~E. Schwartzentruber.
\newblock Large-scale molecular dynamics simulations of normal shock waves in
  dilute argon.
\newblock {\em Phys. Fluids}, 21(6):066101, 2009.

\bibitem{Vincenti1966Introduction}
W.~Vincenti, C.~Kruger, and T.~Teichmann.
\newblock Introduction to physical gas dynamics.
\newblock {\em Phys. Today}, 19(10):95--95, 1966.

\end{thebibliography}
\end{document}